\documentclass[12pt,preprint]{aastex}

\shorttitle{COSMIC ACCELERATION FROM CAUSAL BACKREACTION}
\shortauthors{BOCHNER}

\begin{document}

\title{COSMIC ACCELERATION FROM CAUSAL BACKREACTION 
	   IN A SMOOTHLY INHOMOGENEOUS UNIVERSE}

\author{Brett Bochner}
\affil{Department of Physics and Astronomy, 
Hofstra University, Hempstead, NY 11549}
\email{brett\_bochner@alum.mit.edu, phybdb@hofstra.edu}

\begin{abstract}
A phenomenological formalism is presented in which the apparent 
acceleration of the universe is generated by large-scale 
structure formation, thus eliminating the coincidence and 
magnitude fine-tuning problems of the Cosmological Constant 
in the Concordance Model, as well as potential instability 
issues with dynamical Dark Energy. The observed acceleration 
results from the combined effect of innumerable local 
perturbations, due to individually virialized systems, 
overlapping together in a smoothly-inhomogeneous adjustment 
of the FRW metric, in a process governed by the causal flow 
of inhomogeneity information outward from each clumped system. 
We discuss several arguments from the literature claiming to 
place sharp limits upon the strength of backreaction-related 
effects, and show why such arguments are not applicable in a 
physically realistic cosmological analysis. A selection of 
simply-parameterized models are presented, including several 
which are capable of fitting the luminosity distance data from 
Type Ia supernovae essentially as well as the best-fit flat 
$\Lambda$CDM model, without resort to Dark Energy, any 
modification to gravity, or a local void. Simultaneously, 
these models can reproduce measured cosmological parameters 
such as the age of the universe, the matter density required 
for spatial flatness, the present-day deceleration parameter, 
and the angular scale of the Cosmic Microwave Background to 
within a reasonable proximity of their Concordance values. 
We conclude by considering potential observational signatures 
for distinguishing this cosmological formalism from 
$\Lambda$CDM or Dark Energy, as well as the possible 
long-term fate of such a universe with ever-spreading spheres 
of influence for its increasingly superposed perturbations. 
\end{abstract}


\section{\label{SecIntroMotiv}INTRODUCTION AND MOTIVATIONS} 

During the past decade or so, a number of complementary 
observational techniques have come together to point towards 
the existence of an exotic cosmic ingredient known as 
Dark Energy (DE). Luminosity distance measurements from Type Ia 
supernovae \citep{PerlAccel99,RiessAccel98} bear evidence of 
an apparent cosmic acceleration, which seems to indicate the 
existence of a substance capable of violating the strong energy 
condition \citep{HawkingEllis}, and perhaps even the 
dominant energy condition \citep{CaldOrigPhant,CaldPhantom}. 
Observations of the Cosmic Microwave Background (CMB) indicate 
a spatially flat universe \citep{WMAP5yrCosmInterp}; 
yet various lines of evidence regarding large-scale structure 
formation and other measurements \citep[e.g.,][]{TurnerCaseOm0pt33} 
indicate a total dynamical matter content of only 
$\Omega _\mathrm{M} \sim 0.3$, far less than the value of unity 
needed for flatness. The standard conclusion from this, barring 
modifications to gravity, is that there must exist a smooth 
(non-clumping) component to the inventory of the universe 
supplying the remaining $\Omega _\mathrm{DE} \sim 0.7$, 
which is capable of accelerating the universe without joining 
in the process of localized structure formation. 

Within the current resolution of the data, combinations of 
observations are still consistent \citep[e.g.,][]{WMAP5yrCosmInterp} 
with the simplest Dark Energy model: flat $\Lambda$CDM with a 
Cosmological Constant (vacuum energy) as the Dark Energy `substance'. 
A nonzero $\Lambda$ density would clearly avoid participating in 
spatial clustering, while also possessing an equation of state, 
$w \equiv w_{0} = -1$, that would begin to drive the acceleration 
once $\Omega _\Lambda$ had become the dominant cosmic component.

There are well-known aesthetic problems with $\Lambda$ 
as the missing piece to the puzzle, however. This includes a 
fine-tuning problem in which one must explain why $\rho _{\Lambda}$ 
is some $\sim$$120$ orders of magnitude smaller than what would 
naively be expected from the Planck scale \citep[e.g.,][]{KolbTurner}. 
And perhaps even more relevant to issues of cosmic evolution, 
there is the ``Coincidence Problem" \citep[e.g.,][]{ArkaniHamedCoinc}, 
which questions why, given that 
$\rho _{\Lambda}/ \rho _{\mathrm{M}} \propto a^{3}$, that 
observers {\it today} should just happen to live right around the 
time (within a factor of $\sim$$2$ in scale factor) when 
$\Lambda$ began dominating the cosmic evolution. After all, if 
$\rho _{\Lambda}$ had been just a little smaller, then cosmologists 
in this epoch would have had no hope of detecting it; and if 
$\rho _{\Lambda}$ had been just a little larger, then the acceleration 
would have prevented structure formation from ever having occurred, 
and no observers would exist to measure it at all. 

Without delving deeply into anthropic issues, one possible way of 
avoiding such problems is to give the Dark Energy an evolution in 
$w(z)$, where such coincidences are eliminated or moderated; an example 
being tracker quintessence solutions \citep{ZlatevQuint,AlbrechtQuint}, 
where there are natural reasons for Dark Energy to `pop up' at a 
propitious, recent (relative to the Planck scale) time -- such as being 
triggered by a phase transition like the onset of matter domination, 
for example. Moving away from a Cosmological Constant by giving the 
Dark Energy an evolving equation of state, however, also opens the 
door to other possible dynamics -- such as the ability for the 
Dark Energy to be mobile and become spatially inhomogeneous 
\citep{CaldDDEsNotSmooth}. A dynamical Dark Energy (DDE) of this type 
is liable to join in structure formation to an observationally 
unacceptable degree, thus ruining it as a candidate to serve as the 
smoothly-distributed missing cosmic ingredient. 

An under-appreciated aspect of the notion of Dark Energy is what it 
really means when something is a `negative pressure' substance. 
Thinking of the acceleration of the cosmic expansion, it is common 
to refer to the DE with words like ``repulsive", ``antigravity", 
and so on \citep[e.g.,][]{HeavensRepuls}, as if negative pressure 
substances were materials that {\it naturally} did not possess 
attractive forces or tend to clump on local scales, thus making them 
natural candidates for being a smooth cosmic component. But this is 
the exact opposite of the truth: a simple look at the $1^{\mathrm{st}}$ 
Law of Thermodynamics, 
$P = - \partial [\mathrm{Energy}] / \partial [\mathrm{Volume}]$, 
shows that for a negative pressure ($P < 0$) substance, an increase 
in system volume requires an increase in its energy content 
($\partial E / \partial V > 0$). Consequently, since the environment 
must do work upon a local patch of DE to make it expand (and gets work 
back when letting it contract), a material with negative pressure 
should be {\it self-attractive}, not self-repulsive. Thus a
true negative pressure species which is mobile (i.e., not $\Lambda$), 
able to move around and clump due to its own self-attractive 
forces, would naturally be expected to not only {\it fail} to remain 
smooth on galaxy cluster scales, but due to its extreme 
($w \equiv P/ \rho \sim -1$) negative pressure, it would clump 
{\it relativistically}, making it the most strongly clumping material 
in the universe! 

Being self-attractive would seem to be a counterintuitive property to 
have for a substance that is invoked in order to make the universe 
increase its expansion rate and thereby spread out faster, but 
in fact it is due to a sign peculiarity in the FLRW acceleration 
equation \citep[e.g.,][]{KolbTurner}: 
\begin{equation}
\frac{\ddot{a}}{a} = - \frac{4 \pi G}{3} (\rho + 3 P) ~ ,
\label{EqnFRWAccel}
\end{equation}
where $a(t)$ is the cosmic scale factor. From the negative sign on 
the right-hand side of this equation, we see that negative pressure 
does in fact cause a cosmic acceleration, not a deceleration; and thus 
we have the seemingly paradoxical (but true) result that a force which 
attracts {\it locally}, repels {\it globally}. This qualitative view 
is supported by the work of \citet{MaorLahavVirial} and 
\citet{MaorSphCollVirial}, where they found that for a two-component 
system made up of matter and dark energy, the virialization radius will 
increase when the Dark Energy is allowed to dynamically participate in 
the virialization, a sign of an additional attractive force (due to the 
increased virial kinetic energies occurring in a deeper potential well). 
One implication of this is that signs of a cosmologically recent slow-down 
in the growth of clustering \citep[e.g.,][]{VikhDEevo}, findings which are 
viewed as strong evidence in favor of Dark Energy, really only support 
a Cosmological Constant form of DE, not any more general DDE material 
with locally-attractive negative pressure. It therefore seems that a 
negative pressure substance like dynamical Dark Energy would be exactly 
the wrong prescription when looking for a cosmic component that can 
remain smoothly distributed in order to avoid enhancing the small-scale 
clustering of material -- thus indicating an essential conflict between 
the dual requirements of smoothness and acceleration for just about 
anything other than the immobile, coincidence-prone $\Lambda$. 

This potential instability of DDE to spatial perturbations 
is well known by experts in the field 
\citep[e.g.,][]{TurnerWhiteSmooth,HuGDM98,HuDEint05}; 
and various forms of support pressure may be proposed in order to subvert 
the natural behavior of negative pressure substances towards small-scale 
clumping, thus allowing a DDE to remain smooth on scales large enough to 
evade limits on $\Omega _\mathrm{M}$ from observations of large-scale 
structure formation. But beating the laws of thermodynamics is a 
tricky business to be in, and often one errs by implicitly introducing 
a source of {\it positive} pressure somewhere in the problem -- perhaps 
within the Dark Energy itself, such as by considering relativistic 
particles (with their ordinary, thermodynamic positive pressure) as the 
DDE \citep{TurnerWhiteSmooth}; or by using the (positive) degeneracy 
pressure of long-wavelength particles like neutrinos to keep them 
unclumped \citep[e.g.,][]{WigmansDegen}. 
Or alternatively, positive pressure can be generated by an 
auxiliary material invoked in combination with the DDE to keep the 
latter from clustering \citep[e.g.,][]{BjSchremppDEstability}. 
Either way, a positive pressure source sufficient for supporting the 
DDE against collapse must invariably contribute a {\it decelerating} 
contribution to the cosmic expansion (cf. Equation~\ref{EqnFRWAccel}) 
that counteracts the accelerative properties for which the Dark Energy 
was recruited in the first place. 

Even without any implicit, possibly inadvertent introduction of a 
positive (adiabatic) pressure source, it could still be argued that 
some form of {\it nonadiabatic} pressure within the DDE might succeed 
at preventing its small-scale clustering without ruining the 
actual cosmic acceleration. For a scalar field, it is well known 
\citep{GrishNEGwPOScs,CaldDDEsNotSmooth} that the sound speed 
is scale-dependent; and despite being imaginary (thus unstable 
against collapse) with $c^{2}_{s} < 0$ for long-wavelength 
perturbations of a $w < 0$ scalar field, for small wavelengths 
the sound speed instead approaches $c^{2}_{s} \simeq c^{2}$, 
thus seemingly solving the instability problem automatically when 
considering small-scale clustering. But the situation is not 
quite so simple, however, when considering the detailed processes 
of inhomogeneous structure formation. For example, in a study modeling 
the spatial evolution of scalar field DDE in the presence of a 
spherically-collapsing matter overdensity, \citet{MaorDuttaDEvoids} 
find that the initial response of the DDE is to join the matter in 
entering into a collapsing phase, thus adding to the growing central 
overdensity. This collapsing phase does not last long, since the 
slower Hubble expansion in the central matter overdensity allows the 
scalar field to locally roll down its potential and lose energy, 
actually creating a weak 
($\delta \rho _{\phi} / \rho _{\phi} \sim -0.02$) {\it void} of 
DDE there, thus appearing once again to solve the small-scale DDE 
clustering problem. But recalling the $1^{\mathrm{st}}$ Law of 
Thermodynamics ($P = - \partial E / \partial V$), a material 
that loses energy when the universe expands is more of a 
positive pressure substance than a negative pressure substance; 
and sure enough, \citet{MaorDuttaDEvoids} find that the 
effective equation of state $w$ of their DDE does become substantially 
less negative as the system evolves, particularly in the center where 
the DDE void is developing. Thus we see that the less a DDE engages 
in small-scale clustering, the less it behaves like a negative-pressure 
substance capable of accelerating the universe; and it is not unreasonable 
to suppose that these two properties are fundamentally connected in a 
thermodynamically essential way. Now, we should not overstate the case 
that can be made from those results, since the scalar fields 
examined in \citet{MaorDuttaDEvoids} never actually lost their 
accelerative nature (i.e., $w \geq -1/3$ is never reached): the equation 
of state only increases from $w \equiv -1$ (everywhere) at $z = 35$ to 
$w \simeq -0.8$ (at the center) at $z = 0$ for their simplest scalar field 
potential (and increases only to $w \simeq -0.945$ for their more 
theoretically-motivated double exponential potential). But then again, 
their numerical analysis only studied the system within the regime of 
linear perturbations; and it would be interesting to see what happens to 
the averaged equation of state of the total cosmic contents (matter plus 
DDE) -- in order to evaluate the actual ability of such a cosmic mixture to 
effectively accelerate the universe -- when a significant fraction of the 
DDE is located within regions of highly nonlinearly collapsing matter 
overdensities. (While \citet{MotaSilkNonlinDEclust} do perform a analysis 
that is fully nonlinear for the matter perturbations, they do not evaluate 
any effective changes to $w$ which would occur as a result. Furthermore, the 
assumption made in both of these papers of a very small mass scale for the 
DDE scalar field, $m_{\phi} \sim O(H_{0})$, may inadvertently be recruiting 
degeneracy or relativistic pressure within the DDE itself to help keep it 
unclumped on small scales; this might explain the seemingly odd result in 
\citet{MotaSilkNonlinDEclust} that the DDE cannot clump onto pre-existing 
overdensities by itself, but instead stops increasing its density contrast 
once the nonlinear matter overdensity pulling the DDE in stabilizes its own 
collapse through virialization of the matter alone.) 

And even beyond the question of changes to the effective equation of state, 
$w$, there is a more subtle issue involved here as well, which we pose 
as the following question: if the {\it clustering} properties 
of a substance with nonadiabatic pressure do not solely depend upon $w$, 
then why would its {\it accelerative} properties depend solely upon $w$? 
In the textbook derivation \citep[e.g.,][]{KolbTurner} of the acceleration 
equation, Equation~\ref{EqnFRWAccel}, one proceeds by first assuming a 
perfect-fluid form for all cosmic constituents; but a DDE with its 
inhomogeneities controlled by nonadiabatic pressure would clearly fail 
to behave as a perfect fluid, placing the very simple form of the FLRW 
acceleration equation itself into question (as well as muddying the 
physical meaning of the equation of state parameter, $w \equiv P / \rho$). 
If one cannot trust the DDE to behave as a perfect fluid in one sense, 
then there is less reason to trust it to behave as a perfect fluid in 
any other sense; and thus it seems quite reasonable to believe that a 
dynamical substance sufficiently exotic as to avoid clumping despite 
$P < 0$, might {\it also} fail to generate $\ddot{a} > 0$ despite 
$P < - \rho / 3$. Rather than simply quoting a sufficiently negative 
value of $w$ as demonstrating a claim that some non-clustering DDE is 
capable of accelerating the universe, a fully general-relativistic 
treatment of all of the inhomogeneous-DDE-induced accelerative and/or 
decelerative effects within a fair cosmological sample (clusters and 
voids) would need to be done to properly substantiate that claim. 

While the foregoing discussion does not constitute actual proof 
against the possibility of achieving acceleration with a negative 
pressure, mobile substance that somehow fails to clump, it should 
at least introduce a degree of skepticism into an idea that has 
generally been accepted at face value. And if these concerns about 
coincidences and negative-pressure instabilities do in fact cause 
one to abandon both $\Lambda$ and DDE, then explaining the cosmic 
acceleration would seem to present a conundrum. Fortunately, 
however, the self-attractive nature of negative pressure presents 
us with the seed of an alternative idea: in its own way, 
{\it normal gravitational attraction} represents a form of negative 
pressure -- specifically, the very same gravitational attraction which 
is involved in the growth of galaxies and galaxy clusters in large-scale 
structure formation. One might therefore make a virtue of necessity, 
and try to recruit structure formation itself as the driver (not just 
the trigger, as in some tracker quintessence models) of the cosmic 
acceleration. Such a step would completely solve the Coincidence 
Problem, since the remarkably contemporary onsets of both the 
acceleration and the existence of galaxies (and hence planets, life, 
and astronomical observers) could be viewed as nothing more 
coincidental than finding two neighboring apples that have fallen 
from the same tree. 

It was the realization of the paradoxically accelerative yet locally 
attractive nature of $P < 0$ for Dark Energy that initially led this 
author to search for a structure-formation-induced solution to the 
cosmic acceleration problem \citep{Bochner21Texas}. During the past 
few years, a number of other researchers have also proposed a variety 
of methods \citep[e.g.,][as a few 
examples]{SchwarzFriedFail,RasanenFriedFail,KMNRinhomogExp,WiltInhomogNoDE} 
of trying to obtain the observed cosmic acceleration from the breaking of 
FLRW homogeneity and the formation of structure, without requiring the 
intervention of non-standard gravitational or particle physics, or the 
non-Copernican approach of placing us near the center of a large cosmic 
void. This type of effort has generally become known as ``backreaction". 
The backreaction paradigm, in the consensus view, has so far been unable 
to completely replace Dark Energy as the source of the apparent 
acceleration (we will discuss some of the reasons for this belief 
in the next section); and the arguments which this author originally 
presented in \citet{Bochner21Texas} were advanced without the benefit 
of a plausible mechanism for realistically generating an acceleration. 

In this paper, however, we will introduce a physically plausible 
mechanism for producing an observed acceleration; and though our 
method here is strictly phenomenological, it serves as the basis 
for a formalism that we can use to not only reproduce the cosmic 
acceleration (as indicated by observations of standard candles 
such as Type Ia supernovae), but also, simultaneously, to reproduce 
several of the other key features of the apparent $\Lambda$CDM 
concordance. 

This paper will be organized as follows: in 
Section~\ref{SecTheorCounters}, we discuss several of 
the oft-quoted limitations of the backreaction mechanism, 
countering them, and then use such arguments as guideposts 
towards developing our formalism in Section~\ref{SecFormalism}; 
in Section~\ref{SecResults}, we describe our specific models 
and present our simulated Hubble curves; in 
Section~\ref{SecNewConcord}, we show how to extend these 
results to the formulation of a new ``Cosmic Concordance''; 
in Section~\ref{SecTesting}, we evaluate the success of this 
alternative concordance, and then suggest potential methods for 
observationally testing our paradigm in order to distinguish it 
from $\Lambda$CDM (and perhaps also from DDE); in 
Section~\ref{SecFate}, we discuss the possible long-term futures 
(`fates') of a universe dominated by our ``causal backreaction'' 
mechanism; and we conclude with a summary of these ideas and 
results in Section~\ref{SecConclude}.

\section{\label{SecTheorCounters}VARIETIES OF ``BACKREACTION" 
AND THEIR COSMOLOGICAL IMPACT} 

There are several distinct ways of achieving an observation 
that looks like a cosmic acceleration, and one may categorize 
the different approaches or scenarios in various different ways 
\citep[see, for example,][]{BisManNotAppAccel,BisNotLTBneglig}. 
Here we will find it useful to characterize the various forms of 
observed acceleration according to the $(q_{1},q_{2},q_{3},q_{4})$ 
terminology from \citet{HirataSeljak} -- where $q_{1}$ refers 
to an `actual' acceleration as described by an increase in the 
volume expansion $\theta$; $q_{2}$ refers to an `averaged' 
acceleration for inhomogeneous spacetimes, an effect strictly 
due to nonlinearities in general relativity (GR) when one 
has departures from Newtonian metric perturbations 
\citep[i.e., the ``fitting problem",][]{EllisFit, EllisStoegerFit}; 
and $q_{3}$, $q_{4}$ refer to an {\it apparent} acceleration, 
due to the limitations that one always experiences in the act of 
observation, attempting to infer the general cosmic evolutionary 
behavior from a circumscribed set of luminosity distance measurements. 
(The difference between $q_{3}$ and $q_{4}$ has to do with how 
they perform angle-averaging for an anisotropic universe -- for 
example, regarding the possibility of finding `acceleration' 
in some directions, and `deceleration' in others.) 

In general, $q_{3}$ and $q_{4}$ may be strongly dependent 
upon the observer's position in space, in relation to 
inhomogeneities, voids, etc., as well as being affected by 
anisotropic observations. For our formalism, however, which we 
will be describing as ``smoothly inhomogeneous", the observed 
acceleration will depend neither upon the position of the observer 
nor the direction in which one looks, but will instead depend solely 
upon the {\it time} of the observation. Technically speaking, only 
$q_{2}$ -- which relies directly upon the nonlinear character 
of GR -- was originally termed ``backreaction", although common  
usage has generally broadened the term to mean any form of observed 
or apparent acceleration from cosmic inhomogeneity. In this paper, 
we will use the term in its broadest sense, to mean any or all of 
$(q_{1},q_{2},q_{3},q_{4})$. In the full complexity of GR for an 
inhomogeneous universe, it is not always a simple matter to 
disentangle these different terms in a clear-cut fashion; but our 
formalism presented here is likely best described as being the 
ultimate result of an averaged $q_{1}$-type volume creation, perhaps 
in combination with $q_{3} / q_{4}$-type effects due to the fact 
that gravitational perturbation information coming in from far away 
-- and thus from earlier look-back times, when the universe was less 
clustered -- actually allows the late-time/small-distance cosmic 
expansionary behavior to be appear quite different from what we see 
from far away, but only in a purely observational sense (i.e., 
not due to any real local differences), and in a way that {\it any} 
observer would see for their own local cosmic neighborhood. 

At this present moment in the general theoretical consensus, 
backreaction is not widely considered to be a viable method of 
achieving the observed acceleration 
\citep[e.g.,][]{SchwarzBackReactNotYet}. There are a number of 
important reasons for this, which seem on the surface to be 
strong arguments; but we will explain here why despite being 
technically correct, several of those arguments do not properly 
apply to a physically realistic cosmological model.

\subsection{\label{Sub1NoHolesBarred}Smoothly-Inhomogeneous 
Backreaction: No Voids, Holes, or Special Boundary Conditions} 

One simple way of achieving an apparent acceleration via 
FRW-violating inhomogeneities, is to imagine ourselves located 
within an underdense void \citep[e.g.,][]{TomitaSNeVoid,
MoffatCMBAccelInhomog,AlAmGronVoidAccel}. Such a void, 
expanding at a faster rate than the average cosmic expansion 
overall, would imprint Hubble curves with what looks like the 
signs of a recent cosmic acceleration, as incoming light signals 
from standard candles cross over from slower-expanding regions 
farther away from us, to the faster-expanding regions inside our 
local void. Such a void solution has difficulties, however -- 
including a coincidence problem of its own, regarding our fairly 
suspicious (if random) position very close to the center of the 
void, as would be needed to maximize `acceleration' effects while 
minimizing the resulting anisotropies imposed upon the observed 
CMB radiation; as well as the necessary existence of a 
substantially large and deep void that would be capable of 
achieving sufficient apparent acceleration 
\citep{AlexBisNotVaidVoid}. The existence of such a void would 
eventually become detectable using luminosity distance data 
over intermediate redshifts \citep{ClifFerrLandVoidTest}; and 
though it has been argued that a variety of cosmological 
observations remained compatible with the possibility of such a 
large local void \citep{GarBelHaugVoidOK}, recent Type Ia supernova 
data is consistent with a determination that no so-called 
``Hubble bubble" exists \citep{ConstitutionSNeCosm}, a rejection 
of the void hypothesis further substantiated by new Cepheid data 
used for calibrating such supernovae \citep{RiessCephNoVoid}. 

Perhaps fortunately, therefore, our formalism does not require the 
existence of a special local void centered upon us, but instead 
uses the overall, averaged effects of {\it many} inhomogeneities, 
all summed together, in order to achieve the desired observational 
result. But the question necessarily arises as to how to represent 
a situation with multiple inhomogeneities via an appropriate 
cosmological model. One answer to this difficulty that has been 
studied in detail is a class of models known as ``Swiss-Cheese" 
solutions, in which one starts with a homogeneous FRW background 
(the ``cheese"), and carves out spherical regions (the ``holes") 
in which matter can be distributed inhomogeneously -- though still 
spherically symmetrically -- in such a way as to represent many 
independent clumped structures of matter. One can use any 
spherically symmetric (dust-only) metric that one wishes for 
modeling the holes, such as the static (fully collapsed) 
Schwarzschild metric, as used in the original Einstein-Straus 
solution \citep{EinStraus}; or the radially-evolving 
LTB metrics \citep{LeMaitreLTB,TolmanLTB,BondiLTB}, as used in 
various recent studies \citep[e.g.,][]{BisNotLTBneglig}. The key 
here is that by retaining spherical symmetry -- thus recruiting 
Birkhoff's Theorem \citep[e.g.,][]{WeinbergGravCosmo} -- and by 
carefully matching the mass-energy content of the hole to the 
exterior cheese at the hole boundary, one aims to completely zero 
out the effects of the existence of the hole on its surroundings. 
Thus the exterior regions continue to expand exactly as they would 
in an entirely homogeneous FRW cosmology, as if no holes existed, 
while giving one the ability to produce exact solutions for 
detailed study which do include deviations from homogeneity. 
Since each hole is presumed to have no effect on anything exterior 
to itself, one can include any number or variety of holes (as long 
as they do not intersect one another), even to the extent of filling 
the universe {\it entirely} with holes (and thus with essentially 
no cheese anywhere) in a fractal-like pattern, such as the 
``Apollonian gasket" \citep{KolbMarraSwissLTB}. Such a model would 
have the same average expansion as a FRW-everywhere model, despite 
actually being FRW virtually nowhere.

Despite being mathematically convenient for study, however, such 
models have serious disadvantages. One key shortcoming is that 
since they obey FRW dynamics on average, and since the holes have 
no effect on the exterior cheese, there is by definition zero 
actual ``backreaction": all effects causing apparent deviations 
from matter-dominated FRW expansion are purely observational, 
due to effects upon the photons passing through the holes. 
Obviously, this will inhibit the ability 
of the inhomogeneities to mimic a cosmic acceleration; and 
\citet{BisNotLTBneglig} find that such effects are indeed strongly 
suppressed, by a factor $(L/R_{H})^{3}$ for passage through a hole, 
where $L$ is the size of the hole and $R_{H}$ is the Hubble radius. 
Even if one integrates the paths of light rays through many holes, 
this is clearly too small of an effect to replace the Dark Energy. 

Such suppression of backreaction effects is entirely artificial, 
however, since the Swiss-Cheese models themselves are artificial 
-- they are, in fact, entirely inadequate for plausible studies of 
the real universe, since they lack so many key physical features. 
First, the spherical symmetry of the holes makes them unable to 
incorporate the virialization and stabilization of forming structures 
(a point noted by \citet{BisNotLTBneglig}, and which will take on 
more importance below). Second, the prescription of taking all 
effects of the hole back upon the exterior universe and setting them 
to {\it zero} is entirely unphysical -- when condensed structures 
form in the real universe, rather than leaving exterior regions alone, 
they instead tend to become overdense attractors which keep growing in 
mass and perturbative cosmological influence; consider for example our 
acceleration towards the Great Attractor/Shapley Concentration 
\citep[e.g.,][]{LuceyAttractor,BolejkoAttractorCMB,SarkarUnion2Aniso}, 
and other evidence of large-scale bulk flows \citep{CosFlowHumeFeld}. 
Third, the notion of disjoint, non-intersecting holes -- a necessary 
assumption for producing exact metric solutions, requiring special and 
implausible boundary conditions at the hole/cheese interfaces -- is a 
fatal failure of the model, since in the real universe, the spheres of 
influence of individual inhomogeneities always continue to expand as 
they pull in more material (and have more time to spread their causal 
influences), until the separate perturbations eventually {\it overlap 
and merge}. The superposition of gravitational perturbations from many 
different inhomogeneities is a phenomenon that Swiss-Cheese models 
cannot represent at all. Simply speaking, if we recall the 
earlier suppressive factor of $(L/R_{H})^{3}$ -- and now consider that 
each hole (overlapping with all of the other holes) is about the size 
of the observational horizon out to which its gravitational pull can be 
causally felt -- then this factor is much closer to unity, and not much 
of a `suppression' after all. 

The challenge, of course, is how to represent such a model 
mathematically; since once the holes all merge together, it seems 
like we are back to the homogeneous FRW model again, in some averaged 
sense. The metric that one uses, in any case, cannot depend upon 
spatial position, since we are only considering averaged effects. 
But what we will justify below, in this section -- and present a 
phenomenological model for, in Section~\ref{SecFormalism} -- is that 
the perturbations can indeed be perpetually nontrivial functions of 
time (and of time alone); and the fact that we live merely in an 
averaged FRW universe (instead of in a truly homogeneous FRW universe) 
can be modeled as alterations to the behavior of the average scale 
factor, $a(t)$, in such a way that it looks like we live in a FLRW 
universe with matter and Dark Energy, rather than dust alone. This 
model, in which we re-establish the mathematical homogeneity and 
isotropy of the Cosmological Principle, while retaining the average, 
integrated effects of many overlapping inhomogeneities (incorporated 
as a perturbation to the Friedmann expansion), is what we refer to as 
a ``smoothly-inhomogeneous" universe. As will be shown in 
Section~\ref{SecResults}, these effects are capable (under realistic 
conditions) of reproducing the entire observed acceleration; and an 
important benefit of the smoothly-inhomogeneous model is that there 
is nothing special about our own position, and so {\it all} observers 
(barring strong local influences) would see the {\it same} apparent 
acceleration at the same cosmic time.

\subsection{\label{Sub1Vortex}Vorticity as a Large-Scale Player} 

During the evolution from `truly homogeneous' to 
`smoothly inhomogeneous', a strong phase transition takes 
place throughout the entire universe. The question is whether or 
not the effects of this phase transition would have large 
`accelerative' effects -- or as some have argued, whether or not 
it can have {\it any} accelerative effects. Sharp limits have 
been placed upon such mechanisms by arguments in the literature, 
and so we must explain here why those limits are not applicable 
to realistic situations. 

It has been argued, for example \citep{AlAmGronNoLTBAccel}, that 
cosmologies dominated by pressureless dust do not permit acceleration 
(i.e., true volumetric acceleration, in the sense of $q_{1}$). 
This conclusion was based upon a study of LTB inhomogeneity, 
which as we have noted is limited to spherical symmetry, and lacks 
the ability to model virialization; but, what is the true importance 
of virialization? Consider the $q_{1}$ acceleration ``no-go" theorem 
of \citet{HirataSeljak}, that was based upon the 
Raychaudhuri equation \citep{HawkingEllis}, which assuming 
perfect fluids can be written as: 
\begin{equation}
\frac{d\theta}{dt} = - \frac{\theta^{2}}{3} 
- \sigma_{\mu \nu} \sigma^{\mu \nu} 
- 4 \pi G (\rho + 3 P)
+ \omega_{\mu \nu} \omega^{\mu \nu} ~ ,
\label{EqnRaychaudPerfFluid}
\end{equation}
where $\theta$ is the volume expansion, $\sigma_{\mu \nu}$ is 
the shear tensor, $\omega_{\mu \nu}$ is the vorticity tensor, 
and $\rho$ and $P$ are, respectively, the density and the 
isotropic pressure of all material contained within the 
stress-energy tensor $T_{\mu \nu}$.
From this expression, it is obvious that if the vorticity is zero 
(or more precisely, if 
$\omega^{2} \equiv \omega_{\mu \nu} \omega^{\mu \nu} / 2 = 0$), 
and if the strong energy condition (SEC) is obeyed 
($\rho + 3 P \geq 0$) for all species, then clearly 
$d\theta / dt \leq 0$, and no acceleration is possible. On the 
other hand, any violation of the SEC due to negative pressure 
such that $\rho + 3 P \leq 0$ (i.e., $w \leq -1/3$), strong 
enough to lead to $d\theta / dt > 0$, would necessarily be 
classed as a form of Quintessence, Dark (or Phantom) Energy. 
Therefore, in a matter-dominated universe with negligible 
pressure ($P \approx 0$), with this dust being largely 
irrotational ($\omega^{2} \approx 0$), all of the remaining 
nonzero terms would provide only negative (decelerative) 
contributions to ensure that $d\theta / dt \leq 0$. 

Here, however, is one of the key questions of this paper: 
{\it Why is the vorticity set to zero?} This seems to be an 
especially unusual prescription in a universe where nearly 
everything is rotating and/or revolving about something. 
After all, all virialized structures in the universe larger than 
individual solid objects\footnote{Solid objects are supported by 
their internal pressure gradients and body forces, which also 
contribute (though likely on a cosmologically small level) 
to $d\theta / dt > 0$, but which have been dropped even before 
getting to Equation~\ref{EqnRaychaudPerfFluid} as a result of the 
perfect fluid approximation.} are stabilized against collapse by 
{\it some} version of vorticity or velocity dispersion. Henceforth 
referring to all varieties of such mechanisms as simply ``vorticity", 
this includes both the organized rotational motions of solar system 
planets and spiral galaxies, as well as the less organized orbital 
motions of stars in elliptical galaxies, and of individual galaxies 
in galaxy clusters. The virialization mechanism itself 
\citep[e.g., violent relaxation;][]{LyndenBellVioRelax,ShuVioRelax} 
achieves such stabilizations by generating or concentrating all of 
this vorticity. The value of $\omega^{2}$ within galaxies and 
galaxy clusters must obviously be large -- and in fact, physically 
dominant -- since it is clearly sufficient to counter the 
gravitational attraction of the matter (including Dark Matter) 
to produce static regions of space with $d\theta / dt = 0$. 

It is certainly a bad physical approximation to take the dominant 
physical force (in opposition to gravity) that regulates cosmic 
structure formation\footnote{All of this vorticity requires great 
amounts of motion -- i.e., kinetic energy. This energy comes from 
the gravitational potential energy of inhomogeneities originally 
pulled apart by the expanding universe. Thus the matter is `pumped' 
by the expansion; and this accumulation of energy taken from 
the expanding universe is how the gravitationally-self-attractive 
cosmic matter effectively behaves like a `negative pressure substance'.}, 
and set it equal to zero. We must therefore 
consider the reasons for which this is typically done. Mathematically, 
zero vorticity is necessary for the existence of the synchronous 
comoving gauge \citep[e.g.,][]{HirataSeljak}, such as is used for 
the homogeneous and isotropic FLRW cosmologies; and vorticity causes 
problems with the meshing together of the hypersurfaces of proper 
time \citep[e.g.,][]{RasanenAccelStruct}. Thus one cannot define 
an idealized ``Hubble flow" unless one sets the rotation to vanish; 
and so to use this common and useful simplification, it must be 
assumed that on large enough scales, one is justified in averaging 
any cosmological fluid to an irrotational, comoving state. In essence, 
therefore, incorporating vorticity explicitly would cause extreme 
complications for the entire mathematical machinery of how 
cosmological evolution and structure formation are usually studied 
in practice\footnote{There are of course a few analyses which do not 
immediately drop cosmological vorticity at the outset of calculations: 
e.g., \citet{EhlBuchNewtCosmFound}, \citet{KasaiVorticityUpInClumps}, 
\citet{ChristoPertVort}, and \citet{RasanenWithVort}, as a few 
examples. But this does not guarantee that vorticity is being 
included to the degree necessary for generating causal backreaction. 
For example, \citet{RasanenWithVort} finds the relevant contribution 
due to vorticity to be simply a total divergence, leading only 
to a negligible surface term; but they restrict their analysis 
to first order in velocity $v$ -- and as will be made clear below 
in Section~\ref{Sub1NewtBackReact}, an analysis at least up to 
$O(v^{2})$ is necessary in order to incorporate the causal flow of 
gravitational information.}. Such complications, barring the unlikely 
possibility of rotation on supercluster or even larger scales, 
would appear to be physically unnecessary to keep track of on a 
truly cosmological level. Thus vorticity has in many cases been 
relegated to lesser status as a ``small-scale player" 
\citep[e.g.,][]{BuchertDEstructStatus}, relevant only for 
cosmic averages performed over domains that are on or below 
the scales of galaxy clusters.

\begin{figure}
\begin{center}
\includegraphics[scale=0.75]{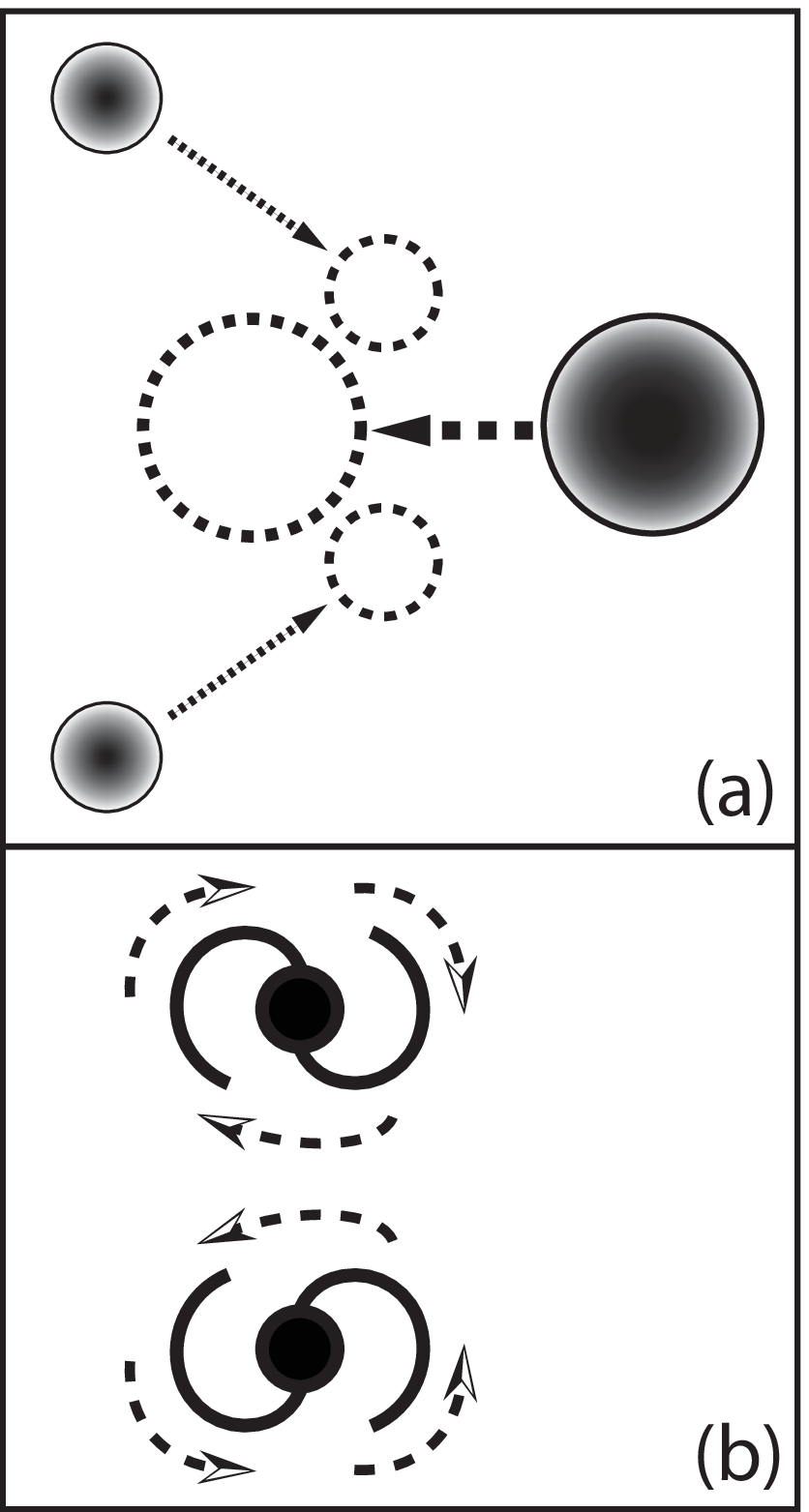}
\end{center}
\caption{Simplified example of a physically permissible 
(though inelastic and nonadiabatic) merger of three 
nonrotating objects (a) into two rapidly spinning 
objects (b). (Not shown: ejected material containing 
mass and energy but zero total angular momentum.) 
Here, the angular momentum averaged over the volume of the 
combined system, $\langle${\boldmath{$L$}}$\rangle$, 
remains $\approx$$0$, though $\langle L^2 \rangle$ goes 
from $\approx$$0$ to a large value.} 
\label{FigOppositeSpins}
\end{figure}

The problem with this approach is that it is not actually the 
vorticity tensor, $\omega_{\mu \nu}$, which appears in the 
Raychaudhuri equation as a positive contributor to $d\theta / dt$, 
but the {\it square} of the vorticity, $\omega^{2}$. And no matter 
what scale one goes up to in their averaging, the square of a 
quantity will {\it never} average away; in fact, the total 
amount of integrated $\omega^{2}$ in a volume will be strictly 
proportional to volume, if the clustering and stabilization of 
mass is essentially the same everywhere in space. As 
\citet{BuchertEhlers97} put it, cosmic averages of positive 
semi-definite quantities like $\langle \omega^{2} \rangle$ 
(or in opposition to it, $\langle \sigma^{2} \rangle$) get 
``frozen" at the size of ``typical subdomains"; and thus they 
cannot be made to go to zero by averaging over larger domains, 
even if the spatially averaged value of the parent tensor, 
$\langle \omega_{\mu \nu} \rangle$, itself becomes negligible on 
large scales due to collective cancellation from neighboring local 
subregions in the domain. Furthermore, the evolving cosmic value 
of $\langle \omega^{2} \rangle$ is not subject to any particular 
conservation law (since $\omega_{\mu \nu}$ throughout the early 
universe was not exactly zero), nor is $\omega^{2}$ limited from 
growing as large as it needs to be in any location in order to 
virialize a self-stabilizing region of clustered mass -- formally 
speaking, \citet{KasaiVorticityUpInClumps} find that vorticity is 
coupled to density contrast, becoming strongly amplified in locally 
collapsing regions, contrary to the expectations from standard linear 
perturbation theory. Systems demonstrating significant 
vorticity-dependent behavior should be quite natural in the real 
universe, as depicted generically in Figure~\ref{FigOppositeSpins} 
(with angular momentum and its squared value, {\boldmath{$L$}} and 
$L^2$ respectively, serving as a proxy for $\omega_{\mu \nu}$, 
$\omega^{2}$ here). We can even observe the results of this kind of 
behavior in our nearby cosmic neighborhood by looking at the 
Local Group, in which the Milky Way and Andromeda galaxies each 
support themselves against collapse via their own individual spins; 
and yet Andromeda and the Milky Way, like many pairs of large 
spiral galaxies, are counter-rotating with respect to one other 
\citep[][p.174]{SchaafMWAndromSpins}, thus partially 
canceling $\langle \omega_{\mu \nu} \rangle$ for the Local Group as 
a whole (though not necessarily implying that it is small or conserved 
per se -- e.g., \citet{DunnLafLocGrpAngMom}), without reducing its 
total integrated value of $\langle \omega^{2} \rangle$.

\placefigure{FigOppositeSpins}

Given the potential significance of vorticity in regards to 
both astrophysics and volume expansion, general warnings against 
neglecting it are given (multiple times in fact) by 
\citet{RasanenAccelStruct} (and by various other authors), in the 
context of acceleration via structure formation; but aside from 
such caveats, for typical studies in the literature the issue of 
vorticity is nearly always dropped before actual calculations are done, 
and models which completely lack the ability to represent vorticity 
or virialization (such as LTB) are usually used for the estimates 
of backreaction-related effects of inhomogeneities on the observed 
cosmic expansion. 

As \citet{HirataSeljak} note, the absence of vorticity plays 
a key role in all of the ``no-go" theorems used to prove the 
impossibility of accelerated expansion in the absence of some kind 
of SEC-violating Dark Energy. And therefore, due to the crucial 
importance of vorticity (and thus a substantially nonzero $\omega^{2}$) 
in structure formation for all of the virialized inhomogeneities 
pervading the universe, we can conclude that {\it all acceleration 
``no-go" theorems are cosmologically inapplicable.} A 
smoothly-inhomogeneous cosmology with no Dark Energy, but with the 
overall effects of pervasive virialization, can indeed (as we will 
see) be able to produce enough of a backreaction to create (at the 
very least) an apparent acceleration, if not a full-fledged 
volumetric (`real') acceleration. Vorticity and velocity dispersion 
effects are not in fact a small-scale player in the physical universe, 
but represent the major player (along with gravity), and the only 
question that remains is to figure out how to properly incorporate them. 

The difficulties involved in explicitly including vorticity within 
cosmological models are very real, however, and it would be a highly 
nontrivial problem to try to precisely compute the detailed time 
evolution of (and volume expansion from) a realistically virializing 
structure. Fortunately though, as we contend here, there is no need to 
solve this exact problem in order to get a good approximation of the 
averaged effects of backreaction upon the observed expansion rate. 
Instead, we can exploit the fact that both the beginning state (nearly 
perfectly smooth FRW matter) and the ending state (a reasonably random 
distribution of discrete Newtonian mass concentrations) are extremely 
simple, making the net results of the phase transition to inhomogeneity 
very straightforward to estimate. Only the general evolution of how much 
of the cosmic matter has completed the phase transition to clumpiness as 
a function of cosmic time is important, not the detailed time dependence 
of the virialization of any individual object; and for this paper we 
will use a set of reasonable heuristic models to empirically determine 
which ones are best suited for fitting the observational data. This 
phenomenological formalism will be motivated and presented below, 
in Section~\ref{SecFormalism}.

\subsection{\label{Sub1NewtBackReact}Newtonian-Level Backreaction: 
                   Not Suppressed, Not Small, and Not Slow} 

There remain debatable issues regarding the possible {\it magnitude} 
of such backreaction effects, however, given the fact that we 
continue to rely here upon (individually) Newtonian perturbations, 
and do not explicitly recruit the effects of nonlinear-strength 
gravitational fields. Powerful theoretical arguments have been 
advanced by previous researchers, claiming that cosmological 
backreaction from Newtonian-level perturbations must necessarily 
be very small, if not actually suppressed to {\it zero}, in a 
practical sense. Given that the development of Newtonian 
perturbation terms will provide positive contributions to the 
spatial metric components (to $g_{r r}$, in particular, for the 
spherically-symmetric case), it is unclear why new volume expansion 
is not considered an inevitable result of those newly-developing 
Newtonian perturbations; but it is this assumed restriction 
for backreaction, apparently requiring one to appeal to higher-order 
gravitational terms, which is what has forced researchers to resort 
to desperate measures in order to achieve a significant backreaction. 
This author's original attempt at backreaction \citep{Bochner21Texas}, 
for example, postulated a toy model utilizing black holes, in order 
to generate a strong enough effect to make a difference in the 
cosmic expansion (a clearly unworkable model for several reasons, 
but no alternative was obvious at the time). Using another 
exotic approach, other researchers have attempted to recruit 
super-Hubble-sized density perturbations -- first on their own 
\citep[e.g.,][]{KMNRinhomogExp,KMNRAccelFromInflat}, and then in 
tandem with sub-Hubble terms \citep{KMRCosAccWoDE} -- in order to 
create a usefully large inhomogeneity-induced backreaction, 
in spite of the causality-related difficulties \citep{HirataSeljak} 
of trying to produce physically measurable effects from perturbation 
modes larger than the observable horizon. \citet{KolbNoPertNewt} 
find it necessary to go to the extent of arguing that a perturbed 
conformal Newtonian metric is not even an appropriate description 
for a universe observationally well-described by a $\Lambda$CDM model, 
when considering deviations from an Einstein-de Sitter dust model as 
the unperturbed background. It is clear, therefore, that the 
permissibility of a Newtonian description of inhomogeneous 
perturbations is commonly viewed as an insurmountable impediment for 
backreaction. We must therefore examine the reasons for this belief, 
and show how it comes about from an unwarranted oversimplification 
of cosmological physics. 

From the extensive work of Buchert and collaborators 
\citep[e.g.,][]{BuchertEhlers97,BuchertKerschSicka}, it is well known 
that the entire Newtonian backreaction, $Q_{N}$, can be expressed 
mathematically as a total divergence: 
$Q_{N} \equiv Q_{Div} \equiv \nabla \cdot${\boldmath{$q_{N}$}}. 
We know from the divergence theorem \citep[e.g.,][]{JacksonEM} that 
for any such function integrated within a volume V (with boundary 
$\delta V$), we can relate the volume-integral of $Q_{N}$ to the 
boundary-normal surface integral of {\boldmath{$q_{N}$}}: 
$ \int_{V} Q_{N} 
= \oint_{\delta V}${\boldmath{$q_{N}$}}$\cdot${\boldmath{$n$}}. 
This surface integral will vanish if the universe 
is topologically closed (i.e., if there {\it is no} surface), or if 
periodic boundary conditions are assumed (as is typical for 
large-scale Newtonian cosmological simulations -- e.g., 
\citet{VIRGOsims}), meaning that the total integrated Newtonian 
backreaction ($\int_{V} Q_{N}$) would consequently have to vanish. 
But even if it does not completely vanish, \citet{KMRCosAccWoDE} 
argue that the overall effect will in any case be tiny, 
because the surface term involves the peculiar velocity 
{\boldmath{$u$}}, which should be small in Newtonian situations; 
and perhaps another limiting factor here is that an integral 
over a surface of radius $R$ will only increase as $\sim$$R^{2}$, so 
that the volume average will go like $\sim$$R^{2}/R^{3} \propto 1/R$, 
becoming negligibly small for arbitrarily large $R$. Thus the ability 
to represent Newtonian (i.e., gravitationally linear) backreaction as 
a total divergence, if valid, would act as a strong suppression of the 
magnitude of the effect. 

It is interesting, at this point, to consider the basic equations at 
the heart of the formalism used for placing these limits upon the 
Newtonian-level backreaction -- specifically, Equations 1a-1d in 
\citet{BuchertEhlers97} -- and to compare them with Maxwell's Equations 
(along with the Lorentz force law and continuity equation) for 
electromagnetism \citep[e.g.,][]{JacksonEM}. 
A careful examination of these expressions 
shows that for the former, something is missing: the magnetic fields. 
The Buchert formalism, and the various theorems derived from it, have 
all been done having set what one may call the ``gravitomagnetic" 
fields \citep[e.g.,][]{MashoonGEM} to zero. Now, the neglect of the 
`magnetic' fields in electrodynamics (or in Newtonian gravity) 
happens to be a very significant simplification, as is obvious from 
a quick examination of Amp$\grave{e}$re's Law, 
$\nabla \times${\boldmath{$B$}}$ = (4 \pi /c)${\boldmath{$J$}}$ 
+ (1/c) \partial${\boldmath{$E$}}$/ \partial t$. If {\boldmath{$B$}} 
is set to zero, then the time rate of change of {\boldmath{$E$}} in 
a region is solely dependent upon the local current/momentum flow, 
{\boldmath{$J$}}, inside that region. In other words (say, for 
gravitation), the contributions to the local Newtonian gravitational 
potential and fields by distant masses outside the local volume are 
{\it completely frozen-in}, with any possible causality-driven 
dynamics being entirely suspended. This semi-static formulation 
(i.e., considering only the motions of {\it local} masses) is 
entirely in accord with the common view \citep{BuchertKerschSicka} 
that: ``In the Newtonian approximation the expansion of a domain 
is influenced by the inhomogeneities inside the domain." The main 
thrust of our arguments in this paper is to point out that this 
view is unacceptably incomplete, and in fact we rejected the 
Swiss-Cheese approach earlier on this very same basis, that it is 
physically unrealistic to bar different regions of spacetime from 
communicating with one another. While such an approximation may be 
reasonable according to what researchers usually think of as a 
`Newtonian' approximation, it is certainly not compatible with 
the dictates of special relativity and cosmological causality. 

For an alternate way of expressing Buchert-like restrictions 
on backreaction, consider the arguments of \citet{IshWald}, 
in regards to the following, ``Newtonianly-Perturbed FLRW" metric: 
\begin{equation}
ds^2 = -(1 + 2 \Phi) dt^2 
+ a^{2}(t) (1 - 2 \Phi) \gamma_{ij} dx^{i} dx^{j} ~ . 
\label{EqnIshWaldNewtMetric}
\end{equation} 
In analyzing this metric, they adopt the various `Newtonian' 
conditions for the potential function, $\Phi$: 
\begin{mathletters}
\begin{eqnarray}
\vert \Phi \vert & \ll & 1 ~ , 
\label{EqnIshWaldLimitsA}
\\
\vert \partial \Phi / \partial t \vert ^{2} & \ll & 
\frac{1}{a^2} D^{i}\Phi D_{i}\Phi ~ , 
\label{EqnIshWaldLimitsB}
\\
(D^{i}\Phi D_{i}\Phi)^2 & \ll & (D^{i} D^{j} \Phi) D_{i} D_{j} \Phi ~ .
\label{EqnIshWaldLimitsC}
\end{eqnarray}
\label{EqnIshWaldLimitsTot}
\end{mathletters}
They point out that even in the regions of very large local density 
variations, $\delta \rho / \rho$, the metric perturbation $\Phi$ due 
to these spatially-varying matter concentrations should always remain 
small everywhere (with the exception of extreme regions, such as near 
black holes); and that given these conditions, the nonlinear 
corrections to the expansion will remain small. The linear effects 
are small by hypothesis, and \citet{IshWald} show that the use of 
Equations~\ref{EqnIshWaldLimitsTot}a-c implies that the remaining 
dominant backreaction term for $\Phi$ satisfies an expanding-universe 
version of the Poisson equation, basically reducing it to a total 
divergence and thus subject to Buchert suppression. 

Now, one may concede that spatial variations in $\Phi$ are indeed small 
as per the arguments of \citet{IshWald}; and if the temporal growth 
in $\Phi$ were also small due to Equation~\ref{EqnIshWaldLimitsB}, then 
the perturbation potential $\Phi$ (being negligible in the early universe) 
would thus have remained very small virtually everywhere and at 
all times in cosmic history to this point, so that little backreaction 
could ever have been generated from it. But even dropping the 
idea of large spatial variations -- or of {\it any} spatial variations, 
which we do completely neglect in the case of a smoothly-inhomogeneous 
universe -- why should one assume that Equation~\ref{EqnIshWaldLimitsB} 
is a valid approximation? 

In computing the evolution of structure formation, one typically 
utilizes the Poisson equation, $\nabla^2 \Phi = 4 \pi G \rho$ 
(for now ignoring modifications to it needed for representing the cosmic 
expansion). It would of course be okay to follow this approach (as in the 
Buchert formalism) if it were safe to ignore the gravitomagnetic fields, 
since the ``gravitoelectric" fields thus become curl-free. But this is 
not the rigorously correct formula to use, because the Poisson equation 
is {\it not causal} -- its (unphysical) solution is the integral of 
the {\it instantaneous} Coulomb potential (at that point) due to the 
charge/mass density distributed throughout all of space, integrated out 
to infinity. While such an approach may be appropriate for considering 
a single, localized physical system embedded in an asymptotically 
empty universe, this picture makes no sense for a system embedded 
in an essentially infinite universe with important nonequilibrium 
processes going on everywhere, for which the expanding observational 
horizon is always bringing new and important information (and 
perturbative forces) in towards the observation point of that system.

For a causality-respecting approach, one must instead use the full 
wave equation (for special-relativistically-consistent potential 
function $\Phi _{\mathrm{SR}}$, in analogy with the electric potential 
in the Lorentz gauge, rather than the Coulomb gauge): 
\begin{equation}
\nabla^2 \Phi _{\mathrm{SR}} - 
\frac{1}{c^2} \frac{\partial^2 \Phi_{\mathrm{SR}}}{\partial t^2} 
= 4 \pi G \rho ~ .
\label{EqnPoissonDynamic}
\end{equation}
Now, the usual impulse is to immediately drop the extra term 
in Equation~\ref{EqnPoissonDynamic}, involving 
$\partial^2 \Phi_{\mathrm{SR}} / \partial t^2$ (equivalent 
to dropping the gravitomagnetic terms), because of its resulting 
prefactor of $v^{2}/c^{2}$; this factor would seem to make it very 
small given the (reasonable) assumption of nonrelativistic speeds 
for most matter flows, and thus (assumedly) ensuring it to 
negligible compared to the spatial variations term in any 
backreaction calculation. But this thinking is based only upon 
considerations of individual Fourier perturbation modes, not on 
the overall causal behavior of information flow in the 
structure-forming universe. If we instead {\it keep} all terms, 
and solve Equation~\ref{EqnPoissonDynamic} as-is, then one gets 
\citep[][eq. 6.69]{JacksonEM}: 
\begin{equation}
\Phi_{\mathrm{SR}}({\bf x},t) = 
-G \int^{\infty} \frac{[\rho({\bf x^{\prime}},t)]_{\mathrm{ret}}}
{\vert {\bf x} - {\bf x}^{\prime} \vert} d^{3}x^{\prime} ~ , 
\label{EqnSRpotential}
\end{equation}
where the bracketed numerator is always evaluated at the 
{\it retarded time}, $t^{\prime} = 
t - \vert {\bf x} - {\bf x}^{\prime} \vert / c$.
It is this retarded-time condition which restores causality, 
allowing different regions of the universe to communicate with 
(and gravitationally perturb) one another; and which provides 
the escape route from the Buchert suppression of Newtonian-level 
backreaction, because the backreaction resulting from this 
integrated perturbation potential is {\it not} expressible 
as a total divergence. We will refer to this propagation 
of gravitational perturbation information between distant 
(though communicating) regions as ``causal updating".

Now certainly, the factor of $v^{2}/c^{2}$ multiplying such 
causal updating contributions is typically quite small; but 
there is a crucial difference between ``small" and ``suppressed". 
Many individually small contributions can be added together to 
produce a large overall result, if one has enough of them. And 
the perturbative contributions to the metric at a specific location 
are supplied by {\it every} virializing structure in the universe 
that is within the causal horizon of that location. Consider 
that the gravitational potential (and thus the Newtonian-level 
perturbation effects) of some particular mass at distance $r$ 
decreases only like $\sim$$1/r$, while the number of such masses 
within a spherical shell at that distance goes like $\sim$$r^{2}$. 
Summing up the small but nonzero perturbing contributions of 
all `remote' virializing clumps, in their effects upon any 
particular observer's location, and integrating out to infinity, 
one gets a total effect 
$\propto \int^{\infty} (r^{2}) (1/r) ~ dr 
= \int^{\infty} r ~ dr = \infty$ ! Noting once again that each 
individual contribution is from a strictly {\it Newtonian} 
gravitational potential, we see that the net effect is not only 
non-small, but in fact it is infinite (in this very simplified 
picture), rendering any factors of $v^{2}/c^{2}$ moot. The only 
effects which actually rein in the total integrated perturbation to 
make it finite in the real universe, are the finite causal horizon 
out to which an observer can `see' perturbations (with more distant 
regions of space obviously looking younger and less clumped, due 
to larger look-back times); and the cosmic (Hubble) expansion, which 
dilutes it by continually carrying the virialized, clumped 
structures farther and farther away from the given observation point. 
(This situation is reminiscent of the problem, and the solution, 
of Olbers' Paradox \citep{WeinbergGravCosmo}; except that the 
`infinity' is even stronger here, since the Newtonian gravitational 
potential only fades with distance as $\propto 1/r$, rather than 
$\propto 1/r^2$, as with light intensity.) 

Though this infinity is not physically realized, what the infinite 
behavior of this simple integral actually does alert us to, is that 
the total perturbation may eventually get very large -- ultimately 
defying even the assumed condition $\vert \Phi \vert \ll 1$ from 
Equation~\ref{EqnIshWaldLimitsA} -- meaning that the entire method 
of summing Newtonian potentials and using a Newtonianly-perturbed 
metric will break down completely as the backreaction grows 
strong enough to become general-relativistically nonlinear. 
Thus the summed effect of innumerable, individually-Newtonian 
perturbations is no longer `Newtonian' in total. (We must 
therefore watch the calculated values of $\Phi _{\mathrm{SR}}$ 
produced in our numerically simulated models very carefully, 
recognizing the loss of accuracy of our formalism when this 
quantity approaches a value of order unity.) But even being 
finite and (in many cases) Newtonian in total, it is clear 
that the impact of distant shells of clumpy material upon the 
observer's metric can strongly outweigh the effects of 
inhomogeneities located in the actual vicinity of the observer.

To the extent that a linearized-gravity approximation is valid, 
our smoothly-inhomogeneous approach therefore works by computing 
a time dependent, spatially averaged perturbation potential 
$\Phi _{\mathrm{SR}} (t)$, which we do {\it not} assume is 
slowly-evolving, as is (inappropriately) assumed for perturbation 
theory treatments via Equation~\ref{EqnIshWaldLimitsB}. Realizing 
that the summed effect of a spherical shell of perturbing clumps 
at distance $r$ actually {\it increases} as $\sim$$r$, this means 
that the dominant backreaction contributions will be delivered 
to an observer by masses at the {\it farthest} possible distances 
out to which that observer can still see significant matter 
inhomogeneities; while past that point, the retarded-time 
observations are from a look-back time too early in the 
cosmic history for substantial structure formation to have 
yet occurred, thus making the resulting integral finite. 

Considering the fact that the key metric perturbation function, 
$\Phi _{\mathrm{SR}} (t)$, is predominantly affected by 
perturbation information coming in from distant locations, 
the retarded-time condition of an integrated formula like 
Equation~\ref{EqnSRpotential} (suitably modified for 
cosmological calculations) implicitly gives it the ability 
(contrary to the argument in \citet{GromesSmallBackreaction}) 
to exhibit relativistic behavior in what would otherwise appear 
to be nonrelativistic situations. To see this, consider the following 
argument: suppose that there is some approximate transition time, 
$t_{\mathrm{clump}}$, when the universe evolves (not necessarily 
instantaneously) from being essentially smooth to essentially 
clumpy. Now at any later time, $t > t_{\mathrm{clump}}$, the 
physical radius out to where an observer can `see' this transitional 
epoch is given (again ignoring cosmic expansion here, for simplicity) 
by $r_{\mathrm{clump}} \sim c (t - t_{\mathrm{clump}})$, a radius 
which increases over time, expanding outward from the observation 
point (say, Earth) at the speed of light. The backreaction effects 
get bigger and bigger (due to increasing shell size and thus 
a greater number of contributing perturbations) as this outgoing 
``wave of observed clumpiness" expands outward from us at $c$, 
thus being fundamentally relativistic in nature and defying 
the assumption of small $\partial \Phi / \partial t$ 
\footnote{As an analogy, consider the contact point between the 
two blades of a very long pair of scissors. The rate at which this 
contact point moves outward from the central pivot does not represent 
the physical motion of any real object, and hence is not limited 
by the speed of the material in the blades as they come together.}.

Thus we see that the potential function $\Phi _{\mathrm{SR}}$ can be 
fully (special-)relativistic in character, violating the assumed 
conditions placed upon `Newtonian' perturbations by \citet{IshWald}, 
{\it even if none of the matter in the universe is actually 
moving at relativistic speeds.} This unexpected effect is due to the 
fact that cosmological systems (unlike anything else) are infinite 
in size, meaning that the inherently relativistic act of 
causal observation will end up encompassing volumes which are 
incomparably vast, causing even small effects -- which never 
cancel out in gravitation, unlike in electrodynamics -- to 
sum together to produce a dominant overall influence. Perhaps 
surprisingly, therefore, the apparent acceleration of the universe 
that we observe is not caused by any individually powerful or 
gravitationally exceptional specific objects or regions, but rather 
from the summed influences of the weak-gravity Newtonian tails of 
innumerable mass concentrations, imposing their combined effects 
upon us from extraordinarily large distances.

In this approach, it is clear that we are advocating a 
diametrically-opposed view from that of other authors who have 
sought to define a ``finite infinity'' 
\citep[e.g.,][]{DPGCoxHowFarInfinity}, representing a very limited 
boundary from within which significant effects upon some specified 
local volume can have ever arrived. Such a limitation would restrict 
the history of `important' interactions to be within an effective 
``matter horizon'' \citep{EllisStoeBogusMatterHorizon} of only a 
few Mpc in size, delineated by the timelike world lines traveled by 
pressure-free matter due to scalar perturbations. But though they 
attempt to make the case that effects from outside of this matter 
horizon (yet within the fundamentally causal ``particle horizon'') 
are generally very small, the chief effects actually estimated by 
\citet{DPGCoxHowFarInfinity} include the Weyl terms regarding geodesic 
deviation ($\propto 1/r^3$), and the power radiated in terms of the 
Bondi news function ($\propto 1/r^2$), both of which are vastly smaller 
at cosmological distances than the {\it amplitude} terms ($\propto 1/r$) 
perturbing the metric itself, which are what we study here. Notably, 
it has been shown by \citet{LudvigsenGWGeodesicDev} that even an 
arbitrarily small energy flux due to gravitational waves can result 
in a finite amount of ``geodesic deviation'' (actually, long-term 
positional displacements of observers due to permanent metric changes) 
in the very distant radiation zone. \citet{DPGCoxHowFarInfinity} in 
fact concedes that for galaxies (or clusters, etc.), ``the prospect 
of adequately treating such a diffuse body as isolated is doomed'', 
due to such very-long-wavelength gravitational radiation (incidentally 
demonstrating once again the inadequacy of the Swiss-Cheese approach). 

The point is that even such small metric alterations 
can become significant (even dominant) when summed over the 
very many contributing sources within one's causal horizon. 
Perhaps the most concrete way of demonstrating the validity 
of this idea, is to point out that \citet{DPGCoxHowFarInfinity} 
and \citet{EllisStoeBogusMatterHorizon} both acknowledge the 
Great Attractor/Shapley Concentration as a likely source of our 
bulk flow -- i.e., as the completely dominant influence upon the 
motion of our entire neighborhood of galaxy clusters -- despite 
the clearly contradictory fact (for their claims) that those 
structures which control our bulk motion are exceedingly 
far beyond any reasonable estimate of our so-called matter horizon. 
Thus the assertion in \citet{EllisStoeBogusMatterHorizon} that, 
``the strengths of any other possible long distance influences... 
gravitational waves, or electric Weyl tensor components from sources 
outside that region -- are insignificant compared to local effects,'' 
is simply incorrect. 

Despite the incontestable importance of those very distant mass 
concentrations in determining such `local' motions (and thus necessarily, 
our local metric), there are very straightforward calculations using 
gravitational perturbation theory which are commonly believed to contradict 
all assertions of the significance of backreaction in general (and of 
nonlocally-acting causal backreaction in particular). The problem is that 
despite the standard (almost reflexive) tendency for experts to resort to 
perturbation theory when attempting to evaluate the effects of cosmic 
structure formation, the fact is that perturbation theory is singularly 
ill-equipped for dealing with the most important perturbative effects in 
an unbounded system like the (effectively) infinite universe. The central 
difficulty is that perturbation theory cannot make any predictions unless 
one pre-specifies which approximations to make, and thus which terms to 
drop. For example, \citet{KMNRinhomogExp} neglect information-carrying 
tensor modes generated by nonlinear scalar perturbations 
(i.e., virializing masses) in the final expressions for their analysis 
(as well as assuming irrotational dust, thus neglecting vorticity), 
which leaves one with scalar perturbations only (vector modes are also 
dropped), thus once again limiting the calculation to effects 
coming from within the inappropriately small (as we have seen) matter 
horizon. Alternatively, \citet{RasanenPertPropModes} does comment 
upon the importance of the propagating degrees of freedom due to the 
``magnetic'' gravitational component $H_{\alpha \beta}$ (the same thing 
as our ``gravitomagnetic'' fields discussed above), and notes the 
difficulty of studying backreaction using a post-Newtonian scheme without 
them, given that the usual assumption of a finite and isolated system 
in such calculations is an inappropriate condition to adopt in a 
cosmological setting. Nevertheless, \citet{RasanenPertPropModes} then 
effectively eliminates such terms by re-affirming the approximation that 
the time derivatives of metric perturbations are small -- similar to the 
approximation above in Equation~\ref{EqnIshWaldLimitsB}, which we have 
labored here to refute on the basis of supporting the causal flow of 
gravitational information. Now obviously, if one removes all ``causal'' 
aspects from ``causal backreaction'', then nothing will be left, and 
it is unsurprising for one to then find that backreaction fails to work, 
given that the principal physics responsible for it has once again been 
set to zero. 

The essential problem with perturbation theory in regards to these issues 
relates to its basic program of singling out the `important' physics by 
labeling the amplitude of each term as ``large'' or ``small'', and then 
dropping the small terms in order to focus upon the large ones. But this 
basic procedure is conceptually flawed in the case of causal backreaction, 
which deals with propagating modes, since such modes can bring in 
gravitational perturbation information to a local observer from vast 
spatial volumes, so that a term with a `small amplitude' can actually 
produce an enormous overall effect. For example, both 
\citet{RasanenPertPropModes} (as noted above) and 
\citet{GromesSmallBackreaction} assume that time derivatives of metric 
perturbations are small, leading to the conclusion that velocity-dependent 
terms can safely be neglected; and \citet{GromesSmallBackreaction} 
specifically uses this point as a primary argument for dismissing the effects 
of causal backreaction entirely\footnote{\citet{GromesSmallBackreaction} also 
uses an argument referring to their ``optimal gauge'', in justifying the 
neglect of causal backreaction; but as they themselves note, it can be a 
tricky thing to choose the right gauge when trying to connect theoretical 
expressions to real cosmological observables, and a different gauge would 
lead to a different amount of the kind of backreaction effect which we 
consider here. Thus the permissibility of choosing gauge conditions that set 
components of the backreaction to zero, a procedure which they engage in, 
is a highly nontrivial matter.}. But conflating ``small'' with ``negligible'' 
is a continuing error in perturbation theory analyses of cosmological 
evolution, since the {\it size} of these effects, in terms of amplitude, 
effectively does not matter: it is practically irrelevant how small the 
time-derivative terms may be, when one realizes that such effects are 
being cumulatively contributed by {\it all} of the matter within a 
causal horizon that may be billions of light-years in extent, thus 
multiplying that amplitude by an amount of mass easily large enough 
to overcome its inherent smallness. Furthermore, even if the amplitudes 
of the relevant perturbative terms for causal backreaction were somehow 
`magically' made even smaller than they naturally are, this still would 
not shut off the causal backreaction effect, but merely postpone it to 
begin a little bit later, since a somewhat larger causal horizon of 
self-stabilizing inhomogeneities would then be needed to supply a 
large enough integrated effect to cause an apparent acceleration. 
In that sense, one could {\it never} make the amplitudes of those 
perturbation terms small enough to avoid the eventual dominance of 
causal backreaction, since a big enough causal horizon (containing 
a sufficiently large amount of virializing mass) can always be 
attained after a sufficiently long period of time, which when multiplied 
by even the smallest amplitude, would eventually produce a term of order 
unity that then proceeds to dominate the cosmic evolution. Thus for 
causal backreaction in an effectively infinite universe, 
{\it there is no such thing as ``negligible''.}

\section{\label{SecFormalism}DEVELOPING A PHENOMENOLOGICAL FORMALISM}

\subsection{\label{Sub1NewtPertReality}Observational Horizons and the 
                   Evolving Perturbation Potential} 

As long as the universe remains homogeneous and isotropic enough to 
be considered at least smoothly-inhomogeneous, then as pointed out at 
the end of Section~\ref{Sub1Vortex}, the net result of virializing 
structure formation is quite simple: it consists of the replacement 
of continuous FRW-distributed matter with a discrete collection of 
stabilized clumps, distributed (one assumes) fairly randomly. 
Considering the metric perturbation effects upon a specific volume 
$\mathbf{V}$, a given stabilizing structure $\mathbf{S}$ will only 
produce nonzero backreaction within $\mathbf{V}$ for as long as new 
gravitational information is propagating from the structure to and 
through that volume. Once the contributing gravitational potential 
of the final, stabilized state of $\mathbf{S}$ has had the opportunity 
to completely propagate causally throughout the entirety of $\mathbf{V}$, 
then as per the Buchert limitations on Newtonian backreaction discussed 
above, the backreaction in $\mathbf{V}$ due to $\mathbf{S}$ will be 
over. This variety of backreaction, due to a set of 
individually-Newtonian virializing structures, is therefore just a 
one-shot deal, unlike the kind of self-sustaining acceleration possible 
for truly general relativistic (i.e., nonlinear) backreaction effects. 
On the other hand, a never-ending collection of such one-shot 
contributions can by itself produce a persisting -- and even growing -- 
backreaction effect within $\mathbf{V}$, as structures at ever-greater 
distances come into the causal horizon of $\mathbf{V}$ over time. 

We wish to come up with a simple representation of how the metric within 
$\mathbf{V}$ is altered by such one-shot perturbing contributions. An 
attempt by other researchers to estimate the total backreaction effects 
for a somewhat similar situation -- for that of an infinite lattice of 
compact, static masses -- was done by \citet{GruzKlebanLattice}, which 
once again found the main backreaction effect to be small, of order 
$O(H^{2} l^{2} / c^{2})$, where $l$ is the lattice size. But this result 
is only obtained as the leading-order backreaction term after the 
subtraction of an apparently divergent term due to the classical 
Newtonian gravitational energy of the masses, a `divergence' which they 
assumed to be unphysical and thus removed via a ``bare mass'' 
renormalization step. But we have already seen this apparently infinite 
term above in Section~\ref{Sub1NewtBackReact}, where we argued that the 
term was quite real, and rendered finite by the finite amount of time that 
it takes for inhomogeneities to initially form, and (more importantly) by 
the finite cosmic horizon out to which one can see those inhomogeneities. 
In a calculation using an eternal lattice of static structures, however, 
such a term would appear to be genuinely infinite and unphysical, and 
would thus unfortunately be dropped. 

In order to figure out how to properly implement Newtonian-level 
backreaction with causality in the absence of an exact cosmological 
solution, we do a little thought experiment. 
Consider our local volume $\mathbf{V}$ to be a homogeneous spherical 
region cut out of a matter-dominated, nearly perfectly homogeneous 
early universe. Now, it is well known 
\citep[][pp. 474-475]{WeinbergGravCosmo} that the expansion 
evolution (i.e., the Friedmann equation) for $\mathbf{V}$ can be 
derived -- using nonrelativistic Newtonian equations, in fact -- 
without reference to anything outside of that sphere. Barring 
perturbations, the Friedmann evolution of $\mathbf{V}$ is determined 
completely internally, and all matter outside of it is gravitationally 
unmeasurable, as if the universe outside of $\mathbf{V}$ did not even 
exist. So let us {\it remove} it, leaving an infinite vacuum 
surrounding our spherical region $\mathbf{V}$. 

This perfect isolation cannot last forever, of course, since later on, 
distant stabilized structures will form which certainly do exert real 
effects upon our local volume by becoming attractors which pull mass in 
towards themselves from all directions, including from within $\mathbf{V}$. 
These gravitational effects, unknown to $\mathbf{V}$ before that time, 
will seem to appear as new from the `empty' region outside of it, as if 
such clumped structures were brought in from infinity at some recent time 
during the structure formation epoch, to only then begin gravitationally 
affecting $\mathbf{V}$. 

In most cases, one concerns oneself solely with the spatial 
{\it gradients} of the gravitational potential, $\Phi$, caused by the 
distant perturbations; and so one is able to study overdensities and voids, 
bulk flows, and the like, even perhaps to the extent of trying to achieve 
cosmological backreaction with them. But we have completely dropped those 
spatial gradients in our adoption of a smoothly-inhomogeneous version of 
the Cosmological Principle, since a randomly-distributed collection of 
inhomogeneities would exert forces in all directions, largely (if not 
perfectly) canceling each other out. 

But what does {\it not} cancel out, regardless of how the perturbations 
outside of $\mathbf{V}$ are distributed -- because it is a non-directional 
scalar, rather than a vector -- is their combined contribution to the overall 
level of $\Phi$; that is, its actual magnitude, which does not matter in true 
Newtonian physics (only differences in potential do), but which does matter 
in general relativity. The contribution by all clumped masses to $\Phi$ 
within $\mathbf{V}$ (i.e., $\Phi _\mathbf{V}$) will always be negative 
(as gravitational energy is negative), thus adding together constructively 
and reinforcing their total effect upon our local volume, even for perturbing 
masses located on opposite sides of it. Furthermore, negative contributions 
to $\Phi _\mathbf{V}$ due to overdensities will not be canceled out by 
positive contributions due to underdensities and voids, because the situation 
is not symmetrical (and in fact is biased towards volume creation): 
underdensities always keep expanding, and at a faster rate than the cosmic 
average; but overdensities do {\it not} keep shrinking, because they 
eventually stabilize themselves through vorticity and virialization. And 
these effects go far beyond the specific regions containing the clumped 
masses themselves, since by adding a nonzero net contribution to 
$\Phi _\mathbf{V}$, these exterior perturbations affect both the flow 
of time and the actual physical volume 
(consider the spatial metric terms of Equation~\ref{EqnIshWaldNewtMetric}) 
within our local region of space. The fact that the effects of such 
distant perturbations are individually small ($\propto 1/r$) does 
not matter when one can sum the combined effects of all of the 
masses in spherical shells out to very large $r$. 

Our phenomenological approach is therefore one in which we model the 
inhomogeneity-perturbed evolution of $\mathbf{V}$ with a metric that 
contains the individual Newtonian perturbations to the potential 
$\Phi _\mathbf{V} (t)$ from all clumped, virialized structures outside 
of $\mathbf{V}$ that have been causally `seen' within $\mathbf{V}$ by 
time $t$, superposed {\it on top of} the background Friedmann expansion 
of $\mathbf{V}$. Now, one may object that material outside of $\mathbf{V}$ 
actually falls within the observational horizon of $\mathbf{V}$ long before 
it became inhomogeneous, yet had no effect upon $\mathbf{V}$ whatsoever at 
that time; and thus such mass should not provide any {\it new} contributions 
to $\Phi _\mathbf{V}$ at a time later on, simply because it has spatially 
redistributed itself from being smooth into being clumpy. But our reply 
would be that each individual structure -- only {\it after} it has clustered 
and virialized itself -- exerts its own individual pull upon the mass inside 
$\mathbf{V}$, causing a peculiar-motion acceleration of the mass inside 
$\mathbf{V}$ towards itself, and that this is a new force (representing 
new ``gravitational knowledge") not seen before within $\mathbf{V}$. 
But one cannot feel a gravitational {\it force}, unless one simultaneously 
feels a gravitational {\it potential}; and so this perturbing potential 
must be new within $\mathbf{V}$ as well, approaching and entering 
$\mathbf{V}$ in causal fashion from this clumped object (and from all 
others) as they develop over time, everywhere in the surrounding universe. 

Now from the perspective of a general-relativist, the natural impulse 
would be to declare this entire procedure non-gauge-invariant, and then 
simply transform $\Phi _\mathbf{V} (t)$ away via a new time coordinate. 
But we assert here that this would be a physically improper procedure, 
since the act of making a cosmological observation is itself not a gauge 
invariant process. 
General relativity only guarantees a physical invariance under 
{\it local} coordinate transformations; but cosmological measurements 
are manifestly nonlocal -- most importantly here in regards to the time 
coordinate -- given that the Hubble curves used to demonstrate the existence 
of a cosmic acceleration are not measurements of a metric quantity at a 
single instant of time (or a single location in space), but are produced by 
{\it integrating} the motion of a light ray over its past light cone, an 
integration over timescales that are by design much longer than what could 
possibly be considered a `local' range of $t$, in order to detect 
evolution of the cosmic expansion rate $a(t)$. In cosmology, therefore, 
the time coordinate cannot be transformed to any other $t^{\prime} (t)$ 
time function with impunity, since the original $t$ coordinate has a 
unique physical meaning: it governs the rate of cosmic expansion (e.g., 
$a(t) \propto t^{2/3}$ during matter domination), which is measurable in 
many ways. This cosmic time $t$ thus serves as an absolute clock which we 
can use as a reference for comparison against the rate of local physical 
processes -- for example, measuring how fast light rays travel through 
the universe, versus how long it takes the cosmic scale factor to double 
in size. Therefore, we claim that this $\Phi _\mathbf{V} (t)$ function 
-- which affects local physics such as light-ray propagation, but 
(ironically, for backreaction) does not alter the cosmic expansion rate 
function $a(t)$ -- encapsulates real physics that produces observable 
consequences, and as such cannot be legitimately transformed away. 

The overall picture presented here is clearly a very heuristic one, and 
the growing perturbation potential $\Phi _\mathbf{V} (t)$ would of course 
not represent the {\it literal} metric of the universe; rather, it is 
simply one that makes sense as a qualitative shorthand for the essential 
backreaction effects generated by structure formation throughout the cosmos, 
as they act upon any particular patch of space. What is really happening, 
in a more complete physical sense, is that collapsing overdensities 
stabilize themselves and halt their collapse by concentrating their local 
vorticity; this concentrated vorticity leads to real, extra volume expansion, 
representable (in the final state) at great distances by the tail of 
a Newtonian perturbation potential to the background FRW metric; and 
this Newtonian tail propagates continually outward into space by inducing 
inward mass flows towards the virialized object from farther and farther 
distances, as time passes. The total perturbation at any location in 
space (which will be independent of position, assuming similar structure 
formation rates everywhere) will then be the combined effects of 
innumerable Newtonian tails of this type, coming in all the way from 
cosmological distances, and from every direction. All that our thought 
experiment has allowed us to do is to show that it is reasonable to 
represent all of these complex physical processes in a very simple way, 
as a cosmologically uniform perturbation potential 
$\Phi _\mathbf{V} (t)$ superposed on top of the original FRW expansion, 
where the potential grows in time as gravitational information about 
virialized structures flows in from an ever-expanding 
``inhomogeneity horizon''. It is the changing absolute 
level of this causally-developing gravitational potential function 
(referred to earlier as $\Phi _{\mathrm{SR}} (t)$, to emphasize 
special-relativistic causality) that we argue is the {\it real} main 
effect of backreaction; and we will show it to be sufficient for 
producing an observed acceleration completely on its own, in a 
smoothly-inhomogeneous context that does not require any other 
inhomogeneity-related effects.

\subsection{\label{Sub1CausUpdMetr}Inhomogeneity Evolution and 
                        Causal Integration for Metric Updating}

In order to derive an expression for the evolving perturbation 
function $\Phi _{\mathrm{SR}} (t)$, we start by considering the 
phase transition itself, during which an extremely symmetrical 
and smooth (``unclumped") state during the early universe has 
gradually evolved to an almost entirely ``clumped" state today. 
During the transition, most of the matter in the universe is 
converted from something close to a completely smooth perfect 
fluid (say, pressureless dust), to a discrete but infinite 
collection of randomly-distributed clumps of various sizes. 
We must define a physical quantity as a measure of `cosmic 
clumpiness', to represent the extent to which this phase 
transition has proceeded to completion. One could perhaps 
use the ratio of the typical size of a stabilized mass 
clump divided by the typical distance between clumps as a 
symmetry-breaking parameter; or alternatively, one could define 
an order parameter, given as the fraction of cosmic mass in 
the clumped (i.e., virially-stabilized) state, versus that 
remaining in the smooth (i.e., freely-expanding FRW) state. 
We choose the latter approach; and as the order parameter grows 
large, this indicates the impending breakdown of the 
matter-dominated FRW evolution of a pure-dust model, and the 
growing relevance of inhomogeneities. 

To phenomenologically model the evolution of clustering, we therefore 
define a ``clumping function", $\Psi (t)$, defined (as a fraction of 
the total density) over the range from $\Psi (t) = 0$ (perfectly smooth 
matter), to $\Psi (t) = 1$ (everything clumped). Note here that we are 
assuming a spatially flat background universe everywhere in our 
calculations; and given that our formalism is specifically designed to 
eliminate the need for Dark Energy, this implies that all of the cosmic 
contents are treated as pressureless dust (for convenience ignoring 
relativistic matter and energy, which do exist but in lesser amounts), 
and that such dust adds up to $\Omega _\mathrm{M} \equiv 1$. (This 
latter condition is only in reference to the unperturbed background, 
however, given that the apparent observational value of 
$\Omega _\mathrm{M}$ will evolve to become quite different from 
unity, as we will see.) We will not be rigorously deriving $\Psi (t)$ 
from first principles, but will rather (as described in later 
sections) be utilizing a variety of functional forms and parameters 
motivated by basic physical principles and measurements, and then 
empirically evaluate their performance in light of cosmological 
observations. 

Obtaining a metric which can explicitly represent the full 
dynamics of a self-stabilizing structure, complete with vorticity 
and virialization, is an enormously complex task; it is so challenging, 
in fact, that we have already seen how exact solutions (like Swiss-Cheese 
models) and more general methods (like the Buchert formalism) have both 
failed in this task, due to the overly-restrictive assumptions that they 
were forced to adopt in order to simplify the situation enough to provide 
a tractable analysis. We therefore seek a practical way in which the key 
backreaction effects of clustering (described by clumping evolution 
function $\Psi (t)$) can be simply but effectively modeled, given our 
smoothly-inhomogeneous formalism with causal updating. 

We must begin by considering a single clumped mass (representing a 
virialized, self-stabilized inhomogeneity) that is embedded within 
the expanding universe. It is actually quite easy to embed a single 
clump within an empty, coasting universe by considering the 
``Expanding Minkowski Universe" \citep{RobertsonNoonan} with 
$a(t) \propto t$; this metric is mathematically equivalent to the 
(static) Minkowski spacetime of special relativity, and can be 
transformed to it via the transformation: 
$t^{\prime} = t \sqrt{1 + r^{2}}$, $r^{\prime} = t r$ \citep[][p. 743, 
using $(t^{\prime},r^{\prime})$ instead of their (t,$\chi$) 
coordinates]{MTW}. The inverse transformation is readily obtainable 
as: $t = \sqrt{(t^{\prime})^{2} - (r^{\prime})^{2}}$, 
$r = r^{\prime} / \sqrt{(t^{\prime})^{2} - (r^{\prime})^{2}}$ 
(noting that the entire coasting universe is covered by just 
the $t^{\prime} \ge r^{\prime}$ portion of the Minkowski metric). 
By using the simple trick of applying this 
inverse coordinate transformation to the static Schwarzschild 
metric\footnote{Given the key role of spinning and vorticity in our 
discussion of backreaction, one might expect the use 
of the Kerr metric here; but the vast majority of mass in the 
universe does not possess relativistic amounts of angular momentum 
-- merely a sufficient degree of it, acting continuously for a 
cosmologically-long time, to ensure that the object remains stabilized. 
Thus we treat the basic existence of persistently-stabilized 
Schwarzschild-like point masses as the biggest perturbation here, 
without needing to include the further metric perturbation which 
results from their actual rotation.} 
of a Black Hole \citep[e.g.,][]{WeinbergGravCosmo}, rather than to 
pure Minkowski space, instead of producing a completely empty coasting 
universe we now obtain an (essentially) coasting universe which 
possesses a clumped mass at the origin. There is the drawback that 
such a cosmology is still empty everywhere outside the central mass, 
and thus has a negative global curvature and cannot represent a 
matter-filled cosmology; but a few appropriate approximations and 
ansatzes can readily convert this metric into a spatially flat, 
matter-dominated metric with a clumped mass at the origin. 

If one objects to such extrapolations, then the exact same resulting 
metric can be obtained by linearizing the McVittie solution 
\citep{McVittieBHinFRW}, as can be seen from the perturbed FRW 
expression given in \citet{KaloKlebanBHinFRW}. One thus gets, 
for the Newtonian approximation of a single clumped object of mass $M$, 
embedded at the origin ($r = 0$) in an expanding, spatially flat, 
matter-dominated (MD) universe: 
\begin{equation}
ds^2 \approx - c^{2} [1 + (2/c^{2}) \Phi (t)] d t^{2} 
+ [a_{\mathrm{MD}}(t)]^{2} [1 - (2/c^{2}) \Phi (t)] d r^{2}
+ [a_{\mathrm{MD}}(t)]^{2} r^{2} [d {\theta}^{2} 
                           + \sin^{2}{\theta} d {\phi}^{2}] 
~ , 
\label{NewtPertSingleClump}
\end{equation} 
where $\Phi (t) \equiv \{ - G M / [a_{\mathrm{MD}}(t) r] \}$, 
and $a_{\mathrm{MD}}(t) \propto t^{2/3}$ is the 
unperturbed MD scale factor evolution function. Note that this 
McVittie solution actually includes no mass accretion -- i.e., 
$M$ in the above metric is a constant -- even though such accretion 
would certainly be expected to occur in a realistic cosmology 
due to inflows onto any overdensity embedded within the matter-filled 
universe (and this was indeed one of our main arguments earlier for 
disregarding the Swiss-Cheese limits on backreaction). In any case, 
this ``$M$'' does not simply represent the total quantity of mass 
(i.e., conserved `dust') within some specified volume, but rather 
embodies the amount of {\it clumped and virialized} mass in that 
volume (which grows in time as a reaction to inflows onto 
overdensities); and so our phenomenological model will therefore 
effectively assume a time-dependent expression, $M(t)$, in the above 
metric, where this time dependence is derived from the clumping 
evolution function $\Psi(t)$, as we now work out in detail. 

Assuming that various mass concentrations $M_{i}$ will be randomly 
and (sufficiently) evenly distributed throughout the universe, 
one must integrate over distance (from a particular observer's 
location) for the various clumps, and one must also 
{\it angle-average} over them to get the metric for the 
combined effects that would represent an averaged $ds^2$ typical 
for that observer experiencing displacements in any given 
direction. (Angle-averaging the effects of an ensemble of 
randomly-distributed clumps will also cancel out any nonspherical 
Kerr-type behavior.) Since a displacement of magnitude 
$\vert d \vec{r} \vert$ will have a radial component 
(with respect to any particular mass concentration) of 
$\vert d \vec{r} \vert \cos \theta$, with a randomly-distributed 
value for the angle, the angle-averaging of the perturbation term 
therefore requires one to average over 
$dr^{2} = \vert d \vec{r} \vert ^{2} \cos^{2} \theta 
= (dx^{2} + dy^{2} + dz^{2}) \cos^{2} \theta$. Since 
$\langle \cos^{2} \theta \rangle = (1/3)$ in three spatial 
dimensions, this means that the summed, averaged value of 
the spatial part of the overall metric perturbation due to 
a large number of randomly-located clumps will be 
multiplied by the factor $(1/3)$; though the temporal part 
(i.e., the perturbation to $g_{t t}$) will remain unaffected 
by such angle-averaging. This is due to the fact that only the 
radial projection of a given translation (with respect to a 
particular mass concentration) will `feel' the perturbation potential 
from that clump in its contribution to the interval $ds ^{2}$ 
(cf. Equation~\ref{NewtPertSingleClump}), while the full strength 
of the gravitational potential always contributes to the temporal 
part of the metric, regardless of directional configuration. 

Consider now the observer to be located at the origin, 
surrounded by a set of (roughly identical) discrete mass 
concentrations with {\it total} mass $M$, all located within 
a particular spherical shell at coordinate distance $r^{\prime}$ 
from the origin, but randomly distributed in $(\theta, \phi)$.
From the above arguments, we can write the total, {\it linearly} summed 
and angle-averaged metric for this observer in `isotropized' fashion, 
as\footnote{Note that this result for the overall summed perturbation 
is mathematically no different from what one would get by integrating a 
smooth spherical shell of continuous matter; but by thinking of mass 
$M$ as being an assortment of many individual, vorticity-stabilized 
discrete sources, it becomes clearer to understand how the same actual 
matter -- considering its mass {\it plus} its dynamical effects -- can 
be responsible both for the background matter-dominated FRW metric, 
{\it and} for the perturbative metric terms superposed on top of it.}: 
\begin{equation}
ds^{2} = 
- c^{2} \{ 1 - [ R_{\mathrm{Sch}}(t) / r^{\prime} ] \} ~ dt^{2} 
~ + ~ [a_{\mathrm{MD}}(t)]^{2} 
\{ 1 + (1/3) [ R_{\mathrm{Sch}}(t) / r^{\prime} ] \} 
~ \vert d \vec{r} \vert ^{2} ~ ,
\label{EqnAngAvgBHMatDomWeakFlat}
\end{equation} 
where 
$R_{\mathrm{Sch}}(t) \equiv \{ (2 G M / c^{2} ) / 
[a_{\mathrm{MD}}(t)] \}$, and 
$\vert d \vec{r} \vert ^{2} 
\equiv (d r^{2} + r^{2} d \theta ^{2} 
+ r^{2} \sin^{2}{\theta} d \phi ^{2} ) 
= \vert d \vec{x} \vert ^{2} 
\equiv (dx^{2} + dy^{2} + dz^{2})$. 

Looking at this expression, it seems physically reasonable 
to regard the term multiplying $\vert d \vec{r} \vert ^{2}$ 
as a `true' increase in spatial volume, when comparing the 
volume of a spatial hypersurface at two different cosmic times; 
and to regard the $g_{t t}$ term as an `observational' term, 
slowing down the perceptions of observers -- relative to the 
expansion of the universe, governed by cosmic time $t$ -- at all 
times after inhomogeneities have begun to develop (including at 
the current time, $t = t_{0}$). Significantly, even if the spatial 
term by itself may not be enough to generate a real volumetric 
acceleration, once it is coupled with the temporal term the 
entire perturbation may indeed be enough to create a so-called 
``apparent acceleration" that is sufficient to explain 
all of the relevant cosmological observations. 

The factor of $(1/3)$ in the perturbative term for the spatial 
metric components (relative to that for $g_{t t}$) may seem 
intriguing because it resembles the initial claim (later retracted) 
by \citet{BeanFact3GRdev} of $(1 / \eta) \equiv (\psi / \phi) 
\equiv (\Phi_\mathrm{Time} / \Phi_\mathrm{Space}) \simeq 3$ 
from their analysis of the weak lensing shear field of the 
Hubble Space Telescope COSMOS data \citep{HSTMasseyCOSMOS}, 
appearing (briefly) to indicate a deviation from general 
relativity, which predicts $\psi = \phi$ in the absence of 
anisotropic stress. But this factor of $(1/3)$ in our 
Equation~\ref{EqnAngAvgBHMatDomWeakFlat} is not `fundamental', 
but merely the result of approximating the linearized sum of many 
individual `Newtonian' solutions; this effectively spreads the 
total spatial perturbation among all three spatial metric terms, 
rather than confining it solely to $g_{r r}$, as is usual when 
considering a single inhomogeneity. Thus the general relativistic 
expectation of equal temporal and spatial potentials -- 
i.e., $\psi = \phi$ -- is not really violated here, and no actual 
new physics or modified gravity is implied by it. 

Implications aside, Equation~\ref{EqnAngAvgBHMatDomWeakFlat} just 
represents the perturbations to the metric due to masses at some 
specific coordinate distance $r^{\prime}$ -- and thus from a 
specific look-back time $t^{\prime}$ -- as seen from some particular 
observational point. The total metric at that spacetime point must 
be computed via an integration over all possible distances, out to 
the distance (and thus look-back time) at which the universe had 
been essentially unclustered. Finally, a light ray reaching us from 
its source (e.g., a Type Ia supernova being used as a standard candle) 
travels to us in a path composed of a collection of such points, 
where the metric at each point must be calculated via its own 
integration out to its individual ``inhomogeneity horizon''; and 
only by calculating the metric at every point in the pathway from 
the supernova to our final location here at $r = z = 0$ can we 
figure out the total distance that the light ray has been able to 
travel through the increasingly perturbed metric, given its 
emission at some specific redshift $z$. 

\begin{figure}
\begin{center}
\includegraphics[scale=0.75]{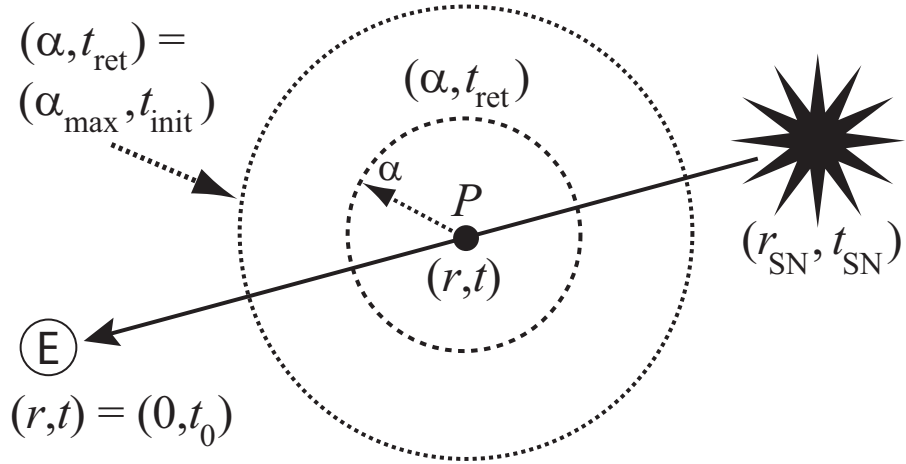}
\end{center}
\caption{Geometry for computing the inhomogeneity-perturbed 
metric at each point along the integrated path of a light ray 
from a supernova to our observation point at Earth.}
\label{FigSNRayTraceInts}
\end{figure}

Consider a light ray emitted by a supernova at cosmic 
coordinate time $t = t_{\mathrm{SN}}$, which then propagates 
from the supernova at $r = r_{\mathrm{SN}}$, to us at $r = 0$, 
$t = t_{0}$. We refer here to the geometry depicted in 
Figure~\ref{FigSNRayTraceInts}.

\placefigure{FigSNRayTraceInts}

For each point $P \equiv (r,t)$ of the trajectory, the 
metric at that point will be perturbed away from the background 
FRW form by all of the virialized clumps that have entered its 
causal horizon by that time. Consider a sphere of (coordinate) 
radius $\alpha$, centered around point $P$, with coordinates 
$(\alpha, t_{\mathrm{ret}})$ (where $t_{\mathrm{ret}} \leq t$ 
is the retarded time), defined such that the information 
about the state of the clumping of matter on that sphere 
at time $t_{\mathrm{ret}}$ will arrive -- via causal 
updating, traveling at the speed of null rays -- to 
point $P$ at the precise time $t$. To compute the 
fully-perturbed metric at $P$, we must integrate over the 
clumping effects of all such radii $\alpha$, from 
$\alpha = 0$ out to $\alpha _{\mathrm{max}}$, the farthest 
distance from $P$ out from which clumping information can 
have causally arrived since the clustering of matter had 
originally begun in cosmic history.

To really determine the relationships between 
$(\alpha, t_{\mathrm{ret}})$, $\alpha _{\mathrm{max}}$, 
and $(r,t)$ with precision, we would need to compute the 
propagation of null rays from such 
$(\alpha, t_{\mathrm{ret}})$ to $(\alpha = 0, t)$ 
recursively or iteratively, since the propagation time of 
a null ray carrying perturbation information would itself 
be affected by all of the other perturbation information 
coming in to cross its path from everywhere else, during 
all times prior to arrival. In other words, causal updating 
is itself slowed by the metric perturbation information 
carried by causal updating, creating an operationally 
nonlinear problem. But out of necessity for this initial, 
proof-of-principle paper, we avoid this ``recursive" 
(as opposed to gravitational) nonlinearity through the 
convenience of assuming an \textit{unperturbed}, flat, 
matter-dominated FRW cosmology for computing all causal 
updating effects for the metric at $(r,t)$. This crucial 
simplification -- in addition to the simplification of 
ignoring gravitational nonlinearities as well, since we 
compute the total metric perturbation terms via a 
trivial summation of individual contributions -- is 
necessary for the purposes of this paper, though a more 
complete analysis (considering both nonlinearities) must 
eventually be employed in order to meet the standards 
of Precision Cosmology. For now, however, what we can do 
is to check at the end of calculations that these 
nonlinearities have not grown too severe as $z \rightarrow 0$ 
(they should not be for most of our simulation runs, below), 
and to note that the output cosmological parameters and fits 
from our calculations here will unavoidably possess some 
systematic theoretical uncertainty as a result of these 
approximations.

Now, for a FRW metric with $a(t) = a_{0} (t/t_{0})^{2/3}$, 
the coordinate distance traveled by a null ray in the cosmic 
time span from $t_{1}$ to $t_{2}$ will be $\alpha \equiv 
(c/a_{0}) \int^{t_{2}}_{t_{1}} (t/t_{0})^{-2/3} dt = 
[(3 c / a_{0}) (t_{0})^{2/3} (t_{2}^{1/3} - t_{1}^{1/3})]$. 
Defining $a_{0} \equiv 3 c t_{0} = 2 c / H_{0}$, 
and with $t_{2} \equiv t$, $t_{1} \equiv t_{\mathrm{ret}}$, 
we thus have: 
$\alpha = [(t/t_{0})^{1/3} - (t_{\mathrm{ret}}/t_{0})^{1/3}]$. 
We then turn this into a prescription for computing 
$t_{\mathrm{ret}}$ as a function of $t$ and $\alpha$ (and 
implicitly of the present time, $t_{0}$), as follows:

\begin{equation}
t_{\mathrm{ret}} (t, \alpha) = 
t_{0} [(t/t_{0})^{1/3} - \alpha]^{3} ~ .
\label{EqntRet}
\end{equation}

Similarly, we can determine $\alpha _{\mathrm{max}}$, 
given some initial time $t_\mathrm{init}$ at which 
structure formation can be reasonably said to have 
started (i.e., 
$\Psi(t \le t_\mathrm{init}) \equiv 0$):

\begin{equation}
\alpha _{\mathrm{max}} (t, t_\mathrm{init})
= [(t/t_{0})^{1/3} - (t_\mathrm{init}/t_{0})^{1/3}] ~ .
\label{EqnalphaMax}
\end{equation}

Now, how the metric is affected at $P$ by a spherical 
shell of material at coordinate radius $\alpha$ depends 
upon the state of clumping there at the appropriate 
retarded time: $\Psi [t_{\mathrm{ret}} (t, \alpha)]$. 
The total effect is computed by integrating all shells 
from $\alpha = 0$ out to 
$\alpha = \alpha _{\mathrm{max}} (t, t_\mathrm{init})$; 
but in order to compute the metric perturbation from 
each shell quantitatively, it is first necessary to 
relate this clumping function to an actual physical 
density of material. 

As discussed above, we define the $\Psi (t)$ function 
as representing the dimensionless ratio of matter which 
can appropriately be defined as `clumped' at a given time, 
expressed as a fraction of the total physical density. 
With the convenience of assuming a flat FRW cosmology as 
the initially unperturbed state, the total physical density 
at all times will merely be an evolved version of the 
unperturbed FRW critical closure density from early 
(pre-perturbation) times. 

Recalling Equation~\ref{EqnAngAvgBHMatDomWeakFlat}, 
we have the perturbation term 
$[ R_{\mathrm{Sch}}(t) / r^{\prime} ] = 
\{ (2 G M / c^{2} ) / [r^{\prime} ~ a_{\mathrm{MD}}(t)] \}$, 
with $a_{\mathrm{MD}}(t) = a_{0} (t / t_{0})^{2/3} 
\equiv c [18 t^{2} / H_{0}]^{1/3}$. 
The value of $M$ to use here is given by the 
clumped matter density at coordinate distance $\alpha$, 
times the infinitesimal volume element of the shell. 
The clumped matter density at time $t$, as implied above, 
will equal $[ \Psi (t) \rho _{\mathrm{crit}}(t) ]$; 
and the volume element in the integrand for that shell 
is given by: $4 \pi R_{\mathrm{phys}}^{2} d R_{\mathrm{phys}} 
= 4 \pi [a_{\mathrm{MD}}(t) ~ \alpha]^{2} 
[a_{\mathrm{MD}}(t) ~ d \alpha]$. (Note that any 
density-dilution effects in 
$[ \Psi (t) \rho _{\mathrm{crit}}(t) ]$ due to volume creation 
by virializing inhomogeneities will be precisely canceled by the 
corresponding volume increase of that spherical shell, leaving 
its (backreaction-effective) differential mass element, ``$dM$'', 
unchanged. On the other hand, there {\it should} be an extra 
distance factor multiplying the denominator of 
$[ R_{\mathrm{Sch}}(t) / r^{\prime} ]$, due to 
$\Phi (t)$, $R_{\mathrm{Sch}}(t) \ne 0$; but we must 
drop it here in this simplified treatment in which 
we neglect ``recursive nonlinearities''.) 

Collecting these terms (and letting 
$t^{\prime} \equiv t_{\mathrm{ret}} (t, \alpha)$), 
the integrand will thus be equal to: 
\begin{mathletters}
\begin{eqnarray}
[ R_{\mathrm{Sch}}(t) / r^{\prime} ]
_{r^{\prime} = \alpha \rightarrow (\alpha + d \alpha)} & = & 
\{ (2 G / c^{2} ) ~ d M ~ / 
~ [a_{\mathrm{MD}}(t) ~ \alpha] \} 
\\ 
& = & \{ (2 G / c^{2} ) ~ 
[a_{\mathrm{MD}}(t) ~ \alpha]^{-1} 
~ [ \Psi (t^{\prime}) ~ 
\rho _{\mathrm{crit}}(t) ] 
~ [4 \pi R_{\mathrm{phys}}^{2} d R_{\mathrm{phys}} ] \} 
\\ 
& = & \{ (8 \pi G / c^{2} ) 
~ \Psi (t^{\prime}) 
~ [a_{\mathrm{MD}}(t) ~ \alpha]^{-1} 
~ [\rho _{\mathrm{crit}}(t) ~ 
[a_{\mathrm{MD}}(t)]^{3}] 
~ {[ \alpha^{2} d \alpha} ]\} 
\\ 
& = & \{ (8 \pi G / c^{2} ) 
~ \Psi (t^{\prime}) 
~ a_{\mathrm{MD}}(t)^{-1} 
~ [\rho _{\mathrm{crit}}(t_{0}) ~ a_{0}^{3}] 
~ {[ \alpha d \alpha} ]\} 
\\ 
& = & \{ (8 \pi G / c^{2} ) 
~ \Psi (t^{\prime}) 
~ [(t_{0} / t)^{2/3} ~ (3 c t_{0})^{-1}]
~ \{ [3 H_{0}^{2} / (8 \pi G)]  
~ (3 c t_{0})^{3} \} 
~ {[ \alpha d \alpha} ]\} 
\\ 
& = & \{12 ~ \Psi (t^{\prime}) 
~ [(t_{0} / t)^{2/3}]
~ {[ \alpha d \alpha} ]\} ~ , 
\end{eqnarray}
\label{EqnIintegrandPrelim}
\end{mathletters}
where for simplification we have used 
$H_{0} = (2/3) t_{0}^{-1}$ and the fact that 
$[\rho(t) a(t)^{3}]$ is constant (or effectively so, 
as explained above), both true for a matter-dominated 
universe. Note also that only $\Psi$ is evaluated at the 
retarded time, $t_{\mathrm{ret}} (t, \alpha)$. The 
{\it strength} of the metric perturbation (at time $t$ 
for point $P$) for a point-like Newtonian perturbation 
embedded in the expansion actually depends upon its 
instantaneous physical distance from $P$ at $t$, 
as is obvious from 
$\Phi (t) = \{ - G M / [a_{\mathrm{MD}}(t) ~ r] \} 
\simeq [- G M / R_{\mathrm{phys}}(t)]$ 
in Equation~\ref{NewtPertSingleClump}. The only 
``relativistic" piece of propagating information 
which is causally delayed is the state of clumping, 
$\Psi [t_{\mathrm{ret}} (t, \alpha)]$, that has 
just then arrived from coordinate distance $\alpha$ 
to observer $P$ at $(r,t)$.

From this result, we can now determine the total 
integrated metric perturbation function due to clumping, 
$I(t)$, as experienced by a null ray (or any observer) 
passing through point $P$ at $(r,t)$: 
\begin{equation}
I(t) = 
\int^{\alpha _{\mathrm{max}} (t, t_\mathrm{init})}_{0}
\{12 ~ 
\Psi [t_{\mathrm{ret}} (t, \alpha)] 
~ [(t_{0} / t)^{2/3}] \} ~ 
\alpha ~ d \alpha ~ ,
\label{EqnItotIntegration}
\end{equation}
with $I(t)$ implicitly being a function of 
$t_\mathrm{init}$ (with $I(t) \equiv 0$ for 
$t \leq t_\mathrm{init}$), as well as of $t_{0}$.

Finally, we can insert this result back into the 
formalism of Equation~\ref{EqnAngAvgBHMatDomWeakFlat}, 
to obtain the final clumping-perturbed metric that 
we will use for all of our subsequent cosmological 
calculations: 
\begin{equation}
ds^{2} = 
- c^{2} [ 1 - I(t) ] ~ dt^{2} 
~ + ~ \{ [a_{\mathrm{MD}}(t)]^{2} ~ 
[ 1 + (1/3) I(t) ] \} 
~ \vert d \vec{r} \vert ^{2} ~ ,
\label{EqnFinalBHpertMetric}
\end{equation}
Representing a smoothly-inhomogeneous universe, 
as it does, this metric depends only upon time, 
and is thus equally good anywhere in the modeled 
universe -- in particular, at every point in the 
trajectory of a light ray from a distant supernova 
to us. 

A few important comments must be made about this 
result. First, note that the integrand for 
$I(t)$ in Equation~\ref{EqnItotIntegration} would 
actually become infinite as $\alpha \rightarrow \infty$, 
were it not limited by the finite causal horizon for 
seeing perturbations (i.e., finite $\alpha _{\mathrm{max}}$), 
and by the lessened degree of clumping as one looks 
back to earlier retarded times, deeper in the past 
(i.e., $\Psi(t \rightarrow t_\mathrm{init}) \simeq 0$).
This implies that the integrated result for $I(t)$ 
can indeed become quite large at late times for 
Newtonian-level perturbations alone, given a large 
degree of clumping, in accord with our discussion from 
Section~\ref{Sub1NewtBackReact}; and also that the 
dominant contribution to $I(t)$ will typically come 
from the largest coordinate distance out to which one 
can still see a substantial degree of clumping at its 
associated $t_{\mathrm{ret}}$, indicating (as we will see) 
that clustering models $\Psi (t)$ with stronger clumping 
earlier on will have a much more powerful overall 
perturbative effect. 

Second, we should reiterate the various approximations 
that have been made in order to obtain this 
smoothly-inhomogeneous cosmological metric -- beyond, 
of course, the core assumption of spatially-random clustering. 
These include the gravitationally-linearized treatment in 
the summing together of independent metric perturbations 
(only valid for $I(t)$ sufficiently less than unity, as 
will be examined later in Section~\ref{Sub2ModelCalcParams}), 
and the dropping of the recursive nonlinearities inherent 
to the physical process of causal updating. 
Such effects would eventually need to be included to produce 
a high-precision model of this type of backreaction, from 
individually-Newtonian, virialized structures. But too-sophisticated 
a model along these lines does not necessarily make sense, since at 
some point the basic assumption of spatial randomness itself breaks 
down, requiring the abandonment of this smoothly-inhomogeneous 
formalism entirely in favor of a fully 3D cosmic structure 
simulation program -- perhaps along the lines of \citet[]{VIRGOsims}, 
for example, with Newtonian-level backreaction effects and 
causal updating added in. Since such a 3D simulation model is 
far beyond the scope of this paper, we will consider it 
sufficient here to stick with Equation~\ref{EqnFinalBHpertMetric} 
as a reasonable first-order approach to the problem, 
with all approximations and caveats kept in mind. 

Lastly for this subsection, we note that the metric 
in Equation~\ref{EqnFinalBHpertMetric} has been made to 
look very much like an ordinary FLRW metric; and it can 
be made to look exactly like one with the transformation 
$d t^{\prime} \equiv \sqrt{1 - I(t)} ~ d t $, 
and with an appropriate redefinition of the scale factor, 
$a^{\prime} (t^{\prime}) \equiv 
\{ a_{\mathrm{MD}}[t(t^{\prime})] 
~ 
\sqrt{1 + (1/3) I[t(t^{\prime})]} \} $. 
One might then be tempted to conclude that observing an 
acceleration in $a^{\prime} (t^{\prime})$, just like in 
the usual FLRW case, still requires some form of Dark Energy 
violating the strong energy condition. But this would be 
incorrect, because the cosmological model that we have 
produced here is not simply a true FLRW metric shown in 
different guises via coordinate transformations, but in 
fact is a very physically different (dynamically inhomogeneous) 
model that is merely being approximated with FLRW-like averaged 
parameters. The detailed astrophysical effects being averaged 
into it, in contributing to the `real' volumetric effects 
in $g_{r r}$ and/or to the `observational' signal-delaying 
effects in $g_{t t}$, are themselves capable of combining 
together sufficiently to create an apparent acceleration in 
$a^{\prime} (t^{\prime})$ -- even without any SEC-violating 
component, or any actual accelerating patch of spacetime 
as might be measured by observers in a local reference frame.

\subsection{\label{Sub1dLumCalcs}Light Propagation, Redshifts, and  
                         Luminosity Distances with Causal Updating}

Due to the altered mathematical form of 
Equation~\ref{EqnFinalBHpertMetric} with respect to the pure 
FLRW case, the observed cosmological evolution and parameters 
will differ from what would normally be expected given some 
usual $a(t)$ function. One must distinguish between parameters 
which merely refer to the underlying theoretical FRW (unperturbed, 
no Dark Energy) model that holds for the pre-clumping universe -- 
that is, the ``bare" parameters -- versus those (``dressed'') 
parameters which are obtained from current-day observations. 
We will use the superscript ``$\mathrm{FRW}$" for the former, 
and the superscript ``$\mathrm{Obs}$" for the latter, and will 
compute the relationships between these different parameter sets 
for observables like redshift, the Hubble Constant, 
the Age of the Universe, and other important variables.

We begin by focusing upon parameters used for computing the 
supernova-based luminosity distance function that most directly 
traces out the cosmic expansion history. First, one must 
calculate how observed redshifts are altered in this model. 
Looking at Equation~\ref{EqnFinalBHpertMetric}, which is 
in the form $ds^{2} = - g_{t t}(t) ~ dt^{2} 
~ + ~ g_{r r}(t) ~ \vert d \vec{r} \vert ^{2}$, 
we note again that one could transform away the 
$g_{t t}(t)$ function with a redefinition of the time 
coordinate, and thus redshifts can be calculated here 
simply by taking $\sqrt{g_{r r}(t)}$ as the new scale factor. 
In other words, given the standard relationship for 
the ``$\mathrm{FRW}$" variables: 
\begin{equation}
z^\mathrm{FRW}(t) \equiv 
\frac{a_{\mathrm{MD}}(t_{0})}{a_{\mathrm{MD}}(t)} - 1 
= (t_{0}/t)^{2/3} - 1 ~ ,
\label{EqnDefNzFRW}
\end{equation}
we can similarly write: 
\begin{equation}
z^\mathrm{Obs}(t) \equiv 
\frac{\sqrt{g_{r r}(t_{0})}}{\sqrt{g_{r r}(t)}} - 1 
= [ \sqrt{\frac{1 + (1/3) I(t_{0})}{1 + (1/3) I(t)}} 
~ (t_{0}/t)^{2/3}] - 1 ~ ,
\label{EqnDefNzObs}
\end{equation}
where we define $t_{0} \equiv t^\mathrm{FRW}_{0}$, 
$t_\mathrm{init} \equiv t^\mathrm{FRW}_\mathrm{init}$; 
and we will mean $t_{x} \equiv t^\mathrm{FRW}_{x}$, 
$z_{x} \equiv z^\mathrm{FRW}_{x}$, for time coordinate 
and redshift values in general, except when expressly 
referring to them in the form $t^\mathrm{Obs}_{x}$ 
or $z^\mathrm{Obs}_{x}$. 

Now, in order to calculate observed luminosity 
distances, we must compute the coordinate 
distance $r$ of a supernova going off at coordinate 
time $t$, for which a light ray would be arriving here 
(at $r = 0$) precisely at $t_{0}$. The modification of 
$r(t)$ imposed here as a perturbation to the FRW result, 
essentially due to the summed Shapiro time delays 
\citep[e.g.,][]{WeinbergGravCosmo} contributed by all 
causally-seen virialized clumps, is the major effect of 
inhomogeneities that we consider in this paper, since we 
neglect other effects like lensing along beam paths 
\citep[e.g.,][]{KantowSwiss03}, and so on, which are 
likely too small to generate an observed cosmic 
acceleration. For a null ray, with $ds^{2} = 0$, and 
considering pure inward radial motion 
($\vert d \vec{r} \vert ^{2} \rightarrow dr^{2}$, 
$dr /dt < 0$), we have 
$dr / dt = - \sqrt{ g_{t t}(t) / g_{r r}(t)}$. 
We can thus compute the coordinate distance of the 
supernova from us, as a function of $t$, as follows 
(with $I(t)$ still as computed via 
Equation~\ref{EqnItotIntegration}):
\begin{mathletters}
\begin{eqnarray}
r^\mathrm{FRW}_{\mathrm{SN}}(t) & \equiv & 
\vert r^\mathrm{FRW}(t_{0}) - r^\mathrm{FRW}(t) \vert 
= \int^{t_{0}}_{t}
\{ \sqrt{\frac{g_{t t}(t^{\prime})}{g_{r r}(t^{\prime})}} \} 
~ d t^{\prime} 
\\
& = & 
\int^{t_{0}}_{t}
\{ \frac{c}{a_{\mathrm{MD}}(t^{\prime})} ~ 
\sqrt{\frac{1 - I(t^{\prime})}{1 + (1/3) I(t^{\prime})}} \} 
~ d t^{\prime} 
\\
& = & 
\frac{c}{a_{0}} \int^{t_{0}}_{t}
\{ (t_{0}/{t^{\prime}})^{2/3} 
~ \sqrt{\frac{1 - I(t^{\prime})}{1 + 
(1/3) I(t^{\prime})}} \} 
~ d t^{\prime} 
~ .
\end{eqnarray}
\label{EqnRofTIntegration}
\end{mathletters}

This coordinate distance function can then be converted 
into an expression for the observed luminosity distance. 
For the perfectly homogeneous FRW case, the luminosity 
distance of a standard candle is given by the current 
physical distance to it, times a redshift factor: 
$d_{\mathrm{L}, \mathrm{FRW}} = 
[a_{0} ~ r_{\mathrm{SN}} ~ (1 + z)]$. For our 
inhomogeneity-perturbed model, all of the 
appropriate time dilation/redshift factors are 
taken care of by using $z^\mathrm{Obs}(t)$ from 
Equation~\ref{EqnDefNzObs}; and the physical distance 
can be given by $r^\mathrm{FRW}_{\mathrm{SN}}(t)$ and 
the modified scale factor, as follows:
\begin{mathletters}
\begin{eqnarray}
d_{\mathrm{L}, \mathrm{Pert}}(t) & = & [ a_{0} 
\sqrt{1 + (1/3) I(t_{0})} ] ~ 
r^\mathrm{FRW}_{\mathrm{SN}}(t) ~ [1 + z^\mathrm{Obs}(t)] 
\\
& = & 
\frac{1 + (I_{0}/3)}{\sqrt{1 + [I(t) / 3]}}
~ 
\frac{c ~ t_{0}^{4/3}}{t^{2/3}} 
~ 
\int^{t_{0}}_{t}
\{ (t^{\prime})^{-2/3} 
~ \sqrt{\frac{1 - I(t^{\prime})}{1 + 
[I(t^{\prime}) / 3]}} \} 
~ d t^{\prime} 
\\
& = & 
\frac{1 + (I_{0}/3)}{\sqrt{1 + [I(t_{r}) / 3]}}
~ 
\frac{c ~ t_{0}}{t_{r}^{2/3}} 
~ 
\int^{1}_{t_{r}}
\{ (t^{\prime}_{r})^{-2/3} 
~ \sqrt{\frac{1 - I(t^{\prime}_{r})}{1 + 
[I(t^{\prime}_{r}) / 3]}} \} 
~ d t^{\prime}_{r} 
~ , 
\end{eqnarray}
\label{EqnDlumDefn}
\end{mathletters}
where $I_{0} \equiv I(t_{0})$, and $t_{r}$, 
$t^{\prime}_{r}$ are dimensionless time ratios 
(e.g., $t_{r} \equiv t / t_{0}$), with no change 
to the essential form of $I(t)$ (i.e., 
$I(t) = I(t_{r} \cdot t_{0}) \Rightarrow I(t_{r})$). 
Note that this $d_{\mathrm{L}, \mathrm{Pert}}(t)$ 
is simply proportional to 
$(c t_{0}) = [(2/3) ~ c / H^\mathrm{FRW}_{0} ]$, 
as would be expected.

With this expression for $d_{\mathrm{L}, \mathrm{Pert}}(t)$, 
and Equation~\ref{EqnDefNzObs} for $z^\mathrm{Obs}(t)$, 
one could in theory combine them analytically to produce 
$d_{\mathrm{L}, \mathrm{Pert}}(z^\mathrm{Obs})$, the 
function actually observed in supernova luminosity 
distance (i.e., Hubble) curves. But this is neither 
analytically nor computationally practical, and so 
we have instead performed our numerical calculations 
using arrays for many discrete points in $t$, 
utilizing the above formulae to evaluate arrays for 
$d_{\mathrm{L}, \mathrm{Pert}}(t)$ and 
$z^\mathrm{Obs}(t)$, which we then simply combine 
together into one array as 
$d_{\mathrm{L}, \mathrm{Pert}}(z^\mathrm{Obs})$. 
It should therefore be noted that all of our subsequent 
plots of luminosity distance curves in this paper for 
different clustering evolution models, though presented 
as smooth curves, are in fact connected dots of points 
evaluated for discrete values of $t^\mathrm{FRW}$.

Our pixelization, which we have rigorously tested to 
ensure accurate results, typically employs $\sim$2150 
discrete time values, most of which ($\sim$1500) are 
concentrated in the more recent times 
($z^\mathrm{FRW} \lesssim 1$) where most of the dynamical 
evolution is happening; fewer pixels are reserved for 
earlier times, going back to before clumping had started 
when the simple FRW model was still correct. Increasing the 
number of discrete time values used, even by a factor of 5, 
results in only a small change (e.g., $\lesssim 0.15 \%$) 
for all cosmological fits and parameters, even those 
requiring up to three derivatives of the luminosity 
distance function. 

Given that this discrete version of 
$d_{\mathrm{L}, \mathrm{Pert}}(z^\mathrm{Obs})$ must in 
fact be differentiated to obtain cosmological parameters 
from it (see Section~\ref{SecNewConcord} below), we do so 
by using the definition of the derivative for each pixel. 
That is, to get the $i^{\mathrm{th}}$ pixel entry (i.e., 
evaluated at $t_{ \{ i \} }$) for the $N^\mathrm{th}$ 
derivative of $d_{\mathrm{L}, \mathrm{Pert}}$ with 
respect to $z^\mathrm{Obs}$, we compute it as follows: 
\begin{equation}
[d^{N \prime}_{\mathrm{L}, \mathrm{Pert}}]_{ \{ i \} } 
= \frac{d}{d ~ z^\mathrm{Obs}} 
[d^{(N-1) \prime}_{\mathrm{L}, \mathrm{Pert}}]_{ \{ i \} } 
\equiv \frac{d^{(N-1) \prime}_{\mathrm{L}, 
\mathrm{Pert}}(t_{ \{ i+1 \} }) 
- d^{(N-1) \prime}_{\mathrm{L}, 
\mathrm{Pert}}(t_{ \{ i \} })}
{z^\mathrm{Obs}(t_{ \{ i+1 \} }) 
- z^\mathrm{Obs}(t_{ \{ i \} })}
 ~ .
\label{EqnPixelDiffDef}
\end{equation}
Note that each subsequent derivative array has one less 
pixel than the prior derivative array, since we lack an 
extra pixel at the (high-$z$) end to compute the 
last pixel of the differentiated array (e.g., if 
$d^{\prime}_{\mathrm{L}, \mathrm{Pert}}$ has 2149 pixels, 
then $d^{\prime \prime}_{\mathrm{L}, \mathrm{Pert}}$ has 
2148 pixels, and so on). 

The calculation of currently observable cosmological 
parameters requires derivatives which are evaluated 
at $z = 0$, $t = t_{0}$. We simply use the first 
(lowest-$z$, highest-$t$) pixel for that derivative 
value: 
\begin{equation}
[d^{N \prime}_{\mathrm{L}, \mathrm{Pert}}]_{
(z \rightarrow 0, ~ t \rightarrow t_{0})} 
\equiv
[d^{N \prime}_{\mathrm{L}, \mathrm{Pert}}]_{ \{ 1 \} } 
 ~ .
\label{EqnFirstEval}
\end{equation}
This appears to be a robust procedure, since the arrays 
of derivative values are well-behaved everywhere (except 
for transient jumps at transitions of our piecewise-continuous 
input $\Psi (t)$ functions at high-$z$), and are smooth heading 
towards $t \rightarrow t_{0}$. Moreover, many simulation 
runs with a variety of different clumping function 
time-dependencies and amplitudes have given results for 
these cosmological parameters which seem to be reasonable 
and mutually consistent.

Ultimately, luminosity distance functions for standard 
candles are plotted as residual Hubble diagrams, where 
one logarithmically plots the ratio of $d_{\mathrm{L}}$ 
to the luminosity distance function for a coasting universe 
of the same current expansion rate. More precisely, 
we are interested in the function 
$\Delta (m - M) = 5 \{ \mathrm{Log}_{10} 
[d_{\mathrm{L}, \mathrm{Data}} (z)] - 
\mathrm{Log}_{10} [d_{\mathrm{L}, \mathrm{Coast}} (z)] \}$, 
where $d_{\mathrm{L}, \mathrm{Coast}} (z) 
= [(c/H_{0}) (1+z) \mathrm{Ln} (1+z)]$. But, 
\textit{which} $H_{0}$ does one use here for subtracting 
off the coasting universe from our numerically 
simulated perturbed-universe models? Although we see 
from Equation~\ref{EqnDlumDefn} that 
$d_{\mathrm{L}, \mathrm{Pert}}$ is indeed proportional 
to $t_{0} \propto (1 / H^\mathrm{FRW}_{0})$, that is not 
sufficient to tell us the real observed expansion rate that 
is asymptotically approached as $z \rightarrow 0$ 
(i.e., $H^\mathrm{Obs}_{0} \ne H^\mathrm{FRW}$), 
since the integral expression could evaluate to 
practically anything, given the appropriate 
clumping evolution function. 

To determine the proper value to use for the 
$z \rightarrow 0$ expansion rate, consider 
Equation 2.51 of \citet{KolbTurner}, from which 
one obtains the general approximation 
$d_{\mathrm{L}} \simeq c z / H_{0}$ for very small $z$. 
This gives an operational definition 
for the observed expansion rate, $H^\mathrm{Obs}_{0}$, 
via: 
\begin{equation}
d^{\prime}_{\mathrm{L}, \mathrm{Pert}, 0} \equiv 
\{ d [d_{\mathrm{L}, \mathrm{Pert}}] / d [z^\mathrm{Obs}] 
\} _{z \rightarrow 0} 
\equiv c / H^\mathrm{Obs}_{0} ~ . 
\label{EqnEffObsHubbConst}
\end{equation} 
(We will later use this general method to explicitly relate 
$H^\mathrm{Obs}_{0}$ to $H^\mathrm{FRW}_{0}$, 
and to find similar relationships for other 
cosmological observables, in Section~\ref{SecNewConcord}.) 
Using this prescription, we can finally define the 
residual Hubble diagram function for our 
clumping-perturbed model, as follows: 
\begin{equation}
\Delta (m - M)_{\mathrm{Pert}} (z^\mathrm{Obs}) = 
5 ~ \{ ~ 
\mathrm{Log}_{10} 
[d_{\mathrm{L}, \mathrm{Pert}}(z^\mathrm{Obs})] ~ 
- ~ \mathrm{Log}_{10} 
[(d^{\prime}_{\mathrm{L}, \mathrm{Pert}, 0}) 
(1 + z^\mathrm{Obs}) \mathrm{Ln} (1 + z^\mathrm{Obs})] 
~ \} ~ . 
\label{EqnResHubbDiagBHpert}
\end{equation}
A straightforward series expansion demonstrates that 
we have correctly specified the formula for the 
observed expansion rate, $H^\mathrm{Obs}_{0}$, in order 
to match the expansion velocities of the coasting and 
clumping-perturbed models at $t_{0}$. 

Using Equation~\ref{EqnResHubbDiagBHpert} and the rest of the 
expressions in this section, we are now able to convert any 
clumping evolution function $\Psi (t^\mathrm{FRW})$ into a 
residual Hubble diagram that can be compared with any 
theoretical FLRW model that one chooses 
(e.g., $\Lambda{\mathrm{CDM}}$ models), as well as with real 
standard candle data from Type Ia supernova observations.

\section{\label{SecResults}MODELS, PARAMETERS, AND RESULTS}

\subsection{\label{Sub1ClumpModels}Selection of Clumping  
                                      Evolution Functions}

In order to produce cosmological predictions with our formalism, 
one must first choose a set of $\Psi (t^\mathrm{FRW})$ 
functions to evaluate which are reasonable models of how the 
fraction of cosmic matter in the clumped state has evolved 
over time. Choosing likely functions is not as trivial as it 
may seem, however. For example, while the contrast of a 
density variation will evolve 
as $\delta \rho / \rho \propto a(t) \propto t^{2/3}$ in the 
linear regime for a matter-dominated universe 
\citep[e.g.,][]{KolbTurner}, this only represents the linear 
evolution of a single clump; it says nothing about the 
initial development of new clumps (often due to collisions), 
or about the nonlinear regime and virialization for very 
dense clumps. Our clumping evolution functions, however, 
must serve as proxies for emulating all of these effects 
combined together.

As one possibility, one could perhaps look at clumping 
evolution functions that are output from large cosmological 
structure formation simulations \citep[e.g.,][]{VIRGOsims}. 
But since such models basically use the instantaneous 
Poisson Equation formalism 
(recall Equations~\ref{EqnIshWaldLimitsTot}-\ref{EqnSRpotential}), 
without causal updating, they lack the realism necessary 
to properly simulate the evolution of clustering without 
the use of a Dark Energy ``fudge factor", which complicates 
the interpretation of such results. 

This author's first approach in defining $\Psi (t^\mathrm{FRW})$ 
was to consider observational data directly, for guidance, 
treating the star formation rate ($SFR$) as a tracer of the rate 
of the increase in clumping -- i.e., $d[\Psi (z)] /d z \propto SFR(z)$. 
This method of choosing likely $\Psi (t)$ functions was indeed 
capable of finding some which reproduced the supernova Hubble curves 
with reasonable success (results not shown here); but given the large 
uncertainties in the observationally measured $SFR$ power-law 
parameters \citep[e.g.,][]{GlazebrookSFR}, and the large number of 
arbitrarily-tunable parameters in the $\Psi (t)$ functions adapted 
from such data, the actual statistical significance of a `good fit' 
was too difficult to meaningfully determine. 

Given these difficulties, we decide here to opt for simplicity 
in choosing which clumping evolution models to use for the 
main runs of our numerical simulation program in this analysis. 
Obviously, there is nothing simpler to use than 
$\Psi(t) \propto t$, a fairly sensible choice to 
begin with, since one would assume that the amount of clumping 
that can occur should to some degree depend directly upon how much 
time is available for that clumping to develop. For some alternatives, 
we have also chosen to use models with $\Psi(t) \propto t^{2/3}$, 
as this is proportional to the linear density contrast evolution 
in a matter-dominated universe; as well as models with an 
`accelerating' clumping rate, $\Psi(t) \propto t^{2}$, to test 
whether that would possibly help in creating an observed acceleration. 
This latter time-dependency also potentially corresponds to the 
final {\it nonlinear} evolution of a density perturbation 
\citep[][p. 322]{KolbTurner}. 

Quantitatively, we define our three different classes of 
clumping evolution models as follows:
\begin{mathletters}
\begin{eqnarray}
\Psi _{\mathrm{Lin}} (t) & \equiv & 
\Psi _{0} ~ 
(\frac{t - t_\mathrm{init}}{t_{0} - t_\mathrm{init}}) 
\\
\Psi _{\mathrm{MD}} (t) & \equiv & 
\Psi _{0} ~ 
(\frac{t - t_\mathrm{init}}{t_{0} - t_\mathrm{init}})^{2/3}
\\
\Psi _{\mathrm{Sqr}} (t) & \equiv & 
\Psi _{0} ~ 
(\frac{t - t_\mathrm{init}}{t_{0} - t_\mathrm{init}})^{2}
~ , 
\end{eqnarray}
\label{EqnClumpModels}
\end{mathletters}
where $t_\mathrm{init}$ represents the beginning of 
clumping, such that 
$\Psi(t \le t_\mathrm{init}) \equiv 0$ for all models; 
and $\Psi _{0} \equiv \Psi(t_{0})$ represents 
the current state of clumping today. Note that all of these 
functions are defined in terms of $t^\mathrm{FRW}$ 
(and hence $z^\mathrm{FRW}$), rather than in terms of 
$t^\mathrm{Obs}$ or $z^\mathrm{Obs}$; the latter would 
of course be preferable, though it is not possible here 
because $t^\mathrm{Obs}$ and $z^\mathrm{Obs}$ depend 
recursively upon $\Psi(t)$.

For these three different classes of models, 
we have two physically meaningful parameters 
to vary: $\Psi _{0}$ and $t_\mathrm{init}$ (though 
we will usually express the beginning of clumping in terms 
of $z_\mathrm{init}$, which can be obtained from 
$t_\mathrm{init}$ via Equation~\ref{EqnDefNzFRW}). 
Now, while this gives us a fairly wide region of parameter 
space to explore through in trying to match the observed 
supernova data, results that succeed in matching the data 
are still meaningful, since these model input parameters 
are constrained by astrophysical considerations; and also 
because (as will be seen below) these classes of models 
have characteristic behaviors that quite naturally look 
very much like $\Lambda{\mathrm{CDM}}$ cosmologies over a 
wide range of redshifts and model parameter choices.

\subsection{\label{Sub2HubbDiags}Numerical Results: 
               Residual Hubble Diagrams and Supernova Data Fits}

In designing a suite of simulation runs to test 
the effectiveness of our model at reproducing the 
observed cosmic acceleration, we use established 
observational data as our guide for specifying 
interesting choices for parameters $z_\mathrm{init}$ 
and $\Psi _{0}$.

With $z_\mathrm{init}$ representing the beginning 
of the cosmic phase transition from smooth to clumped, 
it seems reasonable to associate $z_\mathrm{init}$ 
with another symptom of the beginning of structure 
formation: the epoch during (or slightly before) the 
onset of cosmological reionization. 

The five-year WMAP Data analysis 
\citep{WMAP5yrLikeliParams} supported a value 
of $z_\mathrm{reion} \simeq 11$ in the case of an 
instantaneous reionization; but the data also suggests 
the possibility of an extended period of partial 
reionization, perhaps beginning as early as 
$z \sim 20$, and extending no later than $z \sim 6$. 
To broadly cover this range (and to bracket it, 
to be conservative), we have chosen values of 
$z_\mathrm{init} = (5,10,15,25)$ for our simulation 
runs, with the larger values of $z_\mathrm{init}$ 
generally being more astrophysically interesting 
as starting times for the transition to clumping.

Specifying appropriate values of $\Psi _{0}$ is 
a more subtle task, though, since one has to 
quantitatively characterize the much more complex 
dynamical situation of late-time clustering with a 
single number. Furthermore, we must rely a great deal upon 
our assumption of a smoothly-inhomogeneous universe, since 
the existence of a local void or bubble would seriously 
alter the limiting behavior of $\Psi (z \rightarrow 0)$. 
Nevertheless, we are able produce a range of 
parameter values which make general astrophysical sense, 
and also turn out to produce good cosmological results. 

First, consider that while observations may tell us 
that, say, $\Omega^{\mathrm{Obs}}_{\mathrm{M}} \equiv 
1 - \Omega^{\mathrm{Obs}}_{\Lambda} \sim 0.27$ 
with $\Omega^{\mathrm{Obs}}_{b} \sim 0.04$ 
(and thus $\Omega^{\mathrm{Obs}}_{\mathrm{DM}} \sim 0.23$), 
our model here is one of a flat, apparently accelerating 
universe, with {\it no} dark energy. Thus we exploit the 
fact (details to be given in Section~\ref{Sub2LSS} below) 
that $H^\mathrm{Obs}_{0} \neq H^\mathrm{FRW}_{0}$, 
in order to achieve 
$\Omega^{\mathrm{FRW}}_{\mathrm{M}} \equiv 1$ 
without ever changing the actual physical matter density 
($\omega _{\mathrm{M}} \propto \rho _{\mathrm{M}}$) 
that can be more directly determined through 
other observations (growth of structure, cluster mass 
measurements, etc.). Furthermore, since the physical baryon 
density $\omega_{b}$, and its relationship to 
$\omega_{\mathrm{DM}}$, are fairly well tested by the 
CMB peak height ratios \citep{WMAP1yrRes} regardless of 
whatever model is used to produce acceleration at later 
times, it is therefore wise not to change this ratio, 
$\omega_{b} / \omega_{\mathrm{DM}}$. A simple scaling up 
to flatness ($\Omega^{\mathrm{FRW}}_{\mathrm{Tot}} \equiv 1$) 
thus gives us 
$\Omega^{\mathrm{FRW}}_{b} \sim [ 0.04 (1.0/0.27) ] 
\simeq 0.15$, and $\Omega^{\mathrm{FRW}}_{\mathrm{DM}} 
\simeq 1 - \Omega^{\mathrm{FRW}}_{b} \simeq 0.85$.

We must then determine reasonable estimates for how much 
of each species, Dark and baryonic matter, might be clumped 
at the present time, noting 
(as per Section~\ref{Sub1CausUpdMetr}) that $\Psi _{0}$ is 
given simply as a dimensionless fraction of the total matter 
density.

At one extreme, we may treat the universe as `completely 
clumped' at the present time -- i.e., $\Psi _{0} = 1$. 

At the other extreme, we may consider almost all baryonic 
matter to still be unclumped, due to `gastrophysics' like 
shock heating and its resultant thermal pressure; exceptions 
being only that amount of baryonic matter clumped into stars, 
and perhaps (depending upon how to appropriately define 
`clumped') the amount of gas bound into virialized galaxies. 
Dark matter, on the other hand, not being subject to ordinary 
thermal pressure (and being able to virialize gravitationally), 
would be almost entirely clumped (for Dark Matter clustering, 
see, e.g., \citet{GilmoreDMclust}), 
{\it except} for some small portion of it that 
would actually be Hot Dark Matter (neutrinos). 

Getting precise numbers for these quantities is not trivial, 
but estimates are available: for example, 
\citet{TurnerCosmoSense} estimates $\sim$$1/8$ of the baryonic 
matter to be contained in stars, so that (for our model) 
$\Omega_{\mathrm{stars}} \sim 0.0185$. Also, 
\citet{WMAP5yrBasicRes} limits the neutrino physical density 
to being less than either $\sim$$12-13 \%$ of the total dark 
matter density (WMAP data only), or $\sim$$5-6 \%$ of it 
(WMAP+BAO+SN combined data), which corresponds (respectively) 
in our model to 
$\Omega_{\nu} \lesssim 0.1$ or $\Omega_{\nu} \lesssim 0.05$. 

Thus, summarizing the total density budget, one ends up with 
$\sim$$2 \%$ locked up into stars, $\sim$$13 \%$ remained as 
either clumped or unclumped baryonic gas, $\lesssim 10 \%$ 
(or $\lesssim 5 \%$) as unclumped neutrinos, 
and $\gtrsim 75 \%$ (or $\gtrsim 80 \%$) as mostly clumped 
Cold Dark Matter. 

We can therefore classify $\sim$$77-82 \%$ of the mass as 
`probably clumped', $\lesssim 5-10 \%$ as `probably unclumped', 
with $\sim$$13 \%$ or so as some mix of both. With this 
information, and to space out our parameters fairly evenly, 
we have chosen values of 
$\Psi _{0} = (0.78,0.85,0.92,0.96,1.0)$ for our simulation runs, 
with the mid-range values of $\Psi _{0}$ likely being the 
most astrophysically sensible.

With four different values of $z_\mathrm{init}$, five different 
values of $\Psi _{0}$, and three different clumping evolution 
models, this gives us $4 \times 5 \times 3 = 60$ simulation runs 
in total. Residual Hubble diagrams have been computed for all 
of these runs, and are plotted below, with 
Figure~\ref{FigLinPlots} showing the results for the 
$\Psi _{\mathrm{Lin}}$ runs, Figure~\ref{FigSqrPlots} depicting the 
$\Psi _{\mathrm{Sqr}}$ runs, and Figure~\ref{FigMDPlots} depicting 
the $\Psi _{\mathrm{MD}}$ runs. Each figure includes four panels, 
with the panels representing $z_\mathrm{init} = (5,10,15,25)$ 
in order from top to bottom. Within each panel itself, the FLRW 
cases of the flat SCDM model ($\Omega_{\mathrm{M}} = 1$) and a 
Concordance $\Lambda$CDM model 
($\Omega_{\Lambda} = 0.73 = 1 - \Omega_{\mathrm{M}}$) are 
shown for comparison against five of our models with different 
degrees of clumping: $\Psi _{0} = (0.78,0.85,0.92,0.96,1.0)$ 
in order from the lowest curve to highest.

A detailed quantitative analysis of the best of these runs 
in terms of quality of fit to real supernova data, and with 
several important cosmological parameters being computed for 
each run, will be presented below in Section~\ref{SecTesting}. 
For now, though, we make the important qualitative observation 
that this formalism {\it works.} Not only does it yield a 
selection of several models possessing a perturbative effect 
strong enough to reproduce the observed (apparent) acceleration, 
but it produces curves that clearly do behave very much like 
$\Lambda$CDM -- in particular, the $\Psi _{\mathrm{Sqr}}$ runs 
look like flat $\Lambda$CDM with $\Omega_{\Lambda} \sim 0.3-0.4$, 
the $\Psi _{\mathrm{Lin}}$ runs look like flat $\Lambda$CDM with 
$\Omega_{\Lambda} \sim 0.5-0.8$, and the $\Psi _{\mathrm{MD}}$ 
runs look like flat $\Lambda$CDM with 
$\Omega_{\Lambda} \sim 0.65-0.97$.

(Ironically, the `accelerated' clumping models, 
$\Psi _{\mathrm{Sqr}}$, produce the weakest observed 
acceleration effect, because they have less clumping at 
early times. Clumping is more important the earlier it occurs, 
because of a geometric effect: a longer look-back time for 
the beginning of clumping results in a much larger horizon 
out to which an observer can see substantial inhomogeneities; 
and it is the huge volume of this outer shell of perturbations 
that produces the strongest effect on the observer, as pointed 
out towards the end of Section~\ref{Sub1CausUpdMetr}.)

Most importantly, even without doing a search over our 
parameter space for `best-fit' models -- just by choosing 
a set of simple clumping evolution models and
astrophysically-reasonable parameters for input into 
our model -- we find that a significant number 
($\sim$$10$ or so) of these 60 simulation runs manage to 
fairly precisely reproduce the Concordance $\Lambda$CDM 
Hubble curve. Furthermore, as will be shown shortly, 
these runs are able to fit the supernova data 
essentially as well as (and in certain cases, with some 
model parameter optimization, even better than) 
such best-fit flat $\Lambda$CDM Hubble curves that have 
heretofore been used to argue for the existence of 
Dark Energy.

%
%
\placefigure{FigLinPlots}
%
%
%
%
\placefigure{FigSqrPlots}
%
%
%
%
\placefigure{FigMDPlots}
%
%

\section{\label{SecNewConcord}FORGING A NEW CONCORDANCE 
                              FOR A SMOOTHLY-INHOMOGENEOUS UNIVERSE}

The qualitative ability of our formalism to reproduce 
$\Lambda$CDM-like Hubble curves was shown in the previous 
section, and a more quantitative analysis demonstrating how 
these models are able to fit the observed supernova data 
will be given later on. 

But beyond just succeeding at explaining the apparent acceleration seen 
in data like that from Type Ia supernovae (SNe), it is widely recognized 
that a true model of the universe must satisfy the constraints imposed 
by several different, complementary cosmological data sets, while 
simultaneously producing a set of cosmological parameters that 
are consistent with all other relevant astronomical observations. 
Only such a fully consistent, astronomically-correct cosmological 
solution would attain the status of a `concordance', sufficient to 
replace the well-known ``Cosmic Concordance" representing the 
range of $\Lambda$CDM models that appear already to be cosmologically 
consistent, given the state of the data at this time.

In order to extract the appropriate cosmological parameters 
from our simulated Hubble curves, we must compute each of the 
relevant cosmological observables from 
$d_{\mathrm{L}, \mathrm{Pert}}(z^\mathrm{Obs})$ as it was 
defined previously in Equation~\ref{EqnDlumDefn} (and 
subsequent discussion), with all derivatives (with respect to 
$z^\mathrm{Obs}$) and limits of this (discrete) simulation 
output array performed as indicated via 
Equations~\ref{EqnPixelDiffDef}-\ref{EqnFirstEval}. Brief 
derivations will be presented below of the expressions needed 
for converting model simulation results into observable parameters. 

With such expressions in hand, we will be able to determine 
how these (``dressed'') observables relate to unperturbed 
(``bare'') model parameters such as $t^\mathrm{FRW}_{0}$, 
$H^\mathrm{FRW}_{0}$, and $\Omega^\mathrm{FRW}_{\mathrm{M}}$, 
for different choices of $\Psi _{0}$, $z_\mathrm{init}$, and 
clumping evolution model $\Psi _{\mathrm{Lin}}$, 
$\Psi _{\mathrm{MD}}$, or $\Psi _{\mathrm{Sqr}}$. 
These relationships are interesting not only because they
give us hard numbers to use for comparison with real observations, 
but also because they reveal how much causal backreaction due to 
structure formation has altered the evolution of our universe 
from that predicted by purely homogeneous FRW models.

While it is beyond the scope of this paper to construct a 
completely new concordance in all of its aspects, in the 
subsections below we will show how to calculate several important 
cosmological parameters from our numerical simulations. Then, in 
Section~\ref{SecTesting}, we will demonstrate the observed 
consistency of our best-fitting models with several of the 
key observational parameters of the $\Lambda$CDM Cosmic Concordance, 
without the use of any negative pressure species like Dark Energy.

\subsection{\label{Sub2HubbConstAge}$H_{0}^{\mathrm{Obs}}$, 
Cosmic Proper Time, and the Age Problem} 

As discussed earlier in regards to  
Equations~\ref{EqnEffObsHubbConst}-\ref{EqnResHubbDiagBHpert}, 
an operational definition for the observed Hubble constant 
in terms of our simulation results can be given 
as $H^\mathrm{Obs}_{0} 
\equiv c / d^{\prime}_{\mathrm{L}, \mathrm{Pert}, 0}$. 
For a normal, matter-dominated flat SCDM cosmology with 
no Dark Energy (and neglecting radiation), one has 
$H^\mathrm{FRW}_{0} \equiv [\dot{a}(t) / 
a(t)]_{t \rightarrow t_{0}} = (2/3) t^{-1}_{0}$. 
We can put this together to get the straightforward result: 

\begin{equation}
H^\mathrm{FRW}_{0} = H^\mathrm{Obs}_{0} ~ 
\{ \frac{2}{3} ~ \frac{1}{c t^\mathrm{FRW}_{0}} ~ 
d^{\prime}_{\mathrm{L}, \mathrm{Pert}, 0} \} ~ . 
\label{EqnHubbObsVsHubbFRW}
\end{equation}

Now, we recall from Equation~\ref{EqnDlumDefn}c that 
$d_{\mathrm{L}, \mathrm{Pert}}(t)$ (and hence 
$d_{\mathrm{L}, \mathrm{Pert}}(z^\mathrm{Obs})$, and all 
of its derivatives with respect to $z^\mathrm{Obs}$) are 
simply proportional to $c t^\mathrm{FRW}_{0}$; thus 
any dependence upon the parameter $t^\mathrm{FRW}_{0}$ 
cancels out (as it must) from 
Equation~\ref{EqnHubbObsVsHubbFRW}, and this formula 
merely gives us a dimensionless ratio between the 
observed and unperturbed Hubble Constants, 
$H^\mathrm{Obs}_{0}$ and $H^\mathrm{FRW}_{0}$, the 
value of which depends upon the result of that 
numerical integration. 

Note that it is $H^\mathrm{FRW}_{0}$ which we place on 
the left hand side of Equation~\ref{EqnHubbObsVsHubbFRW}, 
as the `unknown' parameter; the value of 
$H^\mathrm{Obs}_{0}$ is set from real Hubble recession 
measurements (getting, for example, 
$72 ~ \mathrm{km} ~ \mathrm{s}^{-1} \mathrm{Mpc}^{-1}$), 
which one can then translate into $H^\mathrm{FRW}_{0}$ 
for any particular clumping evolution model, in order to 
obtain its `true' cosmic expansion rate, as would have 
been observed if perturbations had never altered the 
observed expansion rate from its FRW value.

The importance of this relationship is that the expression 
in braces in Equation~\ref{EqnHubbObsVsHubbFRW} will be less 
than unity (since $I(t) > 0$) when perturbations exist, 
thus resulting in $H^\mathrm{FRW}_{0} < H^\mathrm{Obs}_{0}$. 
This permits one to have a low value of $H^\mathrm{FRW}_{0}$ 
(say, in the 40's), while still retaining 
$H^\mathrm{Obs}_{0} \approx 72$. This possibility explains why 
several cosmological measurements may be quite concordant 
with a low Hubble constant, while removing the contradiction 
that such a result would seem to create for late-time measurements 
with standard candles \citep[e.g.,][]{FreedHubbKey} that clearly 
indicate a high $H^\mathrm{Obs}_{0}$. (For low-$H_{0}$ 
discussions, see for example \citet{BlanchardAltConcord}, 
\citet{WMAP1yrCosmo}, and \citet{HuntSarkarGlitches}; and 
also see Figure 14 of \citet{WMAP7yrLikeliParams}, showing the 
consistency of WMAP-only data with $\Omega _{\mathrm{M}} = 1$ 
models for $H_{0}$ in the $\sim$30's-40's.) 

Having $H^\mathrm{FRW}_{0} \neq H^\mathrm{Obs}_{0}$ 
also makes other apparent conflicts go away, such as the 
classic Age Problem/Crisis in cosmology 
\citep[e.g.,][]{KolbTurner,TurnerCosmoSense}, 
in which a matter-dominated SCDM universe appears to be 
younger than some of its oldest constituents 
(e.g., globular clusters). The age of such 
a universe is $t_{0} = (2/3) H^{-1}_{0}$, which for 
$H_{0} \simeq 72 ~ \mathrm{km} ~ \mathrm{s}^{-1} 
\mathrm{Mpc}^{-1}$ gives only $t_{0} \simeq 9$ GYr, 
requiring one to assume an accelerating universe 
that had slower expansion in the past, in order to 
lengthen $t_{0}$ to $\sim$$13-14$ GYr.

For our model, on the other hand, we can use the metric 
given above in Equation~\ref{EqnFinalBHpertMetric} 
to relate the `observed' age of the universe, 
$t^\mathrm{Obs}_{0}$, to $t^\mathrm{FRW}_{0}$, as follows:

\begin{equation}
t^\mathrm{Obs}_{0} = 
\int^{t^\mathrm{FRW}_{0}}_{0} \{ \sqrt{1 - I(t)} \} ~ d t ~ . 
\label{EqnCosmicAge}
\end{equation}

Now, even though this results in 
$t^\mathrm{Obs}_{0} < t^\mathrm{FRW}_{0}$, we also have 
$H^\mathrm{FRW}_{0} < H^\mathrm{Obs}_{0}$, 
so that the value 
of $t^\mathrm{FRW}_{0} \equiv (2/3) (1/H^\mathrm{FRW}_{0})$ 
will be much larger than that 
expected from FRW considerations (i.e., significantly 
larger than $\sim$$13-14$ GYr); and thus 
when these two factors are combined together, the result 
for $t^\mathrm{Obs}_{0}$ is able to fall precisely within 
the range necessary to solve the Age Problem, as we 
will show below in Section~\ref{Sub2ModelCalcParams} for 
several of our best simulation runs.

\subsection{\label{Sub2LSS}Spatial Flatness and the 
Observed Matter Density: $\Omega^\mathrm{FRW}_\mathrm{M}$ 
versus $\Omega^\mathrm{Obs}_\mathrm{M}$}

Another major argument used in favor of Dark Energy as part 
of the Cosmic Concordance, is the apparent contradiction 
between CMB peak data showing the universe to be spatially 
flat ($\Omega _\mathrm{Tot} \simeq 1$), while actual searches 
for the required amount of matter persistently turn up short 
on the overall density of clustering mass, looking instead 
like $\rho _\mathrm{M} / \rho _\mathrm{crit} \simeq 0.3$. 
The usual conclusion is that $\Omega _\mathrm{M} \simeq 0.3$, 
and that the gap of 
($\Omega _\mathrm{Tot} - \Omega _\mathrm{M}) \simeq 0.7$ 
is filled by the existence of Dark Energy.

The unstated assumption in this reasoning, however, 
is that it is always accurate to use 
$\Omega _\mathrm{M} = \omega _\mathrm{M} / h^{2} = 
\rho _\mathrm{M} / \rho _\mathrm{crit}$ to relate 
the closure density value of the matter to its 
actual physical density. But because 
$H^\mathrm{FRW}_{0} \neq H^\mathrm{Obs}_{0}$, and thus 
$\rho ^\mathrm{FRW}_\mathrm{crit} \neq 
\rho ^\mathrm{Obs}_\mathrm{crit}$, 
this ceases to be true. Specifically, 
$\Omega^\mathrm{FRW}_\mathrm{M} \propto \omega _\mathrm{M} / 
(H^\mathrm{FRW}_{0})^{2}$ may very well be equal to unity 
{\it despite} the low value of the physical density, 
$\omega _\mathrm{M} \equiv [\rho _\mathrm{M} (8 \pi G /3) 
(100 ~ \mathrm{km} ~ \mathrm{s}^{-1} \mathrm{Mpc}^{-1})^{-2}]$, 
when $H^\mathrm{FRW}_{0} < H^\mathrm{Obs}_{0}$ is properly 
taken into consideration\footnote{Note that we assume the 
{\it physical} matter density from observations to be 
accurate despite causal backreaction -- 
i.e., $\rho ^\mathrm{FRW}_\mathrm{M} \equiv 
\rho ^\mathrm{Obs}_\mathrm{M} \equiv \rho _\mathrm{M}$ -- 
since it can be measured at lower redshifts and in 
less cosmologically-dependent ways than $\Omega _\mathrm{M}$. 
Thus we treat $\omega ^\mathrm{FRW}_\mathrm{M} \equiv 
\omega ^\mathrm{Obs}_\mathrm{M} \equiv \omega _\mathrm{M}$ 
as unchanged in our formalism from the usual FLRW value, defining 
all of the discrepancy to be within $H^\mathrm{Obs}_{0}$ and 
$\Omega^\mathrm{Obs}_\mathrm{M}$.}. 
This is important because the spatial flatness of the universe 
-- particularly as determined using data observed from the ancient 
and very homogeneous CMB epoch -- is dependent upon the value of 
the unperturbed parameter, $\Omega^\mathrm{FRW}_\mathrm{M}$, 
relevant to the very early universe; {\it not} upon the 
observationally-defined parameter, 
$\Omega^\mathrm{Obs}_\mathrm{M} \propto \omega _\mathrm{M} / 
(H^\mathrm{Obs}_{0})^{2}$, reflective of the more recent, 
post-structure-forming epoch. 

To obtain an expression for the relationship 
between $\Omega^\mathrm{Obs}_\mathrm{M}$ and 
$\Omega^\mathrm{FRW}_\mathrm{M}$ in our 
smoothly-inhomogeneous cosmological formalism, 
we begin by assuming that the universe actually is 
spatially flat in terms of its FRW-defined parameters, 
and would have appeared to be so to observers in the past, 
before the onset of the apparent acceleration due to 
clumping. Now, at such an early time, $t_\mathrm{E}$ -- say, 
$t_\mathrm{CMB} \ll t_\mathrm{E} \ll t^\mathrm{FRW}_{0}$, 
for full matter-domination, but yet small inhomogeneous 
clumping -- one had (as functions of $t_\mathrm{E}$): 
$\rho _{\mathrm{M},\mathrm{E}} 
= \rho _\mathrm{crit} [ t_\mathrm{E} ] 
= 3 ( H^\mathrm{FRW} [ t_\mathrm{E} ] )^{2} 
/ 8 \pi G = (1 / 6 \pi G) (t_\mathrm{E})^{-2}$. 
As the universe evolved to the current epoch, the 
volumetric dilution of matter went like $g_{rr} ^{-3/2} = 
\{ [a_{\mathrm{MD}}(t)]^{-3} 
~ [ 1 + (1/3) I(t) ]^{-3/2} \}$ 
(cf. Equation~\ref{EqnFinalBHpertMetric}), yielding a 
matter density today of: $\rho _{\mathrm{M},0} 
= \rho _{\mathrm{M},\mathrm{E}} 
~ \{ ( t_\mathrm{E} / t^\mathrm{FRW}_{0} )^{2} 
~ [1 + (I_{0}/3)]^{-3/2}\} = 
\{ (1 / 6 \pi G) (t^\mathrm{FRW}_{0})^{-2} 
~ [1 + (I_{0}/3)]^{-3/2}\}
= \{ [ 3 (H^\mathrm{FRW}_{0})^{2} 
/ 8 \pi G ] 
~ [1 + (I_{0}/3)]^{-3/2}\}$. 

To turn this formula into a value for 
$\Omega^\mathrm{Obs}_\mathrm{M}$, one must consider 
it in terms of what we {\it think} the critical 
density is today, observationally. Now clearly, 
$\rho ^\mathrm{Obs}_{\mathrm{crit}, 0} \equiv 
3 (H^\mathrm{Obs}_{0})^{2} / 8 \pi G$; and simply by 
taking the ratio $\Omega^\mathrm{Obs}_\mathrm{M} 
\equiv \rho _{\mathrm{M},0} / 
\rho ^\mathrm{Obs}_{\mathrm{crit}, 0}$, we get the 
following result -- along with an alternative 
way of expressing it, including a convenient 
definition for $\Omega^\mathrm{FRW}_\mathrm{M}$: 
\begin{mathletters}
\begin{eqnarray}
\Omega^\mathrm{Obs}_\mathrm{M} & = & 
( H^\mathrm{FRW}_{0} / H^\mathrm{Obs}_{0} )^{2} 
~ \{ [1 + (I_{0}/3)]^{-3/2}\} ~ , 
\\
\Omega^\mathrm{FRW}_\mathrm{M} & \equiv & 
\Omega^\mathrm{Obs}_\mathrm{M} ~ 
[ ( H^\mathrm{Obs}_{0} / H^\mathrm{FRW}_{0} )^{2} 
~ \{ [1 + (I_{0}/3)]^{3/2}\} ] 
~ ^{\phantom{0} !}_{\phantom{.} =} ~ 1 ~ . 
\end{eqnarray}
\label{EqnOmegaMFRW}
\end{mathletters}

Now, the proper way to view these relationships is as 
more of a consistency check, than as an independent 
prediction. If one assumes an initially-flat FRW universe, 
and adopts some favored inhomogeneity clumping evolution 
function $\Psi (t)$, then one can use it to compute $I_{0}$ 
(via Equation~\ref{EqnItotIntegration}) and 
$( H^\mathrm{Obs}_{0} / H^\mathrm{FRW}_{0} )$ (via 
Equation~\ref{EqnHubbObsVsHubbFRW}); in comparison 
with this, one takes the observationally measured values of the 
physical mass density $\omega _\mathrm{M}$ and expansion rate 
$H^\mathrm{Obs}_{0} \equiv 100 ~ h^\mathrm{Obs} ~ \mathrm{km} 
~ \mathrm{s}^{-1} \mathrm{Mpc}^{-1}$, and puts them together to 
form $\Omega^\mathrm{Obs}_\mathrm{M} = \omega _\mathrm{M} / 
(h^\mathrm{Obs})^{2}$. One then checks to make sure that 
Equations~\ref{EqnOmegaMFRW}a,b are satisfied -- i.e., that 
$\Omega^\mathrm{Obs}_\mathrm{M}$ computed numerically from 
the model via Equation~\ref{EqnOmegaMFRW}a matches 
$\Omega^\mathrm{Obs}_\mathrm{M}$ from observations. Or 
equivalently (and more conveniently), as we do below, one may 
adopt some reliable value of $\Omega^\mathrm{Obs}_\mathrm{M}$ 
from observations, and put it together with $I_{0}$ and 
$( H^\mathrm{Obs}_{0} / H^\mathrm{FRW}_{0} )$ from the model, 
in order to use Equation~\ref{EqnOmegaMFRW}b to check that 
$\Omega^\mathrm{FRW}_\mathrm{M} = 1$. If this test is not 
satisfied to some acceptable level of error, then either: 
the early universe was not spatially flat; 
$\Omega^\mathrm{Obs}_\mathrm{M}$ has been poorly estimated 
(via measurement errors in $\omega _\mathrm{M}$ and/or 
$H^\mathrm{Obs}_{0}$); the chosen clumping model $\Psi (t)$ 
is not optimal; or there is a problem with the formalism itself 
(either fundamentally or with its simplifying approximations). 
The goal is therefore to find a $\Psi (t)$ that fits the 
supernova data well, while simultaneously achieving 
$\Omega^\mathrm{FRW}_\mathrm{M} \simeq 1$ for an appropriately 
specified value of $\Omega^\mathrm{Obs}_\mathrm{M}$. 

One last comment about $\Omega^\mathrm{FRW}_\mathrm{M}$, is that 
while having it be equal to unity does represent a (primordially) 
`flat' universe, this does not necessarily represent a `critical' 
universe (i.e., steady expansion at a rate smoothly asymptoting 
to zero), since the future evolution of such a universe will 
strongly depend upon the detailed effects of causal backreaction. 
These issues will be discussed further in Section~\ref{SecFate}; 
though we note here that the ever-increasing strength of 
backreaction over time (i.e., $I(t)$ monotonically increasing 
towards unity for large $t$) would likely require the development 
of a fully gravitationally-nonlinear treatment of causal backreaction 
in order to determine the true fate of the universe.

\subsection{\label{Sub2Params}More Key Cosmological Observables: 
                    $q_{0}^{\mathrm{Obs}}$, $w_{0}^{\mathrm{Obs}}$, 
                    and $j_{0}^{\mathrm{Obs}}$} 

In conjunction with the fitting of cosmological evolution models 
to Hubble plots of the SNe data, it is also useful to extract a 
few standard cosmological parameters that characterize the data 
in a generalized way (e.g., `accelerating' versus `decelerating', 
etc.). Following \citet{VisserJerkSnap}, we give the definitions 
of (respectively) the Hubble, deceleration, and jerk (or jolt) 
functions as: $H(t) \equiv \dot{a} / a$, 
$q(t) \equiv - (\ddot{a} / a) H(t)^{-2}$, and 
$j(t) \equiv (\dot{\ddot{a}} / a) H(t)^{-3}$, where 
$a \equiv a^\mathrm{Obs}(t^\mathrm{Obs})$ is the `observed' 
volumetric scale factor (equal to $g_{rr} ^{1/2}$, as could be 
read off from the smoothly-inhomogeneous metric, 
Equation~\ref{EqnFinalBHpertMetric}) as a function of observable 
time, and overdots represent derivatives with respect to 
$t^\mathrm{Obs}$. 

The limiting values of these functions as $z \rightarrow 0$, 
$t \rightarrow t_{0}$ are the well-known parameters 
$H^\mathrm{Obs}_{0}$, $q^\mathrm{Obs}_{0}$, and 
$j^\mathrm{Obs}_{0}$, which do not depend explicitly 
upon the entire cosmic evolutionary history, but can be 
described mathematically in terms of a series of 
Taylor expansion coefficients of the luminosity distance 
function, $d_{\mathrm{L}}(z^\mathrm{Obs})$, 
defined for low-$z^\mathrm{Obs}$. As given 
in \citet{VisserJerkSnap,RiessGoldSilver}: 
\begin{equation}
d_{\mathrm{L}}(z^\mathrm{Obs}) = 
\frac{c}{H^\mathrm{Obs}_{0}} \{z^\mathrm{Obs} 
+ \frac{1}{2} [1 - q^\mathrm{Obs}_{0}] (z^\mathrm{Obs})^{2} 
+ \frac{1}{6} [-1 + q^\mathrm{Obs}_{0} 
+ 3 (q^\mathrm{Obs}_{0})^{2} 
- j^\mathrm{Obs}_{0}] (z^\mathrm{Obs})^{3} 
+ O [(z^\mathrm{Obs})^{4}] \}
~ . 
\label{EqnLumDistExpansion}
\end{equation}
Extracting these cosmological parameters requires 
multiple derivatives of $d_{\mathrm{L}}$ -- in our case 
`differentiating' (as described above in 
Equations~\ref{EqnPixelDiffDef}-\ref{EqnFirstEval}) the 
simulated $d_{\mathrm{L}}$ curve for each run. 

The observed and unperturbed Hubble Constants, 
$H^\mathrm{Obs}_{0}$ and $H^\mathrm{FRW}_{0}$ respectively, 
have already been defined in terms of one another and 
such derivatives via Equation~\ref{EqnHubbObsVsHubbFRW}. 
The other cosmological parameters can be computed 
independently of those specific values, by taking ratios 
of the derivatives, as follows:

\begin{mathletters}
\begin{eqnarray}
q^\mathrm{Obs}_{0} & = & 1 - 
\frac{d^{\prime \prime}_{\mathrm{L}, \mathrm{Pert}, 
0}}{d^{\prime}_{\mathrm{L}, \mathrm{Pert}, 0}} 
~ ,
\\
w^\mathrm{Obs}_{0} & \equiv & \frac{2}{3} 
(q^\mathrm{Obs}_{0} - \frac{1}{2}) 
~ , 
\end{eqnarray}
\label{EqnDefnq0andw0}
\end{mathletters}
and: 
\begin{equation}
j^\mathrm{Obs}_{0} = -1 + q^\mathrm{Obs}_{0} 
+ 3 (q^\mathrm{Obs}_{0})^{2} - 
\frac{d^{\prime \prime \prime}_{\mathrm{L}, \mathrm{Pert}, 
0}}{d^{\prime}_{\mathrm{L}, \mathrm{Pert}, 0}}
~ . 
\label{EqnDefnj0}
\end{equation}

With Equation~\ref{EqnDefnq0andw0}b, we have also included 
a reference to the usual Equation of State (EoS) function, 
$w(z)$, which is usually interpreted (though obviously not 
in our formalism) as a measurement of the pressure properties 
of the cosmic contents, particularly that of Dark Energy. 
Note, though, that this $w^\mathrm{Obs}_{0}$ above 
represents the observed EoS (or whatever effect mimics it) 
of the current epoch of the universe {\it as a whole}; 
it is not the same thing as a parameter characterizing the 
Dark Energy component {\it alone}, i.e., 
$w^{\Lambda}_{0}$. In particular, 
for a matter plus Cosmological Constant 
($w^{\Lambda}_{0} = -1$) cosmology with a given 
$\Omega _{\Lambda} = (1 - \Omega _\mathrm{M})$ at $z = 0$, 
one has $w^\mathrm{Obs}_{0} = - \Omega _{\Lambda}$.

In terms of evaluating the third derivative of the 
luminosity distance function, we have deliberately 
chosen to characterize its behavior in terms of 
this jerk/jolt parameter, $j^\mathrm{Obs}_{0}$, rather 
than using alternative formulations. This term in 
$d^{\prime \prime \prime}_{\mathrm{L}, 0}$ (i.e., 
the $O [(z^\mathrm{Obs})^{3}]$ term in the expansion) 
is the lowest-order term containing information 
required to characterize changes in the EoS relations 
of the cosmic contents over time, such as might be 
the result of an evolving, quintessence-like Dark Energy. 
Parameterizations are therefore often chosen 
to highlight or simplify such an analysis, 
by defining the Dark Energy EoS as a function of 
$z$ via parameterizations such as 
$w^{\Lambda}(z) \equiv w^{\Lambda}_{0} 
+ w^{\Lambda \prime} z$ 
\citep[e.g.,][]{RiessGoldSilver}, 
or $w^{\Lambda}(z) \equiv w^{\Lambda}_{0} 
+ [w^{\Lambda}_{\mathrm{a}} z/(1 + z)]$ 
\citep[e.g.,][]{KowalRubinSCPunion}. But we do not 
do this here, for two important reasons.

First of all, $j_{0}$ is a purely empirical parameter, 
allowing our analysis to be completely agnostic with 
respect to the physical cause of the apparent acceleration; 
as our formalism does not include any form of Dark Energy, 
it would be less productive to use parameterizations (such 
as [$w^{\Lambda}_{0}$, $w^{\Lambda}_{\mathrm{a}}$]) which 
are optimized to determine the EoS of a Dark Energy which 
is nonexistent in our models. In the language of 
\citet{VisserJerkSnap}, we are choosing a ``retrodictive" 
approach, rather than a ``predictive" one, by considering 
``cosmographic" fits without any prior assumption of 
Friedmann dynamics. And as has been noted in previous 
cosmological analyses 
\citep{CattVisserCosmographSNeFits,RiessDiffParamsPriors}, 
the choice of parameterization can have a significant impact 
upon the best-fit results obtained, especially for data with 
large scatter and uncertainties.

A second useful feature of the jerk parameter, is the property 
that both $\Omega _\mathrm{M} = 1$ SCDM {\it and} Cosmological 
Constant $\Lambda$CDM have $j(t) = j(z) = j_{0} = 1$ for all 
time. That is, spatially flat $\Lambda$CDM models with 
$w^{\Lambda}(z) = -1$, containing only pressureless 
matter (`dust') and vacuum energy, will always have a jerk 
parameter of unity; a condition that will be true for any 
value of $\Omega _{\Lambda} = (1 - \Omega _\mathrm{M})$, 
regardless of whether the cosmology is dust-only, dust-dominated, 
$\Lambda$-dominated, or $\Lambda$-only. (And presumably even 
a slowly-evolving Dark Energy fairly close to $\Lambda$, with 
$\vert d w(z) / d z \vert \ll 1$ and $w(z)$ never too far from 
$-1$, would yield $j_{0} \simeq 1$.) This apparently coincidental 
result -- it is {\it not} the case when significant radiation 
is present, for example -- allows one to conduct a signature 
test of the entire SCDM/$\Lambda$CDM set of cosmologies, for a 
Dark Energy that is anything close to a Cosmological Constant. 
Searching for deviations from $j_{0} = 1$ therefore represents 
an (essentially lowest-order) test of the $\Lambda$CDM (and FLRW) 
paradigm itself, rather than simply narrowing down the parameters 
{\it within} that paradigm. This also represents a falsifiable 
test of our formalism, since our best-fit simulated cosmologies 
(as will be shown below) generally produce strong deviations 
of $j_{0}$ from unity.

\subsection{\label{Sub2CMB}Cosmic Microwave Background 
                  Observations and the CMB Acoustic Scale}

Testing a theoretical model in Precision Cosmology requires the 
comparison of model predictions against data from several different, 
independent observational methods, in order to reduce the effects 
of large measurement uncertainties, and to 
disentangle parameter degeneracies. One of the most powerful probes 
of the universe is the Cosmic Microwave Background, so we consider 
here the characteristic angular scale of the CMB acoustic peaks, 
$l_{\mathrm{A}}$, which roughly controls the positioning of the peaks 
in the CMB power spectrum.

Our formalism of a smoothly-inhomogeneous universe may be described 
as minimally-disruptive for the CMB, in that few details of the 
very early universe are altered from the standard FRW case: the 
primordial spectrum of fluctuations is not changed, and neither 
is the spatial flatness (on average) of the universe, the 
present-day physical density of matter, the ratio of baryonic 
matter to Dark Matter, or just about any parameter affecting 
the physics of the CMB epoch. The only major physical change 
is that made to the angular diameter distance to the 
last scattering surface; plus some other `apparent' modifications 
due to differences between the observed (i.e., dressed) 
cosmological parameters, and the `true' (i.e., bare) parameters. 

The observed acoustic scale, $l^\mathrm{Obs}_{\mathrm{A}}$, is 
determined by the ratio of two values: the `standard ruler' 
provided by the CMB sound horizon ($r_{\mathrm{s}}$), and the 
angular diameter distance ($d_{\mathrm{A}}$) to the last 
scattering surface. As discussed in \citet{EfBondCMBlA}, the 
exact projection of a three-dimensional temperature power spectrum 
to a two-dimensional angular power spectrum is complicated, and depends 
upon the Doppler peak number $m$ and the shape of the primordial 
power spectrum; but to simplify matters, we may settle here for the 
approximate (flat-space) relationship given in their Equation 21a: 
$l_{m} \approx m \pi d_{\mathrm{A}} / r_{\mathrm{s}}$, 
and thus: 
\begin{equation}
l_\mathrm{A} \equiv 
\frac{\pi d_{\mathrm{A}}}{r_{\mathrm{s}}} ~ . 
\label{EqnRatioCMBlA}
\end{equation}
Given this simplification, as well as some others 
(e.g., neglecting the contribution of the early ISW effect 
to the location of the first peak \citep{HuThesis}, 
ignoring radiation in the evolution of the scale factor, 
etc.), our results will therefore not be directly comparable 
to the precise peak positions actually observed in the CMB. 
However, we will always be consistent in comparisons of this 
$l_\mathrm{A}$ parameter {\it between} different models, 
including comparisons of our numerical results against the 
usual SCDM and $\Lambda$CDM models.

Now, for computing $d_{\mathrm{A}}$, we note that the 
angular diameter distance is actually just the physical 
distance to that surface as would be measured at the time 
of the last scattering. So for flat FLRW models, the 
result would simply be: $d_{\mathrm{A, FLRW}} = 
[a(t_\mathrm{CMB}) ~ r_\mathrm{CMB}] = 
[a_{0} ~ r_\mathrm{CMB} / (1 + z_\mathrm{CMB})]$, where 
$r_\mathrm{CMB}$ is the coordinate distance traveled by 
null rays from the CMB to us; i.e., the integral of 
$[c / a(t)]$ from $t_\mathrm{CMB}$ to $t_{0}$, where 
$a(t)$ represents whichever FLRW model one chooses. 

Our inhomogeneity-perturbed model, on the other hand, 
requires several alterations. First, we note that the 
(unperturbed) scale factor, redshift and (coordinate) 
time of the CMB recombination will be changed. Taking 
$z^\mathrm{Obs}_\mathrm{CMB}$ as a given measured 
parameter, and noting that the (pre-clumping) CMB epoch 
has $I(t_\mathrm{CMB}) \simeq 0$, we use 
Equations~\ref{EqnDefNzFRW}-\ref{EqnDefNzObs} to get: 
\begin{equation}
z^\mathrm{FRW}_\mathrm{CMB} = 
\frac{1 + z^\mathrm{Obs}_\mathrm{CMB}}{\sqrt{1 + (I_{0}/3)}} 
- 1 ~ , 
\label{EqnzPertCMB}
\end{equation}
and hence, since $a_{\mathrm{MD}}(t_\mathrm{CMB}) \equiv 
a_\mathrm{CMB} \equiv [a_{\mathrm{MD}}(t_{0}) / 
(1 + z^\mathrm{FRW}_\mathrm{CMB})]$, 
and $a_{\mathrm{MD}}(t) \equiv [a_{0} 
(t^\mathrm{FRW} / t^\mathrm{FRW}_{0})^{2/3}]$: 
\begin{mathletters}
\begin{eqnarray}
a_\mathrm{CMB} & = & \frac{a_{0}}{1 + 
z^\mathrm{Obs}_\mathrm{CMB}} ~ 
\sqrt{1 + (I_{0}/3)}
\\
t^\mathrm{FRW}_\mathrm{CMB} & = & t^\mathrm{FRW}_{0} 
~ [ \frac{\sqrt{1 + (I_{0}/3)}}{1 + 
z^\mathrm{Obs}_\mathrm{CMB}} ] ^{3/2}
~ . 
\end{eqnarray}
\label{EqnAtPertCMB}
\end{mathletters}
Using this last result above, we can now compute 
the coordinate distance to the last scattering 
as $r^\mathrm{FRW}(t^\mathrm{FRW}_\mathrm{CMB})$, 
where this value comes from an integration as defined 
earlier in Equation~\ref{EqnRofTIntegration}, 
given whatever clumping evolution function $\Psi(t)$ 
that one is doing a simulation of.

Noting finally that the physical distance to the 
last scattering, at $t_\mathrm{CMB}$, is given by 
$[a_\mathrm{CMB} ~ 
r^\mathrm{FRW}(t^\mathrm{FRW}_\mathrm{CMB})]$, we 
can now combine these previous results to get the 
final expression for the CMB angular diameter 
distance in our inhomogeneity-perturbed formalism: 
\begin{equation}
d_{\mathrm{A, Pert}} = 
a_{0} ~ 
\frac{\sqrt{1 + (I_{0}/3)}}{1 + 
z^\mathrm{Obs}_\mathrm{CMB}} ~ 
r^\mathrm{FRW}(t^\mathrm{FRW}_\mathrm{CMB}) 
 ~ , 
\label{EqnTotalCMBdA}
\end{equation}
where the arbitrary (though dimensionful) factor $a_{0}$ 
ultimately cancels out due to the factor of $1 / a_{0}$ 
in $r(t)$, as per Equation~\ref{EqnRofTIntegration}c.

Now by using the purely matter-dominated (MD) 
expression for the evolution of the scale factor in 
this above calculation -- e.g., using 
$a(t^\mathrm{FRW}) \propto t^{2/3}$ to obtain 
Equation~\ref{EqnAtPertCMB}b for 
$t^\mathrm{FRW}_\mathrm{CMB}$ -- we have dropped the 
effects of radiation upon the cosmic expansion rate. 
We have also made this same approximation for the 
integration in Equation~\ref{EqnRofTIntegration} for 
$r^\mathrm{FRW}(t)$, and for all other integrations and 
calculations in our formalism. This (greatly simplifying) 
approximation makes little difference for our results, 
since the clumping-related perturbations that we model 
only become important deep into the MD epoch, by which 
time radiation has become a fairly negligible cosmic 
component. The exception to this, however, is for 
calculations in which one goes back to very high $z$ 
-- such as in computing $d_{\mathrm{A, Pert}}$ for the 
CMB, or in quoting a total observed age of the universe. 
But even in those cases, the error is only around the 
$\sim$$1\%$ level, certainly accurate enough for 
a proof-of-principle study in which one is comparing 
different paradigms against one another (as we do in 
this paper), as opposed to a high-precision analysis 
attempting to extract best-fit cosmological parameters 
from the data.

Next, to calculate $r_{\mathrm{s}}$ for spatially flat 
FLRW models, \citet{EfBondCMBlA} give the expression: 
\begin{equation}
r_{\mathrm{s, FLRW}} = \frac{c}{\sqrt{3}} ~ 
\frac{1}{H_{0} \sqrt{\Omega_{\mathrm{M}}}} ~ 
\frac{1}{1 + z_\mathrm{CMB}} 
\int ^{a_{\mathrm{CMB}}}_{0} 
\frac{d (a / a_{0})}{ \{ [(a / a_{0}) 
+ (a_{\mathrm{eq}} / a_{0})] 
~ (1 + R) \} ^{1/2}} 
~ , 
\label{EqnFLRWrs}
\end{equation}
where $a_{\mathrm{eq}}$ represents the scale factor at 
equality between matter and radiation (the latter including 
nearly massless neutrinos), and 
$R \equiv (3 \rho _\mathrm{b} / 4 \rho _{\gamma}) 
\propto (a / a_{0})$ is the ratio determining the sound speed 
of the photon-baryon fluid via $c_{\mathrm{s}} = 
[(c/\sqrt{3}) ~ (1 + R)^{-1/2}]$ \citep{HuSugiyCMBunderImpl}. 
Using numerical estimates for the present-day densities 
and properties of radiation and neutrinos, they then 
give the values: $(a_{\mathrm{eq, FLRW}} / a_{0}) = 
(24185 ~ \omega_{\mathrm{M}})^{-1}$ (for three light 
neutrino flavors), and 
$R_\mathrm{FLRW} = [30496 ~ \omega_{\mathrm{b}} ~ (a/a_{0})]$; 
and it becomes a simple matter to perform the 
integration and calculate $r_{\mathrm{s, FLRW}}$. 

For our model, once again, several modifications must be 
made. Note first that we have already made two alterations 
in Equation~\ref{EqnFLRWrs} from the precise formula given 
in \citet{EfBondCMBlA}, which will be needed for clarity 
in our following calculations. First, we do not 
implicitly normalize all FLRW scale factors 
(e.g., $a_{\mathrm{eq}}$, $a_{\mathrm{CMB}}$) to 
$a_{0} = 1$, but rather normalize them explicitly 
(when necessary) as $a / a_{0}$, etc.; this is 
necessary because $a(t) \neq \sqrt{g_{rr}(t)}$ in our 
formalism when $I(t) \neq 0$, so that setting 
$a_{0} = 1$ no longer properly normalizes $g_{rr} (t_{0})$ 
to unity. And second, while their $r_{\mathrm{s}}$ 
is typically called ``the sound horizon at decoupling", 
what it actually is, is the sound horizon at decoupling 
measured {\it today}. To convert it to the sound 
horizon size {\it then} (for proper comparison to 
$d_{\mathrm{A}}$ above), we have had to multiply it by 
$(a_\mathrm{CMB, FLRW}/a_{0, \mathrm{FLRW}}) = 
(1 + z_\mathrm{CMB, FLRW})^{-1}$.

In order to generalize these results for our formalism, 
all of these ratios of scale factors have to be modified 
to account for inhomogeneity-induced perturbations. Once 
again we have $a_{\mathrm{CMB}} = 
[a_{0} \sqrt{1 + (I_{0}/3)} / 
(1 + z^{\mathrm{Obs}}_{\mathrm{CMB}})]$ as per 
Equation~\ref{EqnAtPertCMB}a, and we can similarly write: 
\begin{equation}
a_\mathrm{eq} = \frac{a_{0}}{1 + 
z^\mathrm{Obs}_\mathrm{eq}} ~ 
\sqrt{1 + (I_{0}/3)}
 ~ , 
\label{EqnAequ}
\end{equation}
where $z^\mathrm{Obs}_\mathrm{eq}$ is the usual redshift 
of equality computed from the observed densities according 
to the FLRW formalism.

Relatedly, the baryon-to-photon ratio as a function of 
scale factor will need to be adjusted. Noting that $R$ 
will evolve here in precisely the same way as it does in 
the FLRW formalism when measured as a function of evolving 
$z^\mathrm{Obs}$ (rather than $a$ or $t$), we write: 
\begin{equation}
R_\mathrm{Pert}(t^{\mathrm{FRW}}) \equiv R_{0} ~ \frac{a}{a_{0}} 
~ \frac{\sqrt{1 + [I(t^{\mathrm{FRW}})/3]}}{\sqrt{1 + (I_{0}/3)}} 
 ~ , 
\label{EqnBaryPhotRatio}
\end{equation}
to represent the proper volumetric dilution, where the 
present-day baryon-to-photon ratio (assumed fixed by 
observations) factors in here as 
$R_{0} \equiv (3 \rho _{\mathrm{b}, 0} / 4 \rho _{\gamma, 0})$. 
We must include the factor $\sqrt{1 + [I(t^{\mathrm{FRW}})/3]}$ 
in the formal definition above to get the correct value 
of $R_\mathrm{Pert}$ at $t^{\mathrm{FRW}}_{0}$, though 
it reduces to unity in these CMB-related calculations because 
$I(t \le t_\mathrm{init}) \equiv 0$ in our models (justifiable 
since $I(t_\mathrm{CMB}) \simeq I(t_\mathrm{eq}) \simeq 0$ in 
the real universe).

With these relations and definitions, we can integrate from 
scratch to find the sound horizon using $c_{\mathrm{s, Pert}} 
= [(c/\sqrt{3}) ~ (1 + R_\mathrm{Pert})^{-1/2}]$, along with the 
normal evolution of a radiation/matter early (FRW) universe. 
Keeping careful track of all perturbation factors, we get the 
result: 
\begin{eqnarray}
r_{\mathrm{s, Pert}} = \frac{c}{\sqrt{3}} ~ 
\frac{1}{H^{\mathrm{FRW}}_{0}} ~ 
\frac{\sqrt{1 + (I_{0}/3)}}{1 + z^{\mathrm{Obs}}_{\mathrm{CMB}}} 
\hspace*{0.1in} 
\times 
\hspace*{1.0in} 
\nonumber 
\\ 
\int ^{a_{\mathrm{CMB}}}_{0} 
\{ (\frac{a}{a_{0}} + \frac{\sqrt{1 + (I_{0}/3)}}{1 
+ z^{\mathrm{Obs}}_{\mathrm{eq}}}) 
~ 
[1 + (\frac{a}{a_{0}}) 
(\frac{R_{0}}{\sqrt{1 + (I_{0}/3)}})] \}^{-1/2} ~ 
d (\frac{a}{a_{0}})
~ . 
\label{EqnPERTrs}
\end{eqnarray}
Note that expressing the prefactor of the integral 
in terms of $H^{\mathrm{FRW}}_{0}$, as opposed to 
$H^{\mathrm{Obs}}_{0}$, automatically eliminates the 
$\Omega_{\mathrm{M}}^{-1/2}$ factor (i.e., 
$\Omega^\mathrm{FRW}_\mathrm{M}$ is implicitly set to 
unity, as required for our spatially flat dust-only model).

Inserting the results of Equation~\ref{EqnPERTrs} (which can 
be integrated numerically) for $r_{\mathrm{s, Pert}}$ and 
Equation~\ref{EqnTotalCMBdA} for $d_{\mathrm{A, Pert}}$ into 
Equation~\ref{EqnRatioCMBlA}, we finally obtain an expression 
for $l^\mathrm{Obs}_\mathrm{A} = (\pi ~ d_{\mathrm{A, Pert}} 
/ r_{\mathrm{s, Pert}} )$ to be calculated for each of our 
simulation run outputs. To get numerical results, however, 
one must assign values to the parameters 
$z^{\mathrm{Obs}}_{\mathrm{eq}}$, 
$z^{\mathrm{Obs}}_{\mathrm{CMB}}$, and 
$R_{0} = [(3/4)~(\omega _{\mathrm{b}, 0} / \omega _{\gamma, 0})] 
= [(3/4)~(\omega _{\mathrm{b}, 0} / \omega _{\mathrm{M}, 0} )~
(1 + z^{\mathrm{Obs}}_{\mathrm{eq}})]$. 
(Note that the direct dependence upon 
$H^{\mathrm{FRW}}_{0} \propto 1/t^{\mathrm{FRW}}_{0}$ cancels 
out of the ratio ($d_{\mathrm{A, Pert}} / r_{\mathrm{s}}$); 
cf. Equations~\ref{EqnRofTIntegration}c,
\ref{EqnDlumDefn}c,
\ref{EqnTotalCMBdA},
\ref{EqnPERTrs}.) 

Using the best-fit (WMAP-only data) cosmological 
parameters given in \citet{WMAP5yrBasicRes}, 
we have $z^{\mathrm{Obs}}_{\mathrm{eq}} = 3176$, 
$z^{\mathrm{Obs}}_{\mathrm{CMB}} = 1090.51$, 
and ($\omega _{\mathrm{b}, 0} / \omega _{\mathrm{M}, 0}) = 
[\omega _{\mathrm{b}, 0} / (\omega _{\mathrm{CDM}, 0} 
+ \omega _{\mathrm{b}, 0})] = (0.02273 / 0.13263)$. 
We take these as actual measurements of physical observables, 
which are not altered by our smoothly-perturbed universe 
model; rather, it is the theoretical (``FRW") model parameters 
which must be modified, in order to match these observed values.

To compare the $l^\mathrm{Obs}_\mathrm{A}$ values from 
our models to those of unperturbed FLRW cosmologies, 
we use a flat $\Lambda$CDM cosmology that is best-fit to real 
SNe data. Further discussion of this best-fit optimization 
step is given below, in Section~\ref{Sub3BestModsSNeChi2}. 

In order to construct a FRW SCDM model to compare these 
CMB (and other) calculations to, we increase 
$\Omega _{\mathrm{M}}$ from $\sim$0.27 to 1.0, we multiply 
$\omega _{\mathrm{b}, 0}$, $\omega _{\mathrm{M}, 0}$, 
$(1 + z^{\mathrm{Obs}}_{\mathrm{eq}})$ (and hence $R_{0}$) 
by $(1.0/0.27)$, and we leave 
$(1 + z^{\mathrm{Obs}}_{\mathrm{CMB}})$ unchanged.

With all of these parameters and formulae, we can now 
calculate the CMB acoustic scale $l^\mathrm{Obs}_\mathrm{A}$ 
(and all of the other cosmological parameters specified 
previously) for each of our numerically-simulated cosmologies, 
and compare them to the (homogeneous) $\Lambda$CDM and SCDM 
cases. These results will be collected and discussed 
soon below, in Section~\ref{Sub2ModelCalcParams}. 

One last remark about the calculations in this subsection, 
though, is that one must not overestimate their precision 
in estimating CMB observables. There is a great distance 
in time and space between our era and the decoupling epoch, 
and there will be many effects in a universe described by 
an inhomogeneity-perturbed formalism like ours (e.g., 
possible modifications to lensing, to the ISW effect, etc.), 
that are not included in these above calculations. Such 
complex effects are useful, since they should ultimately help 
in distinguishing our formalism from FLRW/Dark Energy models; 
but for now, it must be noted that the preceding formulae for 
observed cosmological parameters -- and especially these 
calculations of the CMB acoustic scale -- are being derived 
from a highly simplified and averaged model of the universe.

\subsection{\label{Sub3BestModsSNeChi2}Computing $\chi ^{2}$ 
Values and Fit Probabilities for our Inhomogeneity-Perturbed 
Hubble Diagrams}

To quantitatively assess how well our simulated cosmological 
models (plotted earlier in 
Figures~\ref{FigLinPlots}-\ref{FigMDPlots}) manage to reproduce 
the apparent acceleration, we must analyze how well they fit 
a sample of reliable supernova data. The publicly available 
``SCP Union" compilation \citep{KowalRubinSCPunion} of 
307 SNe Ia (after selection cuts) will serve as the 
fiducial set of supernovae for all of the analyses in this 
paper, except where stated otherwise. 

The output of each of our numerically-simulated models 
can be converted into a distance modulus function via 
$\mu_{\mathrm{Pert}} \equiv (m - M)_{\mathrm{Pert}} = 
\{ 5 \phantom{.} \mathrm{Log}_{10} [d_{\mathrm{L}, 
\mathrm{Pert}}(z^\mathrm{Obs})] + 25 \}$, with 
$d_{\mathrm{L}, \mathrm{Pert}}(z^\mathrm{Obs})$ computed 
according to Equation~\ref{EqnDlumDefn}. It can then be turned 
into a residual distance modulus function, 
$\Delta \mu_{\mathrm{Pert}} \equiv \Delta (m - M)_{\mathrm{Pert}}$, 
by subtracting off a coasting universe model (as per 
Equation~\ref{EqnResHubbDiagBHpert}) for analysis 
and for plotting in residual Hubble diagrams. 

Now since this formula for $\Delta \mu_{\mathrm{Pert}}$ is 
auto-normalized to give a zero $y$-intercept as $z \rightarrow 0$, 
there is no need to specify a particular Hubble Constant in the 
theoretical model. Real data, on the other hand, must be normalized 
by assigning some value of $H_{0}$ to the coasting universe model 
that gets subtracted from the data in order to compute the set of 
$\Delta \mu_{\mathrm{SN}} \equiv \Delta (m - M)_{\mathrm{SN}}$ 
values. (Generally speaking, there will always be an adjustable 
factor $H^{\mathrm{Obs}}_{0}$ needed in order to create a 
comparison between a model, which does not have any 
specific Hubble Constant intrinsic to it, versus the 
observed data, which does.) \citet{RiessGoldSilver} 
state that the chosen value of $H_{0}$ (equivalent to 
the absolute distance scale) is arbitrary, since it only 
shifts the plot of $\Delta (m - M)$ up or down by a 
constant amount, and that their analysis only depends 
upon {\it differences} in magnitude. In practice, 
however, choosing a poor value of $H_{0}$ not only makes 
$\chi ^{2}$ worse for every model, but there is also 
no guarantee that the {\it relative} goodness-of-fit 
between different models will stay the same when this 
constant offset is changed, particularly in the case of 
data with large uncertainties and scatter. Furthermore, 
each different theoretical model requires a somewhat 
different vertical offset (i.e., its own individualized 
$H^{\mathrm{Obs}}_{0}$ value) in order for that specific 
model, and the dynamical cosmological evolution that it 
represents, to fit a particular SN data set as well as 
it possibly can. In our analysis, therefore, we optimize 
$H^{\mathrm{Obs}}_{0}$ separately for each simulation run 
(done simply by adding, and optimizing, a constant offset 
value to $\Delta \mu_{\mathrm{SN}}$, for its comparison 
to that particular $\Delta \mu_{\mathrm{Pert}}$ curve). 
Such optimization of the observed Hubble Constant is 
naturally subject to external constraints due to other 
astronomical observations which limit its acceptable range 
of values; but this is in fact advantageous, since the 
best-fit value of $H^{\mathrm{Obs}}_{0}$ that is output 
for each specific simulation run gives us one more 
independent check upon whether or not that model with those 
input parameters is an acceptable approximation of reality, 
above and beyond its ability to properly fit the SNe data.

One other challenge in comparing our numerical models to the 
SNe data is that our simulated results for $d_{\mathrm{L}}$ 
are only lists of points for discrete values of $z^\mathrm{Obs}$, 
not continuous functions. To evaluate differences between 
$\mu_{\mathrm{Pert}}$ and $\mu_{\mathrm{SN}}$ for any given 
$z_{\mathrm{SN}}$, some form of interpolation is necessary 
to make $\mu_{\mathrm{Pert}}$ continuous. Since our simulations 
use enough points to sample $z$ very finely -- with, for example, 
inter-pixel gaps of $\Delta z \lesssim 10^{-3} - 10^{-2}$ 
for $z^\mathrm{Obs} \lesssim 1 - 3$, and 
$\Delta z \lesssim \mathrm{few} \times 10^{-4}$ 
for $z^\mathrm{Obs} \lesssim 0.1$ -- any decent interpolation 
scheme should give good results. In all of our runs, we compare 
three different functions for interpolating between data points, 
using polynomials of order 1, 2, and 3 in $z^\mathrm{Obs}$; 
and we find that these three interpolation schemes always 
produce $\chi^{2}$ values that are identical to one another 
for at least 5 or 6 significant figures.

Given such a continuous interpolation function 
$\mu_{\mathrm{Pert}} (z^\mathrm{Obs})$ for a particular 
simulation run, we can compute the $\chi ^{2}$ value for 
that theoretical model fit as follows:
\begin{equation}
\chi ^{2} _{\mathrm{Fit}} = \sum _{i} 
\frac{[\mu _{\mathrm{SN}, i} - 
\mu _{\mathrm{Pert}} (z^\mathrm{Obs}_{\mathrm{SN}, i})
]^{2}}{\sigma ^{2} _{\mathrm{SN}, i}}
 ~ , 
\label{EqnChiSquaredVals}
\end{equation}
where $\mu _{\mathrm{SN}, i}$ and $\sigma _{\mathrm{SN}, i}$ 
for each SN are as given in \citet{KowalRubinSCPunion} and 
associated SCP data files. (Note that we do not separately 
fold in additional SNe dispersions due to peculiar velocities, 
lensing, or other intrinsic or systematic effects, as they 
do in order to be conservative in estimating ranges of 
cosmological parameter values; such additions here would 
simply make it harder to distinguish between the quality of 
different theoretical fits, without providing any new 
useful information.) 

Once this $\chi ^{2} _{\mathrm{Fit}}$ value (optimized 
with respect to $H^{\mathrm{Obs}}_{0}$) has been calculated 
for a given inhomogeneity-perturbed model, one may compute the 
likelihood of this $\mu_{\mathrm{Pert}}$ curve by integrating 
the cumulative distribution function for the 
$\chi^{2}$ distribution with $N_{\mathrm{DoF}}$ 
degrees of freedom 
(i.e., $\chi^{2}_{N_{\mathrm{DoF}}} [ X ])$, as follows:
\begin{equation}
P_{\mathrm{Fit}} \equiv 1 - P_{N_\mathrm{DoF}} 
\{0 \le X \le \chi^{2}_{\mathrm{Fit}}  \} 
= 1 - \int^{\chi ^{2} _{\mathrm{Fit}}}_{0} 
\chi^{2}_{N_{\mathrm{DoF}}} [X] ~ dX
 ~ . 
\label{EqnCDFProbs}
\end{equation}
This $P_{\mathrm{Fit}}$ represents the goodness-of-fit 
probability that the given theoretical curve, if it actually 
is a correct description of the universe, would give a value 
of $\chi^{2}$ as high (or higher) than the 
$\chi^{2}_{\mathrm{Fit}}$ that was found. (Although we 
will see that the $P_{\mathrm{Fit}}$ values calculated 
for the models considered here -- including best-fit 
$\Lambda$CDM -- are relatively small ($\sim$$0.3 - 0.4$), they 
are not small enough to be a serious concern, since adding 
in the necessary systematic uncertainties not already folded 
into these $\sigma _{\mathrm{SN}, i}$ values would make all 
of the fit probabilities larger.)

The relevant number of degrees of freedom here is given 
by $N_{\mathrm{DoF}} = (N_{\mathrm{SN}} - N_{\mathrm{Fit}})$, 
where $N_{\mathrm{SN}} = 307$ for the SCP Union data set, and 
$N_{\mathrm{Fit}}$ is the number of model fitting parameters, 
once a particular type of theoretical model has been chosen. 
Now first of all, $H^{\mathrm{Obs}}_{0}$ being individually 
optimized for all theoretical and simulated Hubble curves 
gives us one fitting parameter for every model. Additionally: 
flat $\Lambda$CDM has $N_{\mathrm{Fit}} = 2$, since only the 
value of $\Omega _{\Lambda}$ (along with $H^{\mathrm{Obs}}_{0}$) 
is optimizable; and flat SCDM, with no remaining adjustable 
parameters, has $N_{\mathrm{Fit}} = 1$. Alternatively, our 
$\Psi _{\mathrm{Lin}}$, $\Psi _{\mathrm{Sqr}}$, and 
$\Psi _{\mathrm{MD}}$ models each have $N_{\mathrm{Fit}} = 3$, 
since both $\Psi _{0}$ and $z_\mathrm{init}$ can be varied 
for fitting the data. Thus the number of degrees of freedom 
for evaluating the likelihoods of our 
inhomogeneity-perturbed models is $N_{\mathrm{DoF}} = 304$; 
whereas the ``Concordance" $\Lambda$CDM fit 
($\Omega _{\Lambda} \simeq 0.73 = 1 - \Omega _\mathrm{M}$) 
has $N_{\mathrm{DoF}} = 305$, and flat SCDM has 
$N_{\mathrm{DoF}} = 306$.

One thing that we do {\it not} do in this analysis, 
however, is to compute the ``reduced $\chi ^{2}$'', 
$\chi ^{2} _{\mathrm{Fit, DoF}} \equiv 
(\chi ^{2} _{\mathrm{Fit}} / N_{\mathrm{DoF}})$. 
While it is quite common (e.g., \citet{RiessGoldSilver}, 
and many other sources) to use 
$\chi ^{2} _{\mathrm{Fit, DoF}} \sim 1$ as a proxy for 
indicating a good model fit to the data, it is a poor 
statistical practice for informally estimating likelihoods. 
The use of $\chi ^{2} _{\mathrm{Fit, DoF}}$ implicitly 
assumes the approximation, 
$P_{1} \{0 \le X \le \chi^{2}_{\mathrm{Fit, DoF}} \} \simeq 
P_{N_\mathrm{DoF}} \{0 \le X \le \chi^{2}_{\mathrm{Fit}} \}$. 
But $\chi ^{2} _{1} 
[\chi ^{2} _{\mathrm{Fit}} / N_{\mathrm{DoF}}]$ is a very 
poor representation of $\chi ^{2} _{N_{\mathrm{DoF}}} 
[\chi ^{2} _{\mathrm{Fit}}]$ for large $N_{\mathrm{DoF}}$, 
because the former distribution is much broader 
and more gradual: it has a longer right tail, 
thus leading to a much larger chance of a Type II error 
(incorrect acceptance of a false hypothesis, 
e.g., \citet{RossStats}); and the probability distribution 
increases more slowly as one goes towards lower $\chi ^{2}$ 
values, thus also leading to a larger chance of 
a Type I error (incorrect rejection of a true hypothesis).  
In particular, in the case of the fitting results to be given 
below for our best simulation runs and for the 
$\Lambda$CDM model, in which the results span 
$\chi ^{2} _{\mathrm{Fit, DoF}} \sim 1.02 - 1.05$ for 
$N_{\mathrm{DoF}} \sim 304 - 305$, the use of the ``reduced'' 
$\chi ^{2} _{\mathrm{Fit, DoF}}$ would cause us to 
unknowingly mis-estimate model-fitting probabilities of 
$P_{\mathrm{Fit}} \sim 26\% - 38\%$ as probabilities of 
$P_{\mathrm{Fit}} \sim 30\% - 31\%$ -- a significant loss of 
{\it comparative} information about how well the different 
models fit the data. In this paper, therefore, we will stick 
to the more accurate statistical practice of simply quoting 
$\chi ^{2} _{\mathrm{Fit}}$, $N_{\mathrm{DoF}}$, and 
the resulting $P_{\mathrm{Fit}}$ for each theoretical or 
simulated model.

\section{\label{SecTesting}OBSERVATIONAL TESTS OF THE FORMALISM}

In order to firmly establish our smoothly-inhomogeneous backreaction 
formalism as an acceptable paradigm for understanding the cosmic 
evolution, we must ultimately accomplish three goals: (1) Explaining 
the `already-known' -- i.e., reproducing the most important observational 
results that have formerly been interpreted as signs of Dark Energy; 
(2) Providing `falsifiability' for our formalism, by establishing 
{\it new} predictions that can clearly distinguish our model 
from the conventional $\Lambda$CDM Concordance paradigm; 
and lastly: (3) Convincingly argue plausibility or `naturalness', 
in the sense of showing that our model does not suffer from 
similar fine-tuning problems as the Dark Energy approach. For this 
third point, in particular, while the linking of the onset of the 
apparent acceleration with the beginning of widespread structure formation 
clearly removes the Coincidence Problem of Cosmological Constant 
Dark Energy (recall Section~\ref{SecIntroMotiv}), it still remains to be 
shown {\it why} an alternative, non-$\Lambda$ explanation of the 
cosmic acceleration, such as ours -- even if true -- should happen 
`fortuitously' to look so much like the action of a trivially simple 
(if aesthetically and coincidentally unpleasant) Cosmological Constant. 

The first two of these three tasks will be discussed below, in 
Subsections~\ref{Sub2ModelCalcParams} and \ref{Sub1ModelTesting}, 
respectively. The third task -- addressing the basic necessity, itself, 
of employing an alternative cosmology in the face of current cosmological 
data sets that appear to be fairly consistent with the $\Lambda$CDM 
Concordance Model -- will be reserved for 
a future treatment, both because of space limitations here, 
and due to the fact that this is a generic question for all 
alternative cosmologies, not a unique issue for the causal backreaction 
formalism introduced in this paper. We will therefore present our detailed 
discussion on that topic elsewhere; but in any case, it is likely that all 
of these above issues cannot be resolved to a satisfactory degree without 
much more comprehensive and precise cosmological data, along with 
continued input from the broader expertise of the entire cosmological 
community.

\subsection{\label{Sub2ModelCalcParams}Supernova Fits and 
Cosmological Parameters from our Numerical Simulations: Evaluating 
the New Concordance}

Using the fit probabilities, $P_{\mathrm{Fit}}$, obtainable from each 
of the simulation runs presented earlier in Section~\ref{SecResults}, 
in conjunction with the cosmological parameters calculated from 
each run according to the formulae in Section~\ref{SecNewConcord}, 
we now have a quantitative context for judging our causal backreaction 
paradigm and the clumping evolution functions $\Psi (t)$ that have 
been modeled for this paper. Given those results, we have specified 
an informally-chosen set of `best' runs for further detailed 
discussion here. Of the sixty cosmological models plotted previously 
in Figures~\ref{FigLinPlots}-\ref{FigMDPlots}, twelve of them (six 
$\Psi _{\mathrm{Lin}}$ runs and six $\Psi _{\mathrm{MD}}$ runs) 
appear to this author as being `very good' at replicating the 
apparently accelerating behavior of the universe, while also 
having fairly good cosmological parameters. 

Now, given that these so-called ``best runs" are chosen simply from 
a discrete set of 60 runs performed overall, it would seem likely 
that a truly optimized search over the $(\Psi _{0},z_\mathrm{init})$ 
parameter space might find model runs that do an even better job at 
fitting the SNe data. Such an effort is not really called for here, 
though, given the theoretical uncertainties of our formalism -- 
particularly the neglect of both gravitational and recursive nonlinearities, 
as well as the ad-hoc nature of our chosen $\Psi (t)$ clumping functions, 
themselves -- but such an optimization can indeed be done, producing 
many fits with values of $\chi ^{2} _{\mathrm{Fit}}$ that are just as low 
(and often slightly lower) than that of the best-fit flat $\Lambda$CDM model. 
This is most naturally accomplished by choosing parameters to weaken the 
$\Psi _{\mathrm{MD}}$ runs, since strengthening the `accelerative' effects 
of the $\Psi _{\mathrm{Lin}}$ or $\Psi _{\mathrm{Sqr}}$ runs requires one 
to implausibly resort to $z_\mathrm{init} \gg 25$ or $\Psi _{0} > 1$. 
Exploring the range of models with $\Psi _{\mathrm{MD}}$ clumping evolution, 
there turns out to be an extensive `trench' in parameter space extending at 
least from $(\Psi _{0},z_\mathrm{init}) = (0.89,5)$ to $(0.765,14)$ (and 
likely beyond) for which $\chi^{2}_{\mathrm{Fit}} \lesssim 311.7$ always 
remains true, allowing one to fit the SNe data with these models slightly 
better than one can with $\Lambda$CDM over a wide range of model input 
parameters, thus simultaneously providing good fits and opening up an 
extended range of output cosmological parameters to more flexibly match 
the model results to other external observational constraints. 

For illustration purposes, we choose one of these 
input-parameter-optimized runs as an additional example model for 
comparing to the SNe data and observed cosmological parameters. All 
of the $\Psi _{\mathrm{MD}}$ runs in this phase-space trench are 
virtually degenerate in their $\chi^{2}_{\mathrm{Fit}}$ values (there 
is an extremely weak gradient in $P_{\mathrm{Fit}}$ among them), 
so as a somewhat loosely defined ``semi-optimized" run we select the 
$(\Psi _{0},z_\mathrm{init}) = (0.768,14)$ case, which has the lowest 
$\chi^{2}_{\mathrm{Fit}}$ of any $\Psi _{\mathrm{MD}}$ run that we 
found among those tested (in a non-exhaustive search through the 
parameter space), while also having sufficiently different 
parameters from our discrete sample of 20 $\Psi _{\mathrm{MD}}$ runs 
for it to be of further cosmological interest. We thus add this case 
to our twelve other ``best runs" discussed above, giving us a set of 
thirteen highlighted runs, in total, for use in the more in-depth 
discussion of the results of our simulations which follows now.

Residual Hubble diagrams of these thirteen chosen runs are 
shown in Figure~\ref{FigBestCosmSims}, plotted along with the 
(now SNe-best-fit) SCDM and Concordance $\Lambda$CDM models. 
Shown against these curves are the SCP Union SNe data, with 
the 307 SNe averaged here into 40 bins, equally spaced in 
$\mathrm{Log} [1 + z^{\mathrm{Obs}}]$; binning is necessary 
for clear visualization of this data set, given not only the 
large number of SNe data points, but also the very high scatter 
for those data, and the large error bars for each individual 
supernova magnitude.

\placefigure{FigBestCosmSims}

It is obvious by inspection that these thirteen runs represent  
cosmological models that produce good Hubble curves, being 
visually almost indistinguishable from one another (and from 
Concordance $\Lambda$CDM) in the SNe-data-rich region of 
$z^{\mathrm{Obs}} \sim 0.1 - 1$. This is especially evident when 
plotted on this $y$-axis scale; a scale that clearly shows the 
large separation between all of these models from SCDM, as well as 
the large scatter and error bars of the SNe data themselves -- 
{\it even after} binning and averaging -- when contrasted with the 
tight overlap between all of these models and the 
Concordance $\Lambda$CDM cosmology itself.

For a more quantitative analysis, the comprehensive output data 
for these thirteen runs, along with corresponding output parameters 
from the best-fit $\Lambda$CDM and SCDM runs, are given in 
Table~\ref{TableSimRunsCosParams}.

\placetable{TableSimRunsCosParams}

First, considering the goodness-of-fit of the models to the 
SNe data, we see that the results for this discrete selection 
of inhomogeneity-perturbed models are nearly as good as that 
for $\Lambda$CDM in terms of $\chi^{2}_{\mathrm{Fit}}$; and 
they are comparable in terms of $P_{\mathrm{Fit}}$. Our 
thirteen `best' models range from 
$\chi^{2}_{\mathrm{Fit, Pert}} \sim 311.7 - 319.0$, with the 
``semi-optimized" run (and others near it in the `preferred 
$\Psi _{\mathrm{MD}}$ parameter space trench' discussed above) 
having $\chi^{2}_{\mathrm{Fit, Pert}} < 
\chi^{2}_{\mathrm{Fit}, \Lambda \mathrm{CDM}} = 311.9$; and the 
$P_{\mathrm{Fit}}$ values for most of these thirteen models are 
almost as good as the Concordance $\Lambda$CDM result here, 
with ten of them having $P_{\mathrm{Fit, Pert}} > 0.3$, and five 
of those going as high as $P_{\mathrm{Fit, Pert}} \sim 0.36 - 0.37$ 
(as compared to $P_{\mathrm{Fit}, \Lambda \mathrm{CDM}} = 0.38$ 
for the fully-optimized case of $\Omega _{\Lambda} = 0.713$, 
$H_{0} = 69.96$ $\mathrm{km} ~ \mathrm{s}^{-1} \mathrm{Mpc}^{-1}$). 
In short, it seems fair to conclude that the smoothly-inhomogeneous 
formalism presented in this paper -- even at this `toy model' 
stage, given all of the simplifications discussed above 
-- is already able to successfully reproduce the apparent 
acceleration of the universe essentially as well as the 
more standard $\Lambda$CDM paradigm can do. 

Next, to check on the general validity of these models all the 
way to $t = t_{0}$, we must consider the diagnostic parameter 
$I_{0} \equiv I(t_{0})$, which varies from $I_{0} \sim 0.52 - 0.72$ 
for these thirteen best runs, and stays bounded at $I_{0} \le 0.66$ 
for eleven of them. As discussed in Section~\ref{Sub1CausUpdMetr}, 
where $I(t)$ was formally constructed, this function (representing 
the causally-integrated influence of inhomogeneities) is only valid 
in the linear context of Newtonian gravitational terms, for which 
the metric perturbations from different masses can be simply 
summed together. To the extent that $I(t)$ approaches unity, the 
linearized-gravity approach of our formalism breaks down, and 
full general relativity becomes necessary for describing the 
combined effect of all perturbations, in total. 

Now, seeing $I_{0}$ take on values roughly midway between 
$0$ and $1$ for these runs is in fact a good sign for trusting the 
general predictions of these models. If $I_{0}$ (and thus $d[I(t)]/dt$ 
as well) were too small, then the total perturbative effect of these 
inhomogeneities on the average cosmic gravitational potential would 
be too weak to cause much of an observable effect at all (recall 
Equations~\ref{EqnIshWaldLimitsTot}a,b and the related discussion) 
-- certainly not a reaction strong enough to explain the apparent 
acceleration. For example, the very weak FRW-perturbing behaviors 
of the $\Psi _{\mathrm{Sqr}}$ models (cf. Figures~\ref{FigSqrPlots}), 
which have $I_{0} \lesssim 0.2$, is symptomatic of their 
limited accelerative effects (i.e., $w^\mathrm{Obs}_{0} > -0.5$). 
On the other hand, too large a value -- such as $I_{0} \simeq 0.93$, 
from the strongest of all of our models (the $\Psi _{\mathrm{MD}}$ case 
with ($\Psi _{0},z_\mathrm{init}) = (1.0,25)$) -- is so large that the 
detailed quantitative results of such a model cannot really be trusted. 

The fact that the perturbative effects from our thirteen best-fitting 
model cosmologies do not completely overshoot the linearity 
approximation, and thus remain fairly trustworthy in their output 
results, is not completely a matter of luck: one key feature of the 
cosmic acceleration (and the source of the Coincidence Problem) is that 
it seems to have just recently begun, being due (in the backreaction 
paradigm) to the same structure formation that recently (in cosmological 
terms) has created us, as well. But this `luck' will not hold out in the 
long term, since $I(t)$ should continue to grow in the future for many 
billions of years, until $I(t \gg t_{0})$ eventually becomes so large 
that the evolution of the real universe itself (and not just our 
phenomenological representation of it) should cease to be describable 
by any simply-perturbed FRW/FLRW model. The dynamical behavior of such 
a universe would likely become quite extreme and difficult to predict, 
and we will speculate upon the possible cosmic futures in 
Section~\ref{SecFate}.

Moving on to the observable cosmological parameters for each 
run, we first note the changes to $z^{\mathrm{Obs}}$, so that 
it is now no longer precisely equal to $z^{\mathrm{FRW}}$. 
Table~\ref{TableSimRunsCosParams} gives results corresponding to 
$z^{\mathrm{FRW}} = 1$ (i.e., $[a_{0}/a(t)] = 2$) as a sample 
epoch, demonstrating that while the differences are not huge 
(e.g., $z^{\mathrm{Obs}} \sim 11 - 14 \%$ larger than 
$z^{\mathrm{FRW}}$ for these thirteen runs), they are also 
non-negligible, implying that the backreaction effects of 
inhomogeneities have a real ability to alter the relationships 
between observed cosmological time, volume, and redshift, 
which must be taken into account for any precision understanding 
of the evolution of astrophysical objects.

The effects upon the Hubble Constant, on the other hand, 
are much larger in magnitude, as we have both expected and 
required, in order to create an alternative concordance. First, 
considering the individually-optimized values of the observed 
Hubble Constant for each run, we see that our thirteen best runs 
(along with the flat $\Lambda$CDM model) have a range of 
$H^{\mathrm{Obs}}_{0} \sim 68.8 - 71.8$, which compares well 
with the result of $H_{0} = 70.1 \pm 1.3 ~ 
\mathrm{km} ~ \mathrm{s}^{-1} \mathrm{Mpc}^{-1}$ from 
combined (WMAP+BAO+SN) data \citep{WMAP5yrBasicRes}. 
This is an encouraging cross-check of these runs, in contrast to 
the utter failure of $\Omega _\mathrm{M} = 1$ SCDM (with optimized 
$H^{\mathrm{Obs}}_{0} = 61.35$) to properly fit either the SNe data 
or the Hubble Constant; similar to the tendency of our 
model simulation runs with poor SNe $P_{\mathrm{Fit}}$ values 
(not presented here) to {\it also} have 
observationally discrepant $H^{\mathrm{Obs}}_{0}$ values. 

Next, considering the unperturbed/``FRW" Hubble Constant 
values for these thirteen best runs, we get results of 
$H^{\mathrm{FRW}}_{0} \sim 34.2 - 42.8$, with six of the very best 
values lying within $H^{\mathrm{FRW}}_{0} \sim 36.1 - 41.6$. 
These results are entirely in accord with our discussion from 
Section~\ref{Sub2HubbConstAge}, in which we noted how the findings 
of other authors have indicated that a variety of other cosmological 
measurements are seemingly concordant with $\Omega _{\mathrm{M}} = 1$ 
models that have a low Hubble Constant, only to be stymied by 
direct observations apparently indicating $H_{0} \sim 70$. But here 
we see that the ability of our inhomogeneity-perturbed formalism 
to achieve a value of $H^\mathrm{Obs}_{0}$ in the $\sim$70's, 
despite having a `true' Hubble Constant -- i.e., the one that actually 
mattered during the pre-structure-formation cosmic epoch -- of 
$H^\mathrm{FRW}_{0}$ in the $\sim$30's-40's, completely removes this 
contradiction. Furthermore, it also helps in the establishment of an 
alternative, matter-only concordance in several other ways. 

One important, related aspect of this low value of $H^{\mathrm{FRW}}_{0}$ 
is that we have now {\it solved the Age Problem} in consequence, 
as described above in Section~\ref{Sub2HubbConstAge}, without needing a 
period of recent `real' acceleration that would nominally be provided 
by Dark Energy. Our results for these thirteen runs are 
$t^\mathrm{Obs}_{0} \simeq 13.2 - 14.5$ GYr (and 
$t^\mathrm{Obs}_{0} \simeq 13.4 - 14.2$ GYr for the nine best fits), 
which compares very favorably to the SN-best-fit flat $\Lambda$CDM 
value of $13.64$ GYr (noting again that the time estimates here are not 
exact, since none of these models include radiation; though they are 
all mutually consistent for $\Omega _\mathrm{R} \equiv 0$). 
Now, a cosmic age difference of a $\sim$billion years either way 
is certainly not negligible, and careful observations of objects like 
globular clusters, etc., could be expected to someday strongly 
discriminate between $\Lambda$CDM and our formalism (as well to as help 
optimize parameters {\it within} our formalism, for different clumping 
evolution models). But given the typical state of cosmic age lower-limits, 
taken from the measured ages of astrophysical objects -- for 
example, like that from a review on Cosmic Age in \citet{WMAP1yrCosmo}, 
where the strongest lower limit that they quoted was $12.7 \pm 0.7$ GYr, 
from observations of cooling White Dwarfs -- it seems reasonable to 
conclude that the values given here for $t^\mathrm{Obs}_{0}$ from our 
formalism, even at this `toy model' stage, are good enough to have 
effectively solved the Age Problem possessed by SCDM; especially 
considering that $t_{0, \mathrm{SCDM}}$, in contrast, is as low as 
$\sim$10.6 GYr, as seen in Table~\ref{TableSimRunsCosParams}.

Given $H^{\mathrm{FRW}}_{0}$ (along with $H^{\mathrm{Obs}}_{0}$ 
and $I_{0}$) for each run, we can also now address the question 
of the spatial flatness of the universe, by computing 
$\Omega^\mathrm{FRW}_\mathrm{M}$ via Equation~\ref{EqnOmegaMFRW}b. 
This first requires us to choose some value of 
$\Omega^\mathrm{Obs}_\mathrm{M}$ from cosmological observations. 
Now, this value need not be equal to $(1 - 0.713) = 0.287$, which 
one might infer from the SN-best-fit flat $\Lambda$CDM model above, 
since it would be better to use a value that comes from a more 
comprehensive combination of different data sets; and a value of 
$\Omega^\mathrm{Obs}_\mathrm{M} \simeq 0.287$ would in fact be rather 
high, given most recent estimates of $\Omega^\mathrm{Obs}_\mathrm{M}$. 

The actual amount of matter as a fraction of the apparent 
(i.e., observational) closure density is still fairly difficult 
to pin down with great precision. \citet{WMAP3yrCosmo} 
demonstrated in their Tables 5 and 6 that which best-fit value 
of $\Omega^\mathrm{Obs}_\mathrm{M}$ they 
obtained depended upon which external data set (if any) that 
they chose to combine the ($3^{\mathrm{rd}}$-Year) WMAP data set 
with, with variations of $\Omega _\mathrm{M} \sim 0.226 - 0.299$ 
possible. \citet{WMAP5yrLikeliParams} noted the tension in the 
preferred (high versus low) values of the matter density observed 
by SDSS ($\Omega _\mathrm{M} = 0.265 \pm 0.3$), and 2dFGRS 
($\Omega _\mathrm{M} = 0.236 \pm 0.02$). And while they described 
how the uncertainty in the matter density had dropped with each 
new analysis of the growing WMAP data set, it is also true that 
there has been some oscillation in those estimated values, 
with the $1^{\mathrm{st}}$-Year WMAP analysis giving 
$\Omega _\mathrm{M} = 0.27 \pm 0.04$ \citep{WMAP1yrRes}, 
the 3-Year WMAP mean dropping to 
$\Omega _\mathrm{M} = 0.241 \pm 0.034$, and the 5-Year WMAP 
mean rising back up (despite a max likelihood of only 0.249) 
to $\Omega _\mathrm{M} = 0.258 \pm 0.03$ 
\citep[][Table 2]{WMAP5yrLikeliParams}. Furthermore, 
that 5-Year WMAP-only result actually increases to 
$\Omega _\mathrm{M} = 1 - 0.721 = 0.279$ 
\citep[][Table 6]{WMAP5yrBasicRes} when complementary 
data sets are included (WMAP+BAO+SN). Similarly, 
\citet{WMAP7yrLikeliParams} give 
$(\Omega _{c} + \Omega _{b}) \simeq 0.274$ for their 
(updated) 7-Year WMAP values. 

For the calculations of $\Omega^\mathrm{FRW}_\mathrm{M}$ 
in this paper, we choose to adopt what seems to be a 
generally reasonable value for our search for a new 
concordance: $\Omega^\mathrm{Obs}_\mathrm{M} \equiv 0.27$. 
What must be realized in these calculations of the `true' 
(pre-perturbation) spatial curvature 
$\Omega^\mathrm{FRW}_\mathrm{M}$, however, 
is that although we have adopted one specific value of 
$\Omega^\mathrm{Obs}_\mathrm{M}$ for quoting values of 
$\Omega^\mathrm{FRW}_\mathrm{M}$, we can only trust our 
estimates of spatial flatness (i.e., 
$\Omega^\mathrm{FRW}_\mathrm{M} \rightarrow 1$) as being 
accurate to within the observational uncertainties in 
$\Omega^\mathrm{Obs}_\mathrm{M}$; for example, to within 
a (very approximate) range of, say, 
$\sim$$(1/0.27) \cdot [\pm 0.1] \simeq \pm [0.3 - 0.4]$, 
or so. 

Looking at the $\Omega^\mathrm{FRW}_\mathrm{M}$ results for 
our thirteen best runs in Table~\ref{TableSimRunsCosParams}, 
we can clearly say that to within these uncertainties, our 
smoothly-inhomogeneous cosmological formalism has indeed 
achieved spatial flatness; and that this has been done 
without requiring the incorporation of any ``missing mass" 
or ``missing energy" (beyond the usual Cold/Hot Dark Matter), 
such as Dark Energy, to bring $\Omega _{\mathrm{Tot}}$ 
up to unity. 

Specifically, we see that these $\Psi _{\mathrm{Lin}}$ runs 
yield $\Omega^\mathrm{FRW}_\mathrm{M} \sim 0.90 - 1.04$, and 
that the $\Psi _{\mathrm{MD}}$ runs yield 
$\Omega^\mathrm{FRW}_\mathrm{M} \sim 0.99 - 1.64$ 
(narrowing down further to 
$\Omega^\mathrm{FRW}_\mathrm{M} \sim 0.99 - 1.41$, if we drop 
the worst of these seven $\Psi _{\mathrm{MD}}$ cases). Thus the 
$\Psi _{\mathrm{Lin}}$ runs are especially good at achieving 
flatness. And the $\Psi _{\mathrm{MD}}$ runs, even with their 
stronger effects, are not too far off either: if one (reasonably, 
as we have seen) chooses a lower value of the observed matter 
density -- say, $\Omega^\mathrm{Obs}_\mathrm{M} = 0.24$, instead 
of $0.27$ -- then this drops the range down to 
$\Omega^\mathrm{FRW}_\mathrm{M} \sim 0.88 - 1.25$ for the six best 
$\Psi _{\mathrm{MD}}$ runs, making them even more consistent with 
$\Omega^\mathrm{FRW}_\mathrm{M} \equiv \Omega _{\mathrm{Tot}} = 1$. 
Even given the observational uncertainties, as well as the many 
theoretical simplifications of our formalism, it is a notable 
step towards the achievement of an alternative concordance that 
such a trial-and-error set of sixty simulated cosmological models 
(with astrophysically-motivated parameters) has been found to 
include about a dozen runs that succeed in bringing 
$\Omega ^{\mathrm{FRW}} _{\mathrm{Tot}}$ from $\sim$$0.3$ up to 
a reasonably close range around $1$, using matter alone. 

Next, we consider the quantitative amount of apparent 
acceleration -- i.e., the degree to which 
$q^\mathrm{Obs}_{0} < 0$, $w^\mathrm{Obs}_{0} < (-1/3)$ -- 
that our models have produced. Table~\ref{TableSimRunsCosParams} 
shows that each of these thirteen best runs have produced a 
{\it strong} amount of `acceleration', with 
$w^\mathrm{Obs}_{0} \sim (-0.71) - (-0.82)$ 
[i.e., $q^\mathrm{Obs}_{0} \sim (-0.56) - (-0.73)$] 
for the $\Psi _{\mathrm{Lin}}$ runs; and 
$w^\mathrm{Obs}_{0} \sim (-0.75) - (-1.0)$ 
[i.e., $q^\mathrm{Obs}_{0} \sim (-0.62) - (-1.0)$] 
for the $\Psi _{\mathrm{MD}}$ runs. 

Now, a flat $\Lambda$CDM universe with a given value of 
(Cosmological Constant) $\Omega _{\Lambda}$ will have 
$w^\mathrm{Obs}_{0} = - \Omega _{\Lambda}$. Taking the 
values of $\Omega _{\Lambda} = 0.742 \pm 0.03$ 
(WMAP-only) and $\Omega _{\Lambda} = 0.721 \pm 0.015$ 
(WMAP+BAO+SN) from \citet{WMAP5yrBasicRes} as a guide, 
it seems like we would look for our models to reproduce 
values in the range of, say, 
$w^\mathrm{Obs}_{0} \sim (-0.70) - (-0.77)$, in order to look 
like Concordance $\Lambda$CDM. As we see, four of these 
$\Psi _{\mathrm{Lin}}$ runs, and one of these 
$\Psi _{\mathrm{MD}}$ runs, do in fact fall within this range, 
with the others producing somewhat stronger `acceleration' 
effects. 

While this is already a fairly good success rate for our models 
in reproducing the apparent acceleration, it is important to 
note that the precise value of $w^\mathrm{Obs}_{0}$ which is 
`observed' is actually just the result of some particular best-fit 
procedure, given some assumed cosmological model; and output 
parameters from best-fits depend non-trivially upon the fitting 
function used. For example, \citet{KowalRubinSCPunion} quote a 
mean value (from SN+BAO+CMB data) of $w^{\Lambda}_{0} = -0.969$ 
for a constant-$w^{\Lambda}$ analysis, which in conjunction 
with their $\Omega _{\Lambda} = 0.713$, turns into 
$w^{\mathrm{Obs}}_{0} = (0.713) \cdot (-0.969) = -0.691$. 
Alternatively, in their varying Dark Energy EoS analysis using 
$w^{\Lambda}(z) \equiv w^{\Lambda}_{0} 
+ [w^{\Lambda}_{\mathrm{a}} z/(1 + z)]$, they get 
(Rubin, D. 2008, private communication) a mean value of 
$w^{\Lambda}_{0} = -1.13$ (with $w^{\Lambda}_{\mathrm{a}} = 0.73$), 
which in conjunction with a re-fit value of $\Omega _{\Lambda} = 0.718$, 
yields $w^{\mathrm{Obs}}_{0} = -0.811$; this is clearly a substantial 
change from $w^{\mathrm{Obs}}_{0} = -0.691$, just given a change in 
the fitting assumptions. The conclusion here is that it is not 
necessary to {\it precisely} match the amount of `acceleration' 
(i.e., $w^{\mathrm{Obs}}_{0}$, $w^{\mathrm{Obs}}(z)$) found with 
the best-fit Dark Energy models, in order to say that one has 
reproduced the apparent cosmic acceleration. Rather, all that one 
has to do is to produce an apparent acceleration that is very roughly 
in the correct range of the old Concordance $w^{\mathrm{Obs}}_{0}$, 
while simultaneously producing a fit to the SNe data that is 
essentially as good or better than $\Lambda$CDM. As seen above, we 
have achieved this by obtaining $\sim$$5-10$ runs here with fits of 
a quality comparable to that of $\Lambda$CDM, with nearly all of 
them having an apparent acceleration within the fairly good range 
of $w^{\mathrm{Obs}}_{0} \sim (-0.7) - (-0.9)$. 

One further way of characterizing the cosmic acceleration, 
in particular its `sudden' onset, is to describe the universe 
as having recently experienced a strong, positive ``jerk" 
\citep{RiessGoldSilver}. Given our definition of the jerk/jolt 
parameter in Equation~\ref{EqnDefnj0}, we find our six 
$\Psi _{\mathrm{Lin}}$ runs in Table~\ref{TableSimRunsCosParams} 
to have $j^\mathrm{Obs}_{0} \sim 2.5 - 3.5$, while the seven 
$\Psi _{\mathrm{MD}}$ runs have 
$j^\mathrm{Obs}_{0} \sim 2.6 - 5.5$. Recalling from 
Section~\ref{Sub2Params} that spatially flat cases of both 
SCDM and Cosmological Constant $\Lambda$CDM models (for any value 
of $\Omega _{\Lambda} = 1 - \Omega _\mathrm{M}$) {\it always} have 
$j^\mathrm{Obs}_{0} = 1$, it thus appears that these universally high 
values of $j^\mathrm{Obs}_{0}$ -- universal, that is, for those of our 
models also capable of fitting the SNe data and $w^{\mathrm{Obs}}_{0}$ 
properly -- is the result representing {\it the most discriminating test} 
that we have found so far which could potentially distinguish our 
formalism from traditional Concordance Models with a Cosmological 
Constant (or anything close to it) that are capable of generating the 
apparent acceleration. Given that the quantity $j^\mathrm{Obs}_{0}$ 
-- or its corresponding alternative in `dynamic' parameterizations, 
$w^{\Lambda}_{\mathrm{a}}$ -- comes from the third-derivative term 
in the expansion for $d_{\mathrm{L}}$ in $z^\mathrm{Obs}$ 
(cf. Equation~\ref{EqnLumDistExpansion}), its value is not 
yet observationally well constrained. In that sense, we will regard 
the output values of $j^\mathrm{Obs}_{0}$ from our runs as a 
{\it prediction} of our formalism, rather than as an attempt to 
match known results; a matter which we will explore further 
in Section~\ref{Sub2TestingViaj0} below, where we consider (in brief) 
the observational situation for $j^\mathrm{Obs}_{0}$ as it stands. 

Unfortunately, though measurements of $j^\mathrm{Obs}_{0} > 1$ 
would conclusively rule out $\Lambda$CDM, we cannot conclude with 
certainty, based only upon these above results, that high jerk parameters 
are a robust feature of our formalism. Too many theoretical uncertainties 
exist within our models to make this an `iron-clad' prediction of our 
calculations just yet. A number of different effects should all act to 
moderate the strength of the apparent acceleration as $z \rightarrow 0$, 
$t \rightarrow t_{0}$ -- thus weakening the late-time ``cosmic jerk'' 
-- both within our model calculations and in the real universe. 

First, there is the issue of ``recursive nonlinearities" discussed 
earlier in Section~\ref{Sub1CausUpdMetr}: the fact the pre-existing 
inhomogeneities serve to slow down the causal updating which brings 
in later information about new inhomogeneities, thus softening 
the apparent acceleration effect by some undetermined degree. 
As calculating these effects properly would take an algorithmically 
nonlinear simulation program, we cannot quantitatively estimate 
them here; all that we can say is that they would provide some 
significant amount of damping of the apparent acceleration effects 
due to causal backreaction, especially at later times.

Second, even if we could safely rely upon this and other 
simplifications in our formalism, and even if the functional 
forms of the $\Psi _{\mathrm{Lin}}$ and/or $\Psi _{\mathrm{MD}}$ 
clumping evolution models (cf. Equations~\ref{EqnClumpModels}a,b) 
generally do make sense for the long-term growth of structure in 
the universe, there is still no guarantee that those cluster 
evolution functions would remain meaningful all the way to 
$t \rightarrow t_{0}$. As structure forms, the large-scale 
collapse of material and feedback from star formation act to 
shock heat cosmic baryons to millions of degrees, particularly 
at late times ($z^{\mathrm{Obs}} \lesssim 3$), thus inhibiting 
clumping\footnote{On the other hand, given that simulations 
\citep{HoBahcallClustEllipt} show that galaxy cluster 
ellipticities are still decreasing as $z \rightarrow 0$ due 
to continued relaxation, this means that at least some 
virialization does continue all the way to the current time, 
even for structures which largely are `already clumped' -- thus 
implying a continuing ``backreaction" of the kind which we 
phenomenologically model here. But whether or not this 
continuing level of virialization would be strong enough to 
substantially counter the slowing of such backreaction at 
late times due to shock heating, it would in any case 
represent yet more complexity in determining the most 
physically realistic functional form to use for $\Psi (t)$.} 
by keeping and/or sending a significant 
portion of the baryons into the superheated IGM 
\citep[e.g.,][]{CenOstrikerShockedOb}. Now, our models can 
account for {\it some} of this effect, simply by using a lower 
value of $\Psi _{0}$; but if this shock-heating is strong enough 
to really soften the evolution of growth of cosmic structure 
(thus significantly altering the functional form of $\Psi (t)$, 
itself), then this would be a moderating effect that 
is not taken account of by any of the (highly simplified) 
clumping evolution functions that we use in this paper. 

Any of these effects which weaken the apparent acceleration as 
$z \rightarrow 0$ are generally bad for our $\Psi _{\mathrm{Lin}}$ 
models, since this could make at least some of them too weak to 
produce a new cosmic concordance; alternatively, they are generally 
good for our $\Psi _{\mathrm{MD}}$ models, by reducing their 
computed values of $( H^\mathrm{Obs}_{0} / H^\mathrm{FRW}_{0} )$, 
$\Omega^\mathrm{FRW}_\mathrm{M}$, and $w^\mathrm{Obs}_{0}$. 
In any case, however, such effects would likely lead to a 
reduction in our calculated output values of $j^\mathrm{Obs}_{0}$. 
The magnitude of such a change is not certain, since the 
primary effect of such alterations might simply be to change 
{\it which} choices of model input parameters (i.e., which 
$(\Psi _{0},z_\mathrm{init})$ and/or which type of $\Psi(t)$ 
function) manages to fit the cosmological data best, {\it without} 
changing the general values of $j^\mathrm{Obs}_{0}$ (or of other 
output parameters) that tend to emerge from those models that do 
achieve good SNe fits. But all we can say for certain here, 
examining Table~\ref{TableSimRunsCosParams} once more, is that 
for the seven runs `most acceptable' on every front -- i.e., 
those runs with $\chi^{2}_{\mathrm{Fit}} < 315$, 
$t^\mathrm{Obs}_{0} \sim 13.4 - 14.0$ GYr, 
$\Omega^\mathrm{FRW}_\mathrm{M} \sim 0.97 - 1.3$, and 
$w^\mathrm{Obs}_{0} \sim (-0.75) - (-0.87)$ -- that we have 
the range of $j^\mathrm{Obs}_{0} \sim 2.6 - 3.8$ for the jerk 
parameter. It therefore seems reasonable, at this stage, to 
propose that such a range of $j^\mathrm{Obs}_{0}$ may be a 
generic prediction from our formalism for models which 
succeed in achieving a good alternative concordance -- 
a conclusion subject, of course, to further theoretical 
development of this paradigm. 

Lastly in our evaluation of these thirteen best model runs, 
we consider the acoustic scale of the CMB peaks, 
$l^\mathrm{Obs}_{\mathrm{A}}$, as derived above in 
Section~\ref{Sub2CMB}. As noted there, this parameter is 
the one incorporating the most far-ranging assumptions and 
simplifications, and the one most prone to error due to the 
great look-back time and distance to the last scattering surface. 
The theoretical uncertainties in our results for 
$l^\mathrm{Obs}_{\mathrm{A}}$ are therefore expected 
to be large, though we cannot precisely quantify them here.

Table~\ref{TableSimRunsCosParams} shows that these six best 
$\Psi _{\mathrm{Lin}}$ runs have 
$l^\mathrm{Obs}_{\mathrm{A}} \sim 284.2 - 291.3$, while the 
six best $\Psi _{\mathrm{MD}}$ runs (dropping the 
worst of these $\Psi _{\mathrm{MD}}$ SNe fits) have 
$l^\mathrm{Obs}_{\mathrm{A}} \sim 270.5 - 288.7$. Narrowing it 
down even further, the six runs most acceptable on every front 
(i.e., the seven `most acceptable' runs referred to previously, 
now minus the ``Semi-Optimized" $\Psi _{\mathrm{MD}}$ run) 
have $l^\mathrm{Obs}_{\mathrm{A}} \sim 277.5 - 288.7$. Those 
values succeed very well in bracketing the SN-best-fit 
$\Lambda$CDM value (for no radiation) of 
$l^\mathrm{Obs}_{\mathrm{A}, \Lambda \mathrm{CDM}} = 285.4$, 
while simultaneously producing good values for all of the other 
cosmological parameters computed in this paper, as well. 

Such results, while serving as a fairly good match of our models 
to the CMB acoustic scale, may not quite match the precision of 
the actual $\pm$$\sim$$0.8-0.9$ measurement uncertainties quoted 
in observational CMB parameter estimation results for 
$l^\mathrm{Obs}_{\mathrm{A}}$ \citep[e.g.,][]{WMAP5yrBasicRes}. 
Nevertheless, it is significant to note how much better these 
estimates are than any fit achievable with a 
low-$\Omega _{\mathrm{M}}$ {\it open} CDM model without Dark Energy. 
For example, using the rough rule-of-thumb \citep[][]{TurnerCosmoPar} 
of $l^\mathrm{Obs}_{1} \simeq 200 ~ \Omega ^{-1/2} _{\mathrm{Tot}}$ 
for the first CMB peak, shifting from a flat SCDM universe 
($\Omega _{\mathrm{M}} = 1$) to an oCDM universe 
($\Omega _{\mathrm{M}} \sim 0.3$, $\Omega _{\Lambda} \equiv 0$) 
increases $l^\mathrm{Obs}_{1}$ by $\sim$$200 [(1 / \sqrt{0.3}) - 1] 
\simeq 165 \equiv \Delta l^\mathrm{Obs}_{\mathrm{A}}$. Keeping 
$\Omega^\mathrm{Obs}_\mathrm{M} \sim 0.3$ without some mechanism 
to bring about spatial flatness -- such as Dark Energy, or 
our smoothly-inhomogeneous perturbation formalism -- therefore 
leads to errors in $l^\mathrm{Obs}_{\mathrm{A}}$ of 
{\it over a hundred;} whereas insisting on flat SCDM mostly saves 
$l^\mathrm{Obs}_{\mathrm{A}}$ (see 
Table~\ref{TableSimRunsCosParams}), but at the cost of ruining the 
fit to the SNe data ($P_{\mathrm{Fit}} \simeq 3.4 \times 10^{-22}$). 
Our formalism, on the other hand, preserves the good SNe fit 
(and good values for all of the other cosmological parameters 
calculated here), while producing deviations from the Concordance 
$\Lambda$CDM model of only 
$\Delta l^\mathrm{Obs}_{\mathrm{A}} \lesssim 15$ for twelve of 
these thirteen best runs (and to within 
$\Delta l^\mathrm{Obs}_{\mathrm{A}} \sim 0.2 - 7.9$ for the six 
very best runs), all while maintaining $\Omega _{\Lambda} \equiv 0$. 

To sum up the results of this subsection: from the basic set 
of sixty simulation runs done with our inhomogeneity-perturbed 
formalism using astrophysically-motivated input parameters, we 
have obtained twelve runs that produce quantitatively good fits 
to residual Hubble diagrams of the SCP Union supernovae, many of 
them very close in $\chi^{2}_{\mathrm{Fit}}$ to the best-fit done 
with flat $\Lambda$CDM. In addition to these, we include a 
thirteenth run, taken as one example of a long trench in 
$(\Psi _{0},z_\mathrm{init})$-space containing many 
$\Psi _{\mathrm{MD}}$ runs which produce nearly equal values of 
$\chi^{2}_{\mathrm{Fit}}$, all lower than that achievable with any 
flat $\Lambda$CDM model. These thirteen-plus causal backreaction 
models have therefore successfully reproduced the signs of 
apparent cosmic acceleration that are seen in Type Ia SNe data sets. 
Furthermore, as it is not surprising that good fits to the SNe data 
also imply good values for other parameters, we see that six of 
these runs -- specifically, $\Psi _{\mathrm{Lin}}$ with 
$(\Psi _{0},z_\mathrm{init}) = (1.0,25)$, $(1.0,15)$, 
and $(0.96,25)$, and $\Psi _{\mathrm{MD}}$ with 
$(\Psi _{0},z_\mathrm{init}) = (0.78,10)$, $(0.85,5)$, and 
$(0.92,5)$\footnote{Recalling once more that larger $z_\mathrm{init}$ 
and mid-range $\Psi _{0}$ are the most `astrophysically sensible' 
values for these parameters, generally speaking.} -- not only provide 
excellent SNe fits, but also reproduce several other cosmological 
parameters to within acceptable `Concordance-level' bounds, 
including $t^\mathrm{Obs}_{0}$, $\Omega^\mathrm{FRW}_\mathrm{M}$, 
$w^\mathrm{Obs}_{0}$, and $l^\mathrm{Obs}_{\mathrm{A}}$. 
And beyond just achieving an agreement for already-known observable 
parameters with these models, we also find them to yield 
$j^\mathrm{Obs}_{0} \sim 2.6 - 3.8$, indicating that observations 
of a jerk parameter value significantly greater than unity would 
as of now appear (pending further theoretical refinements of 
these calculations) to serve as a potentially powerful way 
of distinguishing our formalism from Cosmological Constant 
$\Lambda$CDM, as we will examine further next. 

We note once again that all of this is achieved in our models 
{\it without} the incorporation of any Dark-Energy-type 
(i.e., negative pressure) species, but is instead done simply 
with a spatially flat, matter-dominated universe that is 
perturbed by causally-propagating information about 
self-stabilizing inhomogeneities.

\subsection{\label{Sub1ModelTesting}Distinguishing the 
Smoothly-Inhomogeneous Formalism from Concordance 
$\Lambda{\mathrm{CDM}}$}

A candidate paradigm for an alternative Cosmic Concordance, 
such as our formalism, must simultaneously satisfy a variety of 
fairly independent observational constraints, including the 
parameters considered above plus several other types of observations. 

Assuming that such a goal can be fully achieved, it becomes rather 
difficult then to find distinctions which would be capable of 
demonstrating a statistically strong preference for the alternative 
concordance over one achievable with Dark Energy (DE); especially so in 
the case of data with high scatter and uncertainties, and a paradigm 
as malleable as DE with an optimizable equation of state. Nevertheless, 
we will discuss a few potential methods here to distinguish our 
formalism from Dark Energy in general, where possible; and more 
feasibly, from Cosmological Constant DE in particular, which despite 
being aesthetically questionable and more constrained (and thus easier 
to falsify) than evolving DE, still appears consistent with a 
wide range of observations.

\subsubsection{\label{Sub2TestViaInhomogs}Direct and Indirect Effects of 
Inhomogeneities: Observational Anisotropies and Other Cosmological Signatures}

One obvious departure of our formalism from any non-clustering version of 
DE is the importance that our model places upon the combined influence of 
many localized inhomogeneities. Though our calculations are done using 
a ``smoothly'' inhomogeneous ansatz in which spatial variations are averaged 
away, in the real universe these effects will not perfectly average to 
smoothness, and thus it becomes more interesting in our paradigm to look for 
evidence of unexpectedly large anisotropies on a variety of scales. 

Though there seems to be no substantial rejection yet of the 
large-scale homogeneity assumed by the Cosmological Principle and 
typical FLRW cosmologies, some intriguing observational results have 
turned up using various methodologies. A preliminary study by this author 
\citep{Bochner22Texas} of anisotropies in the Riess gold04 SNe compilation 
\citep{RiessGoldSilver} showed some marginal, positive signs of the 
existence of real anisotropies; and studies using SNe, galaxy clusters, 
etc., to map anisotropies (or find enhanced variances) in the Hubble Flow 
by other researchers -- e.g., \citet{LahavSNeAniso,McClureDyerHubbAniso,
SchwarzWeinhHubbAniso,RalstonSNeAniso,BlomqvistSNe,SchwarzBackViaSNeAniso,
SarkarUnion2Aniso} -- more or less show similar results, finding signs 
of anisotropy of varying statistical significance. 

Alternatively, considering inhomogeneities seen via the CMB, there have 
been studies concerning the possibility of significant anisotropies 
\citep{Hansen04,Bernui05}, non-Gaussianities \citep{McEwen05,Tojeiro05}, 
and other unexpected features \citep[e.g.,][]{Larson04} present in the 
WMAP data \citep{WMAP3yrTemp,WMAP5yrLikeliParams}, such as the low 
CMB quadrupole \citep[e.g.,][]{WMAP7yrBasicResults}, the CMB 
Axis of Evil \citep{LandMag05}, and the WMAP Cold Spot 
\citep[e.g.,][]{CruzWMAPColdSpot}. Also, some CMB glitches 
\citep[e.g.,][]{HuntSarkarGlitches}, though wiped out by binning 
\citep{WMAP3yrTemp}, may perhaps be signs of real inhomogeneity 
effects that should not be averaged away. (Though one must not ignore 
counterarguments \citep[e.g.,][]{WMAP7BennettAnomReject} that anomalies 
such as these mentioned here might in large part be due to 
{\it a posteriori} selection effects and technological limitations 
in the observations.) 

Still more directly, one may investigate major cosmological structures 
themselves in detail, such as the Shapley Concentration and/or 
Great Attractor regions \citep[e.g.,][]{ShapleyCoreStudy} and the 
Sloan Great Wall \citep{GottSlnGrtWall}, and use them as part of the 
effort to map the detailed inhomogeneity and velocity flow structure 
of the universe on cosmologically large scales 
\citep[e.g.,][]{LuceyAttractor,BolejkoAttractorCMB,SarkarUnion2Aniso}. 
Such studies are particularly interesting, given the evidence of a 
possible ``Dark Flow" \citep{KashDarkFlow,KashDarkFlowPersists} which 
may be due to a tilt across our entire current observational horizon 
(if not due to some causal-inhomogeneity-driven mechanism), as well 
as other evidence of bulk flows out to $100 h^{-1}$Mpc or more 
\citep{CosFlowHumeFeld}.

Searches like these for cosmic inhomogeneities and/or anisotropies 
are already fully underway through the work of many researchers, 
as a way of determining the fine details of the cosmic evolution. 
Our work here merely serves to add to the impetus behind such 
investigations, given our proposal that inhomogeneities may very well 
determine the {\it average} observed cosmological parameters themselves, 
not just the {\it deviations} from the averaged, so-called best-fits.

Even without direct evidence of inhomogeneities, though, the method of 
apparent cosmic acceleration that we propose here would undoubtedly 
impose changes upon several other cosmological measurements, changes 
which we have not explicitly estimated in this paper. Particularly 
interesting are observations which are not explained by (or which 
directly conflict with) the $\Lambda$CDM paradigm. For example, 
the fact that causal backreaction does not occur in completely 
smoothly-distributed fashion, but is instead concentrated mostly near 
virializing masses, may lead to interesting clustering-induced feedback 
behaviors affecting the formation of stars, galaxies and galaxy clusters. 
Conceivably, this may have some useful application to issues such as 
galaxy downsizing \citep{CowieDownsizing}, the cuspy CDM halo problem, 
and the possible dearth of dwarf satellite galaxies \citep{PrimackCuspy}; 
but such connections are speculative, and require a detailed 
quantitative analysis to see if causal backreaction truly succeeds 
(or significantly helps) in explaining these issues. 

Other examples of interesting processes or observables which might 
be affected by causal backreaction include the Late ISW effect 
\citep[e.g.,][]{ISWcorrelLSSwCMB}; the shape parameter, 
$\Gamma \equiv \Omega _{\mathrm{M}} h$ \citep{TurnerCosmoPar}; 
observations of peculiar velocities constraining 
$\sigma _{8} \Omega ^{0.6} _{\mathrm{M}}$ 
\citep{LahavLiddleParams2010}; large-scale velocity flows 
constraining $\beta \equiv \Omega ^{0.6} _{\mathrm{M}} / b$, 
with $b$ as the linear bias factor for galaxies versus the overall 
matter distribution \citep{FukugitaParams1,FukugitaParams2}; 
observations of weak lensing due to large-scale structure, 
constraining $\sim$$\sigma _{8} \Omega ^{0.53 - 0.64} _{\mathrm{M}}$, 
depending upon the angular scales used \citep{FuBenShear}; 
Baryon Acoustic Oscillations, which constrain the combination 
$A \equiv \{ \sqrt{\Omega _{\mathrm{M}}} \phantom{0} 
[H_{0}/H(z_{\ast})]^{1/3} [r_{\ast} / z_{\ast}]^{2/3} \}$, 
with $r_{\ast}$ being the dimensionless comoving distance 
to sampled data having typical redshift $z_{\ast}$ \citep{MarassiBAO}; 
indications of acceleration from the 
Alcock-Paczy\'{n}ski test \citep{AlPacTest,MarBuzzAlPacResult}; 
and observations of the growth and evolution of large galaxy clusters, 
analyzed in conjunction with several other types of measurements 
\citep[e.g.,][]{MantzRapettiDEevo,VikhDEevo}. 

A serious study of how all of these (and other) observations would 
be modified by our apparent acceleration mechanism goes far beyond 
the scope of this paper, and would likely be premature with these 
calculations being essentially in the toy model stage. But any 
eventual finding that our formalism does as well (or better) than 
simple Dark Energy models at achieving a concordance, based upon a 
variety of such measurements, would be convincing evidence in favor 
of our paradigm over the more physically exotic FLRW approach.

\subsubsection{\label{Sub2TestingViaj0}Testing Our Formalism 
with Estimates of the Jerk Parameter, $j^\mathrm{Obs}_{0}$}

Given the importance of the jerk parameter in potentially falsifying 
$\Lambda$CDM (via $j_{0} \neq 1$) and supporting our formalism (if 
$j_{0}$ significantly exceeds unity), it is useful to provide an estimate 
of the actual value of $j_{0}$ here. Accurately measuring $j_{0}$ is quite 
difficult, however, even with the best of today's Precision Cosmology 
data sets, as one is required to go to third order in the expansion 
of the luminosity distance (cf. Equation~\ref{EqnLumDistExpansion}) for 
cosmographic studies; or equivalently, one must go to fits with four terms, 
$(H_{0}, \Omega_{\Lambda}, w^{\Lambda}_{0}$, $w^{\Lambda}_{\mathrm{a}})$, 
in dynamical studies with an evolving Dark Energy EoS, in order to calculate 
$j_{0}$ as a function of $
(\Omega_{\Lambda}, w^{\Lambda}_{0}$, $w^{\Lambda}_{\mathrm{a}})$. 
The difficulty of producing stable, precise estimates of $j_{0}$ and 
higher-order cosmological parameters is evident in \citet{XuWangUnion2j0}, 
in which they use their own cosmographic methods to obtain values like 
$j_{0} = -4.996^{+7.0293}_{-7.331}$ and $j_{0} = 15.665^{+59.715}_{-33.812}$, 
for two cases using different combinations of external data sets in 
conjunction with the recent SCP Union2 SNe compilation 
\citep{AmanRubinSCPunion2}.  

Several previous estimates of $j_{0}$ by this author and by other researchers 
(details not given here), obtained from different SNe data sets and 
produced with a variety of methods, show a general tendency for $j_{0} > 1$ 
(ranging broadly from $j_{0} \sim 0.9 - 5.5$ in most cases); and indeed most 
data sets and analysis methods give values for $j_{0}$ on the high side -- 
though there are important exceptions, such as the Constitution SNe 
compilation \citep{ConstitutionSNeData}, which prefers the rather low 
value of $j_{0} \sim 0.5$. This latter result reflects the significant 
statistical differences between the first SCP Union SNe compilation 
\citep{KowalRubinSCPunion} and the Constitution set; for the former, 
most of the allowed parameter space (see their Figure 16) prefers the 
Phantom Energy regime ($w^{\Lambda}_{0} < -1$), whereas for the latter, 
Figure 4 of \citet{ConstitutionSNeCosm} shows the Constitution set to be 
more centered within the Quintessence region ($w^{\Lambda}_{0} > -1$) of 
DE parameter space. (Conditions of $w^{\Lambda}_{0} < -1$ tend to prefer 
$j_{0} > 1$, barring a reversal due to large values of 
$w^{\Lambda}_{\mathrm{a}}$ of the wrong sign; and vice-versa for 
$w^{\Lambda}_{0} > -1$ favoring $j_{0} < 1$.) 

The Constitution papers themselves 
\citep{ConstitutionSNeData,ConstitutionSNeCosm} do not an include 
an analysis of DE with a time-varying EoS; and here we will focus 
upon SNe from the original SCP Union compilation (our benchmark 
data set for cosmological model fits), along with the more recent 
SCP Union2 compilation (which supersedes the original Union and 
Constitution data sets), using a couple of straightforward methods 
for calculating $j_{0}$. 

As our theoretical motivation is to estimate the purely observable 
parameter $j^\mathrm{Obs}_{0}$ devoid of any assumptions about a DE 
equation of state, the most direct, cosmographic approach is to do 
polynomial best-fits to the SNe luminosity distance values. Using 
the Taylor series expansion for $d_{\mathrm{L}}(z^\mathrm{Obs})$ 
(cf. Equation~\ref{EqnLumDistExpansion}), it is straightforward to 
invert the coefficients from such a polynomial best-fit in order to 
obtain values for $H^\mathrm{Obs}_{0}$, $q^\mathrm{Obs}_{0}$ (or 
equivalently $w^\mathrm{Obs}_{0}$, cf. Equation~\ref{EqnDefnq0andw0}b) 
and $j^\mathrm{Obs}_{0}$ from the first three terms in the expansion. 

The main difficulty, as explored in a key paper 
\citep{CattVisserCosmographSNeFits} examining earlier SNe data sets, 
is that such estimates of cosmological parameters (even for the 
second-order term $q^\mathrm{Obs}_{0}$) are ``distressingly'' unstable, 
depending very sensitively upon which redshift-distance relation one 
selects (the luminosity distance, $d_{\mathrm{L}}$, is not a unique 
choice), which redshift variable one uses, and how many terms one 
employs in the best-fit polynomial. This parameter instability (even 
when the polynomial expansions used all provide quantitatively very 
good fits) renders standard uncertainty estimates useless, as we will 
have to dismiss cases with `unrealistic' parameters out of hand, 
based upon subjective prior judgments of likely ranges for 
$w^\mathrm{Obs}_{0}$ and $j^\mathrm{Obs}_{0}$. 

First, considering the most standard case -- polynomial fits to 
$d_{\mathrm{L}}(z)$ -- it is obvious from 
Equation~\ref{EqnLumDistExpansion} that we must include at least 
three terms (i.e., up to $O [(z)^{3}]$) in order to obtain a value 
for $j_{0}$. But that does not mean that we must use {\it only} three 
terms; in fact, we could include any number of terms, $N \geq 3$, in 
our expansion, and use only the first three of them to obtain a value 
of $j_{0}$ calculated independently of those higher-order terms (thus 
disregarding their weak information regarding higher-order cosmological 
observables). The advantage of using $N > 3$ expansion terms is that this 
gives the best-fit polynomial additional terms for modeling the high-$z$ 
behavior of the data, without forcing that constraining responsibility 
onto the third term from which we must calculate $j_{0}$. As put by 
D. Rubin (2008, private communication), the $N = 3$ case ``truncates 
the series expansion of a(t) after $j_{0}$, so even a LCDM universe 
will not give a fit of $j_{0} = 1$"; a problem alleviated via $N > 3$. 
The downside of using too many terms, however, is that they become 
far too weakly constrained by the data, making the effects of 
statistical uncertainties worse \citep{CattVisserCosmographSNeFits}, 
so that even empirically good fits will yield more and more unrealistic 
cosmological parameters for larger and larger $N$. In practice, 
we perform polynomial fits using $N = (3,4,5,6)$ terms, with the 
$N = (3,4)$ cases usually producing the best parameter values 
(and often the only sensible ones). 

Applying these polynomial fits to $d_{\mathrm{L}}(z)$ for the original 
Union SNe compilation, we obtain 
$(w^\mathrm{Obs}_{0},j^\mathrm{Obs}_{0})=(-0.746,1.32)$ for the $N = 3$ 
case, and $(w^\mathrm{Obs}_{0},j^\mathrm{Obs}_{0})=(-0.822,2.49)$ for 
the $N = 4$ case. Things appear to start going bad for the $N = (5,6)$ 
cases (with $(w^\mathrm{Obs}_{0},j^\mathrm{Obs}_{0}) = (-0.625,-1.76)$ 
and $(-0.918,7.19)$, respectively); but the $N = (3,4)$ cases make a 
range of $j^\mathrm{Obs}_{0} \sim 1.3 - 2.5$ (or more broadly, 
$\sim$$1 - 3$) for the best-fit values seem quite reasonable. This is 
in decent accord with the varying-$w$ analysis in 
\citep{KowalRubinSCPunion}, where the allowed parameter space depicted 
in their Figure 16 shows the $\Lambda$CDM point (i.e., 
$(w_{0},w^{\Lambda}_{\mathrm{a}}) = (-1,0)$, equivalent to $j_{0} = 1$) 
to lie just around the inside edge of the $1 \sigma$ statistical ellipse. 

Applying these same fits to $d_{\mathrm{L}}(z)$ for the Union2 SNe 
compilation, one gets a startling result: the $N = 3$ case produces 
$(w^\mathrm{Obs}_{0},j^\mathrm{Obs}_{0})=(-0.729,0.991)$, and the $N = 4$ 
case yields $(w^\mathrm{Obs}_{0},j^\mathrm{Obs}_{0})=(-0.733,1.050)$. 
(With the $N = (5,6)$ cases again being less physically reasonable, 
yielding $(w^\mathrm{Obs}_{0},j^\mathrm{Obs}_{0}) = (-0.697,0.187)$ 
and $(-0.419,-8.619)$, respectively.) Such estimates of 
$j^\mathrm{Obs}_{0} = 0.991$ and $1.050$ would appear to be amazingly 
good for Cosmological Constant DE, seemingly virtually to prove 
$\Lambda$CDM. But in fact, while these results are most definitely 
{\it consistent} with $\Lambda$CDM, they are misleadingly precise -- 
there is no way that the Union2 SNe data alone could reliably yield 
estimates of $j^\mathrm{Obs}_{0}$ to within $5 \%$ or less of the 
$\Lambda$CDM value, even if that model is true. Such values are 
impossibly accurate, 
requiring us to dig a little deeper in order to get some sense 
of the real parameter estimate uncertainties. 

Following \citet{CattVisserCosmographSNeFits}, we note that 
$d_{\mathrm{L}}(z)$ is not really a good function to use for a 
Taylor series expansion. First of all, the series will not even 
converge for $z \gtrsim 1$ (a relevant redshift range for at least 
$\sim$20 of the Union2 SNe), rendering the high-$z$ behaviors of the 
best-fit polynomials meaningless. Rather, they recommend the use of 
the ``$y$-redshift'' variable, $y \equiv z/(1+z) = [1 - (a/a_{0})]$, 
which is bounded above by unity for past epochs -- 
i.e., $y \in [0,1)$ for $z \in [0,\infty)$. 
Second, in order to remove the dominating influence of the 
``nuisance'' parameter, $H^\mathrm{Obs}_{0}$, which does not 
encode any of the dynamical cosmological information, they recommend 
instead fitting to a function such as $\ln[d_{\mathrm{L}}(y)/y]$. 
Series expansions of this type of function have a first term (the only 
one containing the Hubble Constant) that is decoupled from the higher 
terms and contains no powers of $y$, and can thus be subtracted off 
as a constant offset. Lastly, they note that there is nothing 
necessarily unique or special about the luminosity distance 
function, $d_{\mathrm{L}}$; and that several other cosmological 
distance functions -- e.g., the ``photon count distance'' 
$d_{\mathrm{P}}$, the ``angular diameter distance'' 
$d_{\mathrm{A}}$, etc. -- can be obtained from it simply by 
dividing by different powers of $(1+z) = 1/(1-y)$. This is 
troublesome because fitting data with large scatter and uncertainties 
(such as the SNe data) to different distance functions will produce 
extremely different values of $(q_{0},j_{0})$ -- especially for fits 
to polynomials with a small number of terms, $N$ -- and because (as 
they claim), ``There is no good physics reason for preferring any 
one of these distance variables over the others.'' 

For our analysis here, we will consider fits of the Union2 SNe to 
$\ln[d_{\mathrm{P}}(y)/y]$, where 
$d_{\mathrm{P}} \equiv d_{\mathrm{L}} / (1+z) = d_{\mathrm{L}} (1-y)$. 
We use $d_{\mathrm{P}}$ for three reasons. First, we argue that 
$d_{\mathrm{P}} = (a_{0} ~ r_{\mathrm{SN}})$ actually happens to be 
the most physically appropriate distance function, being simply equal 
to the present physical distance from the SN to us; it is the only 
distance function without an (artificial) factor of $(1+z)$ in its 
definition, and thus it will be finite for any observed physical object 
(even back to the Big Bang, assuming finite cosmological horizons), 
and goes to zero as $z \rightarrow 0$ simply as the real distance 
to such an object would become zero. Second, as $d_{\mathrm{P}}$ 
is in the middle of the five physically-motivated distance functions 
considered by \citet{CattVisserCosmographSNeFits}, it yields the 
median values of their parameter estimates, and quoting the median 
estimates that one may find would seem to be a conservative strategy. 
Third, as a practical matter, we find that none of our polynomial 
fits to $\ln[d_{\mathrm{L}}(y)/y]$ yield reasonable values for 
$j^\mathrm{Obs}_{0}$ (not to mention $w^\mathrm{Obs}_{0}$); whereas 
our $N = 3$ fit (though only that one) to $\ln[d_{\mathrm{P}}(y)/y]$ is 
more well-behaved. 

The result of this $N = 3$, $\ln[d_{\mathrm{P}}(y)/y]$ fit to the Union2 
data set is $(w^\mathrm{Obs}_{0},j^\mathrm{Obs}_{0})=(-0.750,1.338)$. 
This value of $j^\mathrm{Obs}_{0}$ is also consistent with $\Lambda$CDM, 
while being less unbelievably accurate, and gives us some (very 
qualitative) sense of the uncertainties involved. Simultaneously, it 
remains slightly high of $\Lambda$CDM, as our formalism predicts; and 
if not nearly so high as those values of $j^\mathrm{Obs}_{0}$ from our 
simulated cosmologies given above in Table~\ref{TableSimRunsCosParams}, 
the incorporation of recursive nonlinearities into our models will 
probably bring their calculated values of $j^\mathrm{Obs}_{0}$ 
somewhat lower, as discussed previously. Nevertheless, Cosmological 
Constant $\Lambda$CDM remains entirely consistent with these results. 

One other issue for these cosmological parameter fits relates to the 
``outlier rejection'' process performed on these SNe data compilations, 
done as a final supernova cut, after all other cuts based upon the internal 
quality of the data points themselves have been completed. This outlier 
rejection cut is the removal of SNe data points based upon their poor 
($\geq$$3 \sigma$) fits to flat $\Lambda$CDM reference cosmologies that are 
individually optimized and applied to each of the component data sets that 
go into the Union and Union2 compilations. Extensive arguments are given 
in \citet{KowalRubinSCPunion} and \citet{AmanRubinSCPunion2} justifying 
this outlier rejection process as an important technique for robust 
statistical analysis; and yet, since the reference cosmologies used for 
the cuts are invariably the same type of flat $\Lambda$CDM models that are 
at the heart of the Cosmic Concordance Model, there is the unavoidable 
possibility that such outlier rejection cuts may remove some 
legitimate evidence against $\Lambda$CDM, perhaps thereby introducing 
some pro-Concordance bias into the results. 
It is useful here to provide a simple demonstration of the extent to 
which cosmological parameter estimates can be dependent upon the decisions 
made regarding such data cuts. 

This $3 \sigma$ outlier rejection cut removed 8 SNe from the SCP Union 
compilation, reducing the number of SNe to 307 (i.e., culling $\sim$$2.5 \%$ 
of the data); and removed 12 SNe from the Union2 compilation, reducing it to 
557 SNe (i.e., culling $\sim$$2 \%$ of the data). (Furthermore, 
\citet{AmanRubinSCPunion2} applied a $5 \sigma$ `outlier cut' earlier on in 
their analysis, which had already removed $\sim$$6 \%$ of their data; but 
it is not clear if this is the same type of cut as the cosmology-dependent 
outlier rejection done at the end.) Here we specifically consider the Union2 
compilation with the 12 `outlier' SNe added back in (i.e., 569 SNe in total). 
Above, without these outliers, our $N = 3$ fit to $d_{\mathrm{L}}(z)$ yielded 
$(w^\mathrm{Obs}_{0},j^\mathrm{Obs}_{0})=(-0.729,0.991)$; adding the outliers 
back in\footnote{The twelve outlier points, like the rest of the Union2 SNe 
data, are publicly available at http://supernova.lbl.gov/Union; but the 
outlier points are only available in a table including all SNe with lower 
precision values and uncertainties, and with a different $H_{0}$ normalization. 
Going to lower precision for the standard Union2 compilation of 557 SNe only 
changes the cosmological parameter estimates from polynomial fitting by 
$\sim$$0.3 - 5.5 \%$, so for consistency, we do our entire Union2+Outliers 
analysis (569 SNe) using the lower-precision data table.}, this changes to 
$(w^\mathrm{Obs}_{0},j^\mathrm{Obs}_{0})=(-0.736,1.151)$. Similarly, 
the $N = 4$ fit to $d_{\mathrm{L}}(z)$ changes from 
$(w^\mathrm{Obs}_{0},j^\mathrm{Obs}_{0})=(-0.733,1.050)$ to 
$(w^\mathrm{Obs}_{0},j^\mathrm{Obs}_{0})=(-0.826,2.633)$. (Also, 
the $N = 5$ fit to $d_{\mathrm{L}}(z)$ now gives sensible results, 
$(w^\mathrm{Obs}_{0},j^\mathrm{Obs}_{0})=(-0.775,1.351)$.) Meanwhile, 
the $N = 3$ fit to $\ln[d_{\mathrm{P}}(y)/y]$ changes from 
$(w^\mathrm{Obs}_{0},j^\mathrm{Obs}_{0})=(-0.750,1.338)$ to 
$(w^\mathrm{Obs}_{0},j^\mathrm{Obs}_{0})=(-0.810,2.422)$. Adding the 
12 Union2 outliers back in thus moves the best-fit $j^\mathrm{Obs}_{0}$ 
results much farther away from the $\Lambda$CDM value of $j_{0} = 1$ 
than before (though under the circumstances, not representing a 
statistically convincing rejection). What this may ultimately prove 
is not immediately clear, since it is readily apparent from inspection 
of the wild scatter of the outlier SNe points that the quality of 
those measurements is quite mixed, and several of them undoubtedly 
really are bad data points; 
but it does show how sensitive these cosmological parameter estimates 
can be to decisions made about data handling that are nearly invisible 
in the final quoted results. 

Now, these $j^\mathrm{Obs}_{0}$ estimates from polynomial fits to the 
SNe data alone, while interesting, undoubtedly have uncertainties large 
enough to seriously limit their usefulness. In order to get more precise 
estimates of $j^\mathrm{Obs}_{0}$ -- and to be able to calculate some 
believable number for the statistical uncertainty on that value -- we 
are compelled to combine the SNe data with complementary cosmological 
data sets (e.g., SN+CMB+BAO). It is not trivial to do this in a 
completely cosmographic manner, and so we therefore make the concession 
of considering $j_{0}$ in light of the evolving-EoS Dark Energy analyses 
done in the Union and Union2 papers. 

For these analyses, the SCP collaboration adopts the form: 
\begin{equation} 
w^{\Lambda}(z) \equiv w^{\Lambda}_{0} 
+ w^{\Lambda}_{\mathrm{a}} \frac{z}{1 + z} 
= w^{\Lambda}_{0} + w^{\Lambda}_{\mathrm{a}} y 
 ~ , 
\label{EqnDefnwzSCPunion}
\end{equation} 
widely referred to as the Chevallier-Polarski-Linder (CPL) 
parameterization \citep{ChevPolarCPL,LinderCPLexpFormalism} 
for the Dark Energy EoS. 

Using the cosmic expansionary history that would result 
from such a Dark Energy \citep{LinderCPLexpFormalism} to 
relate these CPL parameters to the cosmographic parameters 
$(w^\mathrm{Obs}_{0},j^\mathrm{Obs}_{0})$ obtained 
from the series expansion of $d_{\mathrm{L}}(z)$ 
(cf. Equation~\ref{EqnLumDistExpansion}), 
a straightforward calculation yields: 
\begin{equation} 
w^\mathrm{Obs}_{0} = w^{\Lambda}_{0} ~ 
\Omega _{\Lambda}
 ~ , 
\label{Eqnw0FnOfw0wa}
\end{equation}
which makes obvious sense; and:
\begin{equation} 
j^\mathrm{Obs}_{0} = \{ 1 + [\frac{9}{2} 
\Omega _{\Lambda} w^{\Lambda}_{0} (1 + w^{\Lambda}_{0})] + 
[\frac{3}{2} \Omega _{\Lambda} w^{\Lambda}_{\mathrm{a}}] \} 
 ~ , 
\label{Eqnj0FnOfw0wa}
\end{equation} 
which gives $j^\mathrm{Obs}_{0} = 1$ for the 
Cosmological Constant case of $(w^{\Lambda}_{0} = -1 , 
w^{\Lambda}_{\mathrm{a}} = 0)$, for {\it any} value 
of $\Omega _{\Lambda}$, as required. Furthermore, 
note that Equation~\ref{Eqnj0FnOfw0wa} is an {\it exact} 
expression for $j^\mathrm{Obs}_{0}$ in terms of 
$(\Omega _{\Lambda},w^{\Lambda}_{0},w^{\Lambda}_{\mathrm{a}})$, 
given how they are all defined, even though it has been 
isolated by doing a comparison of various terms between 
two different series expansions.

For use in applying this to the SCP Union compilation 
study in \citet{KowalRubinSCPunion}, D. Rubin states 
(2008, private communication) that their 
best-fit parameter values for the CPL Dark Energy EoS, in a 
combined analysis of Union SNe, CMB, and BAO data sets, is: 
$w^{\Lambda}_{0} = -1.13 ^{+0.15}_{-0.13}$$^{+0.21}_{-0.19}$ 
and $w^{\Lambda}_{\mathrm{a}} = 
0.73 ^{+0.53}_{-0.69}$$^{+0.67}_{-0.82}$, 
with $\Omega _{\mathrm{M}} = 1 - \Omega _{\Lambda} 
= 0.282 ^{+0.018}_{-0.017}$$^{+0.021}_{-0.020}$, where each 
first set of uncertainties includes statistical errors only, 
and with each latter set including statistical plus 
systematic errors. 

Using these best-fit DE EoS values of 
$(\Omega _{\Lambda},w^{\Lambda}_{0},w^{\Lambda}_{\mathrm{a}}) 
= (0.718,-1.13,0.73)$ in Equation~\ref{Eqnj0FnOfw0wa}, 
one obtains $j^\mathrm{Obs}_{0} \simeq 2.26$, if one 
simply plugs those values into this formula for the 
jerk parameter without regard to the probability distributions 
of those parameters. This result for $j^\mathrm{Obs}_{0}$ is 
fairly similar to the value of $2.49$ obtained earlier from 
our $N = 4$ polynomial fit to this Union SNe data alone, it is 
demonstrably above the flat $\Lambda$CDM case, and is not all 
that far out of the lower end of the range of 
$j^\mathrm{Obs}_{0}$ values quoted previously for our 
best simulation runs. 

Using the uncertainties for each of the parameters given above 
(averaging the $\pm \sigma$ values in each case), in conjunction 
with Equation~\ref{Eqnj0FnOfw0wa}, one could naively 
calculate $1 \sigma$ error bars for $j^\mathrm{Obs}_{0}$ of 
$\Delta j^\mathrm{Obs}_{0} \simeq 0.87$ (stat errors only), 
and $\Delta j^\mathrm{Obs}_{0} \simeq 1.14$ (stat plus syst), 
actually placing $\Lambda$CDM more than $1 \sigma$ away in both cases. 
But calculating a valid uncertainty on $j^\mathrm{Obs}_{0}$ is not so 
simple, however: judging by Figure 16 of \citet{KowalRubinSCPunion}, the 
uncertainties on $w^{\Lambda}_{0}$ and $w^{\Lambda}_{\mathrm{a}}$ are 
clearly not independent; and lacking covariance information, we cannot 
calculate a precise value of $\Delta j^\mathrm{Obs}_{0}$. 

For the Union2 compilation, however, we have both more recent and 
more complete information (D. Rubin, 2010, private communication). 
For their stat-errors-only analysis, they obtain best-fit values of: 
$(\Omega _{\Lambda},w^{\Lambda}_{0},w^{\Lambda}_{\mathrm{a}}) 
= (0.727,-1.046,0.160)$, with a (symmetric) covariance matrix of 
$(c_{\Lambda \Lambda},c_{0 0},c_{\mathrm{a} \mathrm{a}},c_{\Lambda 0},
c_{\Lambda \mathrm{a}},c_{0 \mathrm{a}}) 
= (1.782 \times 10^{-4},1.305 \times 10^{-2},3.053 \times 10^{-1},1.766 
\times 10^{-4},7.254 \times 10^{-5},-5.672 \times 10^{-2})$; 
while their stat+syst analysis yields: 
$(\Omega _{\Lambda},w^{\Lambda}_{0},w^{\Lambda}_{\mathrm{a}}) 
= (0.723,-1.134,0.585)$, with a covariance matrix of 
$(c_{\Lambda \Lambda},c_{0 0},c_{\mathrm{a} \mathrm{a}},c_{\Lambda 0},
c_{\Lambda \mathrm{a}},c_{0 \mathrm{a}}) 
= (2.776 \times 10^{-4},2.794 \times 10^{-2},3.370 \times 10^{-1},-4.747 
\times 10^{-4},5.658 \times 10^{-4},-8.803 \times 10^{-2})$. 
A straightforward calculation (though ignoring the non-Gaussian nature of 
the parameter distributions, particularly for $w^{\Lambda}_{\mathrm{a}}$) 
produces final results of: 
$j^\mathrm{Obs}_{0} \pm \Delta j^\mathrm{Obs}_{0} \simeq 1.331 \pm 0.986$ 
(stat errors only), and $\simeq 2.129 \pm 1.292$ (stat+syst). 
Clearly, the uncertainties on the jerk parameter remain quite large 
(with $\Lambda$CDM now back within $1 \sigma$), even when using the 
Union2 SNe compilation data (without outliers) in conjunction with 
the CMB and BAO data sets. Furthermore, these two results differ from 
each other greatly -- not only in the size of the error bars for 
$j^\mathrm{Obs}_{0}$, but also in its best-fit central value -- 
with the only difference between them being whether or not one 
includes systematic uncertainties in the fitting process. 

Summarizing the numerical estimates from this subsection: different 
analyses of the original SCP Union data set yield a variety of values 
for the jerk parameter, including $j^\mathrm{Obs}_{0} \simeq 1.32$ 
($N = 3$ case) and $\simeq 2.49$ ($N = 4$ case) for polynomial fits to 
$d_{\mathrm{L}}(z)$ for the SNe alone; and $j^\mathrm{Obs}_{0} \simeq 2.26$ 
is obtained from a non-cosmographic analysis of an evolving-DE EoS with 
(SN+CMB+BAO) data. For the SCP Union2 data set, alternatively, such 
polynomial fits -- specifically, the $N = 3$ and $N = 4$ fits to 
$d_{\mathrm{L}}(z)$, and the $N = 3$ fit to $\ln[d_{\mathrm{P}}(y)/y]$, 
respectively -- yield three different best-values for the jerk parameter 
of: $j^\mathrm{Obs}_{0} \simeq (0.991,1.050,1.338)$ for the 
official 557 SNe Union2 compilation with outlier rejection, and 
$j^\mathrm{Obs}_{0} \simeq (1.151,2.633,2.422)$ for the 569 SNe Union2 data 
set with the $\Lambda$CDM-outliers added back in. Finally, an evolving-DE 
EoS analysis of Union2 (SN+CMB+BAO) data (557 SNe) yields 
$j^\mathrm{Obs}_{0} \pm \Delta j^\mathrm{Obs}_{0} \simeq 1.331 \pm 0.986$ 
(stat errors only), and $ \simeq 2.129 \pm 1.292$ (stat+syst). 

The conclusions that we may draw from these results can therefore be 
summed up as follows: (1) It seems likely that the jerk parameter lies 
within a (very approximate) range of, let us say, 
$j^\mathrm{Obs}_{0} \sim 0.35 - 3.4$; (2) Very large uncertainties remain, 
with estimates of $j^\mathrm{Obs}_{0}$ remaining highly unstable and 
sensitive to which fitting functions and variables are used to obtain them 
(with many cosmological parameter fits having to be discarded on a purely 
ad-hoc basis), and are also very sensitive to whether or not systematic 
uncertainties are included in the analysis; (3) Various estimates of 
$j^\mathrm{Obs}_{0}$ are systematically high (somewhat above unity), as 
is favored by our formalism; though the Cosmological Constant requirement 
of $j_{0} = 1$ is within $1 \sigma$ of essentially all reasonable fits 
for the cosmological parameters, and going from the Union to the Union2 
SNe compilation does appear to move the results somewhat more towards 
$\Lambda$CDM; (4) The range of jerk parameter values obtained from the 
$\sim$dozen `best' simulation runs with our formalism, 
$j^\mathrm{Obs}_{0} \sim 2.5 - 5.5$ (and 
$j^\mathrm{Obs}_{0} \sim 2.6 - 3.8$ for the 6 very best runs) 
is somewhat high compared to this most likely range of 
$j^\mathrm{Obs}_{0}$ as determined from current observations; but the 
large scatter and uncertainties in the observational data, and the major 
theoretical simplifications in our calculations (e.g., the lack of 
recursive nonlinearities in our current models) may plausibly account 
for this. With further development of our models, and with enlarged and 
improved SNe data sets, it should be possible in the not-too-distant future 
to use observations of the cosmic jerk parameter to definitively support 
or rule out our causally-propagating perturbation formalism as 
a competitor to Concordance $\Lambda$CDM.

\section{\label{SecFate}THE COSMIC FATE: EXPANSION VERSUS DISORDER} 

One of the signature issues to be addressed by any cosmological paradigm 
is the question of its predictions for the cosmic future: What is 
the ultimate fate of the universe? While the toy model presented in 
this paper for calculating the effects of causal backreaction is 
too simplified to produce a detailed quantitative answer to this question, 
it is useful enough for discussing the competing physical processes that 
are involved in determining the final cosmic fate. 

For standard FRW cosmologies -- without any Dark Energy or Acceleration 
-- the issue of the ultimate cosmic fate is a theoretically simple one, 
depending only upon the total mass-energy density, 
$\Omega _\mathrm{Tot}$ \citep[e.g.,][]{KolbTurner}. 
A closed universe ($\Omega _\mathrm{Tot} > 1$) will recollapse in an 
`infinitely' dense ``Big Crunch''; whereas a universe with insufficient 
closure density will either expand eternally (open universe, 
$\Omega _\mathrm{Tot} < 1$) or asymptotically approach zero expansion speed 
(critical/flat universe, $\Omega _\mathrm{Tot} = 1$), in either case 
approaching a thermodynamic `heat death', with an absolute and final cooling 
of the universe into a ``Big Freeze'' (or ``Big Chill''). Cosmic acceleration 
complicates the issue, making it dependent upon the nature of the Dark Energy. 
If the Dark Energy should be some sort of Quintessence that ultimately loses 
its negative-pressure properties and turns into a form of `normal' matter, 
then the cosmic fate would revert to the standard case of depending solely 
upon $\Omega _\mathrm{Tot}$. On the other hand, if the Dark Energy retains 
its accelerative potency -- such as in the case of the Cosmological Constant, 
$\Lambda$ -- then the acceleration would never stop, and all matter not tied 
to the final large cluster containing all bound mass in our region of space 
\citep[``Milkomeda'', e.g.,][]{MilkomedaCoxLoeb} would fly away out of causal 
contact; the total amount of information (and thus useful energy for doing work) 
available to any observer would therefore be finite even for an eternal universe, 
eventually causing the ability to do computations (and thus the possibility of 
sustaining life) to ultimately fade away \citep{KraussStarkCosmicFade}. Or even 
more severely, if the Dark Energy should actually increase in potency over 
time -- such as would be the case for Phantom Energy ($w^{\Lambda} < -1$) -- 
then the universe could end in a ``Big Rip'', leading to a Cosmic Doomsday 
where all cosmic structures, objects, and even atoms are ultimately 
ripped apart \citep{CaldPhantom}. Other exotic models naturally lead to other 
possibilities, such as a cyclic universe with a ``Big Bounce'' 
\citep[e.g.,][]{VenezBigBounce}, and so on. 

For the causal backreaction formalism that we present here, in which exotic 
contributors such as Dark Energy, Higher Dimensions, etc., are unnecessary, 
the situation is less complex in theory while being more complicated in 
practice. While there is no more need to understand the behavior of any 
independent substance or modified physics that exerts some `external' control 
over the rest of the universe, there is now the necessity of calculating a 
fully self-consistent solution for the evolution of an increasingly 
inhomogeneous cosmos, whose own perturbations are causing itself to accelerate 
(or at least to appear to do so). A real solution of this type would require 
a method of dealing with gravitational nonlinearities, along with fully 
three-dimensional modeling to deal with the ultimate breakdown of our 
smoothly-inhomogeneous approximation. But even without such tools at our 
disposal here we can at least delineate some of the more likely possibilities, 
to see how they compare to the popularly regarded cosmic fates mentioned above, 
and also to see if any new wrinkle appears that modifies these possibilities 
in some interesting way.  

In terms of our clumping-perturbed metric, Equation~\ref{EqnFinalBHpertMetric}, 
the effects of causal backreaction completely take over the cosmic evolution as 
$I(t)$ approaches unity; and the model breaks down for $I(t) \approx 1$ 
(and is completely meaningless for $I(t) > 1$), as gravitational nonlinearities 
cause the failure of our approximation of linearly summing together the individual 
Newtonian gravitational potentials from different virialized mass clusters. Two 
questions naturally arise: (1) Will this $I(t)$ function ever actually approach 
unity? (2) And if so, what really happens to the cosmic expansion as a result? 

Considering the latter question first, to see what the stakes are, it seems 
likely that `real' acceleration due to backreaction, in the sense of $q_{2}$ 
from Section~\ref{SecTheorCounters} above, would finally become cosmologically 
important. In other words, the fitting problem of finding an averaged cosmology 
to represent a very inhomogeneous universe would become a dominant concern, 
and accelerated volume expansion due to the continuing causal backreaction, 
now modified by truly nonlinear gravitational effects, would fully control the 
evolution of the universe. In that case, we suppose that it is possible for the 
result to be a runaway acceleration, as potent perhaps as that from a 
Cosmological Constant (or maybe even as strong as a form of Phantom Energy), 
since the effective value of $w^\mathrm{Obs}$ would be due to gravitational 
energy rather than due to the exotic nature of any particular physical substance, 
and hence not be subject to the bounds (i.e., $w \geq -1$) normally imposed upon 
perfect fluids by the dominant energy condition. (One could even speculate that 
such a process may have had relevance to the pre-inflation very early universe 
-- given its presumably chaotic and inhomogeneous nature -- perhaps accounting 
for some or all of the acceleration usually credited to the Inflaton field(s); 
assuming, of course, that such a process could be shown to be 
theoretically possible given causality restrictions.) 

Alternatively, some authors have argued that backreaction due to matter 
inhomogeneities cannot be responsible for $\Lambda$- or Phantom-like 
acceleration; or that if they can be, temporarily, then such an acceleration 
cannot continue eternally. \citet{KasFutNoGoNonlinBackAccel} use the volumetric 
conservation of pressureless dust to make arguments towards a no-go theorem for 
acceleration from nonlinear backreaction; but their calculations are based upon 
the same cosmologically inapplicable approximations discussed previously in 
Section~\ref{SecTheorCounters}. Also, they use the same function $a(t)$ 
as both their unperturbed scale factor, and as their measure of cosmic volume 
(i.e., $a \propto V^{1/3}$) in the perturbed metric -- two roles for $a(t)$ 
that are not simultaneously valid in a model such as ours, for example, 
since $I(t) , \dot{I}(t) \ne 0$ alters the evolving physical volume of a 
perturbed region away from a simple proportionality to $a^{3}$ (e.g., 
recall Section~\ref{Sub2LSS}). 

\citet{BoseMajEvHorFutureDecel}, on the other hand, argue that whatever 
specific physical process may actually be generating the current cosmic 
acceleration, that it cannot continue eternally, due to backreaction from the 
{\it future} cosmological event horizon which would exist in a universe that 
accelerated forever. Despite the use of the Buchert formalism 
\citep[e.g.,][]{BuchertEhlers97} -- along with its unacceptable assumptions 
for causal backreaction (irrotational dust, zero Newtonian backreaction, 
etc.) -- in their calculations of the backreaction from this future horizon, 
their results are obtained without necessarily presuming 
Buchert-generated acceleration; and they claim that the acceleration must 
eventually stop regardless of the mechanism behind it. Looking closely at 
their results, however, one sees that this claim only really works when one 
considers a large region made up of subregions (two, in their calculations) 
that are each individually decelerating; and in that case, all they have 
proven is that the Buchert-generated acceleration of the combined region 
eventually becomes weaker than the total sum of the individual decelerations 
of the component subregions. If one of the subregions is actually accelerating 
on its own through some non-Buchert mechanism (such as via our causal 
backreaction method, nonzero $\Lambda$, etc.), and possesses a sufficiently 
large initial fraction of the total volume -- or if {\it all} subregions are 
individually accelerating ($\alpha > 1$, $\beta > 1$ in their terminology) -- 
then inspection of their Equations 4 and 9 shows that the overall 
acceleration $\ddot{a} _{D}$ {\it cannot} go to zero for positive, finite 
$t$, unless the `apparent volume fraction' of a component subregion as part 
of the total were to become negative, which is obviously not consistent. 

While the conclusions that one may draw from calculations such as these 
are debatable, and the ideas behind them are still in too early a stage of 
development to be definitive, they still make the important point that there 
may be `global' considerations that either prevent or ultimately shut down a 
long-term cosmic acceleration due to effects that go beyond those contained 
within any simplified model like our formalism. It is therefore important to 
keep such caveats in mind when considering the so-called `ultimate' possible 
fates of the universe, below; but regardless of the {\it eternal} validity of 
any conclusions, at least we can use our model to elucidate some reasonably 
long-term effects of the crucial phase transition which must undoubtedly 
occur when (and if) the causal backreaction becomes nonlinear in strength. 

Now considering our former question -- will $I(t)$ really approach unity? 
And even more seriously, will gravitationally nonlinear effects due to 
causal backreaction come to dominate (and in some sense `permanently' 
control) the evolution of the Universe? In our simulation runs, $I(t)$ 
increases monotonically, and gets quite close to 1 for the 
strongest-`accelerating' cosmological models; it would therefore 
be useful to estimate how large $I(t)$ can get within our formalism. 

In the essential physics of backreaction with causal updating, there are 
several different processes competing against one another -- some enhancing 
the apparent (or ultimately real) acceleration, and some damping it -- and 
which way the balance tilts between these factors will largely (but as we 
will see in the end, not completely) determine the ultimate cosmic fate. 
The most obvious factor enhancing the acceleration is the increasing 
`clumpiness' of the universe, represented in this paper by clumping 
evolution function $\Psi (t)$, which increases monotonically over time. 
In our formalism, however, the extent of clumping must saturate as 
$\Psi (t) \rightarrow 1$, thus limiting its ability to continue driving 
the growth of $I(t)$; and this seems like a physically realistic constraint 
on $\Psi (t)$, representing how virtually all cosmic matter should 
eventually reach a `maximally clumped' state, in which the universe 
consists almost entirely of steadily expanding voids, punctuated only 
by a sparse scattering of fixed-size massive clumps that no longer 
expand with the universe, but which manage to virially support 
themselves against singular collapse. 

But even more important than the increasing clumpiness of any given 
region of space, is how the {\it information} about distant clumpiness 
(i.e., $\Psi [t_{\mathrm{ret}} (t, \alpha)]$) can get to an observer 
from farther and farther distances with increasing $t$, as the volume 
of space with significant clumping within the observer's causal horizon 
($\alpha \leq \alpha _{\mathrm{max}}$, cf. Equation~\ref{EqnalphaMax}) 
grows rapidly with time. This effect is counterbalanced, however, by the 
continuing expansion of the universe, which dilutes such backreaction by 
taking any given mass `clump' farther away from the observer as $t$ 
increases (specifically referring to the factor of 
$1 / [a_{\mathrm{MD}}(t)]$ in $R_{\mathrm{Sch}}(t)$; 
cf. Equations~\ref{EqnAngAvgBHMatDomWeakFlat},\ref{EqnIintegrandPrelim}). 
These opposing factors are represented by $\alpha _{\mathrm{max}}$ and 
$(t_{0} / t)^{2/3}$, respectively, in Equation~\ref{EqnItotIntegration} 
for $I(t)$. Ignoring the time dependence in 
$\Psi [t_{\mathrm{ret}} (t, \alpha)]$ for the moment, one approximately 
gets $I(t) \propto [ (t_{0} / t)^{2/3} ~ \alpha ^{2}_{\mathrm{max}} ]$; 
but since $\alpha _{\mathrm{max}} \simeq (t/t_{0})^{1/3}$ for 
$t_{0} \gg t_\mathrm{init}$ (cf. Equation~\ref{EqnalphaMax}), we see that 
the overall time dependence ultimately cancels out of $I(t)$ at sufficiently 
large $t$. We can interpret this to mean that the increasing backreaction 
effect due to the expanding causal horizon {\it also} saturates, due to 
competition against the cosmic expansion itself, with $I(t)$ approaching 
a constant value at times very late compared to $t_\mathrm{init}$ (and 
very late also compared to the time by which the clumping evolution, 
$\Psi (t)$, has mostly saturated near unity). What we need to know, 
of course, is {\it how large} this asymptotically constant value 
of $I(t)$ will be; an obviously model-dependent question. 

Treating $\Psi [t_{\mathrm{ret}} (t, \alpha)]$ as a step function 
-- equal to $\simeq$1 for $\alpha \le \alpha _{\mathrm{max}}$ 
(with $t_\mathrm{init} \simeq 0$), 
and zero beyond there -- one gets $I(t_{0}) \simeq 6 \gg 1$, a result 
deep into the regime requiring a gravitationally nonlinear treatment; 
but this of course is an unrealistic upper bound on the effects of 
clumping, representing a universe that became fully clumped almost 
instantaneously, saturating at very early cosmic times (i.e., 
$\Psi (t) \rightarrow 1$ at $(t/t_{0}) \ll 1$). Trying more realistic 
(and continuous) functional forms for $\Psi (t)$, with a fairly recent 
saturation of clumping -- in particular, considering the three functions 
used for our clumping evolution models, as defined above in 
Equations~\ref{EqnClumpModels}a-c -- the maximum possible value of $I(t)$ 
for each clumping model can be obtained for $t_\mathrm{init} \simeq 0$, 
$t \rightarrow t_{0}$, and $\Psi (t_{0}) \equiv 1$. Doing this, it is 
easy to show that $I _\mathrm{max} (\Psi _{\mathrm{Sqr}}) = 3/14$, 
$I _\mathrm{max} (\Psi _{\mathrm{Lin}}) = 3/5$, and 
$I _\mathrm{max} (\Psi _{\mathrm{MD}}) = 1$. Thus the `strongest' of our 
clumping evolution models ($\Psi _{\mathrm{MD}}$) can definitely produce 
results strong enough to violate the approximation of gravitational 
linearity; and even stronger models, with more gradual clumping growth 
(that is, an earlier start to significantly large $\Psi$) -- given as 
$\Psi(t) \propto (t / t_{0})^{N}$ with $N < 2/3$ -- would have 
$I _\mathrm{max}$ values that easily exceed unity (e.g., 
$I _\mathrm{max} = 2$ for $N = 1/3$, $I _\mathrm{max} = 6$ for $N = 0$, 
etc.). In short, causal backreaction as prescribed by our formalism can 
clearly be strong enough to not only induce an apparent acceleration, 
but also to break down our linearized gravitational approximation entirely, 
indicating a possible runaway acceleration for the real universe, itself. 
But whether that possibility is realized, or whether the strongest effects 
of causal backreaction fall short of it, depends extremely sensitively 
upon the precise time-dependence of the evolution of mass clumping and 
virialization in the universe. 

One important moderating factor for backreaction has been completely left 
out of these estimations, however: the recursive nonlinearities discussed 
previously, representing how the metric perturbations due to early 
inhomogeneities will slow down the causal updating that brings in new 
inhomogeneity information. In addition, some extra volume expansion is 
created (cf. Equation~\ref{EqnAngAvgBHMatDomWeakFlat} and subsequent 
discussion) that adds to the ordinary FRW Hubble expansion in pulling 
clumped masses farther apart from one another, thus diluting these 
backreaction effects to a degree greater than that predicted by 
Equation~\ref{EqnItotIntegration} and numerically estimated above. 
These effects should not only reduce the value of $j^\mathrm{Obs}_{0}$ 
predicted by our simulations (as mentioned in 
Section~\ref{Sub2TestingViaj0}), but may in a larger sense serve as a 
causal backreaction `regulator', feeding back upon that very backreaction 
to slow down its effects upon the universe whenever the acceleration 
caused by it begins going into a runaway mode. But such a picture of 
self-regulating cosmic acceleration remains purely speculative until 
a more complete formalism and simulation program exist which can 
quantitatively model the effects of these recursive nonlinearities 
into the far future. 

Thus the variety of `ultimate' cosmic fates potentially emerging from 
our causal backreaction paradigm seem to fall into a few fairly 
recognizable categories, depending in detail upon the evolution of 
the growth rate of self-stabilized cosmic structures. On the one hand, 
the effects of causal backreaction may weaken or even saturate, leading 
either to a continuing low level of `acceleration', or to the kind of 
coasting/mildly slowing behavior of open/flat universes -- in either case 
resulting in a similar kind of Big Freeze as would have occurred in the 
normal ($\Omega _\mathrm{Tot} \le 1$) FRW case. Alternatively, if the 
cosmic structure formation evolution leads to a stronger, self-sustaining 
level of backreaction, then it is conceivable that the fading-away 
behavior (the ``Big Fade''?) of the rest of the universe as would occur 
for an exponentially-expanding Cosmological Constant universe may not be 
a bad approximation. Or in extreme cases (probably less likely), some form 
of Big Rip could conceivably occur. Qualitatively speaking though, the 
most reasonable scenario in our paradigm may be some kind of stop-and-go 
accelerating behavior, due to the self-regulating nature of causal 
backreaction with recursive nonlinearities. This could lead to a 
situation where the universe is always `riding the edge' of a runaway 
acceleration, but never quite getting there -- perhaps in a permanent 
state of acceleration that oscillates over time but stays relatively 
close to the level of acceleration that we see today, with the same 
relationship that we currently see between effective FLRW observables (i.e., 
$\Omega^{\mathrm{Obs}}_{\Lambda} \sim \Omega^{\mathrm{Obs}}_{\mathrm{M}}$ 
to within an order of magnitude), with these conditions remaining 
generically true for the foreseeable cosmological future. 

Now, the variety of possible cosmic fates considered here is of course 
not unique to our formalism -- each of these possibilities can easily 
be reproduced by a version of Dark Energy with some appropriately 
evolving equation of state. The difference is that instead of being 
dependent upon the unknown (and perhaps ultimately unknowable, in fine detail) 
theoretical nature of a master Dark Energy field running the rest of the 
universe, the cosmic fate here is due to the normal gravitational 
forces and motions of matter, which -- though very complicated in practice -- 
will depend only upon calculations using known laws of physics (once the 
nature of the Dark Matter is known!), and the long-explored mechanisms 
of general relativity. 

And yet, one more wild card still remains, one more consideration regarding 
the plausibility of these possible cosmic fates, which will tell us which 
one really is the most probable description of the cosmic future; 
and the answer is: {\it none of them!} In the entire discussion of our formalism 
in this paper, there is one `elephant in the room' that has not been analyzed, 
because it cannot be modeled in this way -- and that is the breakdown of our 
smoothly-inhomogeneous approximation itself. As the universe grows older, and 
ever-larger bound structures have time to form, the universe will eventually 
become inhomogeneous and anisotropic on a truly cosmological scale. Thus the 
Cosmological Principle will break down entirely as a usable approximation, and 
vast cosmic structures will eventually exert powerful tidal forces that are 
strong enough even to threaten bound galactic clusters. While this is also not 
a theoretical phenomenon unique to our specific formalism, once again there 
is a key difference: whereas typically one expects the dominant perturbing
influences on local structures to be the closer ones (cosmologically speaking), 
consisting of just one or a few definable, huge objects, our causal updating 
analysis shows that the acceleration effects due to backreaction are dominated 
by the {\it farthest} distance out to which one can causally see clumped 
structure. Thus in the distant future, the gravitationally perturbative effects 
acting upon otherwise `local' structures will be dominated by a multitude of 
different inhomogeneities, located in all directions, that are 
cosmologically-extreme distances away, and that together represent an almost 
unimaginable amount of mass in clumped structures. Such tidal or shearing 
forces will continue to grow and grow -- as long as the causal flow of 
information can continue to come in from ever-more-distant parts of the universe 
-- until gravitationally bound local structures are chaotically ripped apart. 
In far more anisotropic fashion than the Big Rip of Phantom Energy -- if 
perhaps less destructively, since {\it some} form of recursive nonlinearities 
and dilution will eventually slow the process -- this evolution will herald 
the end of the smooth cosmic expansion existing at present, rendering all 
FRW/FLRW models obsolete. The first hints of such behavior in our universe 
may already be making themselves evident, perhaps accounting for the observed 
phenomenon known as the ``Dark Flow" \citep{KashDarkFlow,KashDarkFlowPersists} 
-- or at least accounting for the causal part of it, if the effect indeed 
extends to `superhorizon' scales. One thing is clear, according to our 
causal backreaction formalism: this Dark Flow (assuming it to be 
observationally real and at least partly causally generated) is just the 
beginning of a greater trend of inhomogeneity and anisotropy that will be 
acting upon all local cosmic structures, and doing so much more rapidly, 
violently, and chaotically than would be expected in more conventional 
(i.e., non-causal, Poisson-Equation-based) models of structure formation. 

The {\it real} fate of the universe will therefore not be a Big Crunch, 
a Big Freeze, a Big Fade, a Big Rip, or even a Big Bounce -- instead it will 
end (if one can call it an `end') in something entirely more appropriate 
to the entropy-riddled universe that we all have known: a Big Mess.

\section{\label{SecConclude}SUMMARY AND CONCLUSIONS}

We may sum up the motivations, methods, and results of this paper 
as follows. Our starting point is the search for an alternative to 
Dark Energy as an ingredient in the nascent Cosmic Concordance, 
without resorting to the increased theoretical complexity 
represented by modifications to gravity, or the non-Copernican 
`specialness' implied by a local cosmic void. The problems with 
Dark Energy are well known, including the coincidence and magnitude 
fine-tuning problems for a pure-$\Lambda$ Cosmological Constant, 
and the stability problem for a more dynamical form of Dark Energy 
possessing the ability to cluster spatially. The latter (DDE) case 
requires the ad-hoc addition of some form of new pressure term to 
support it against collapse, given the {\it locally attractive} 
nature of the negative pressure required by the DDE to power the 
cosmic acceleration. Such a new pressure term, if adiabatic 
(e.g., degeneracy pressure), would represent a form of {\it positive} 
pressure contributing to a cosmic {\it deceleration} that partially 
or totally nullifies the acceleration meant to come from the DDE; and 
if non-adiabatic (due to some new effects from the DDE Lagrangian), 
would invalidate the Dark Energy as a perfect fluid, thus calling 
the usual (accelerative) cosmic implications of $w ^{\mathrm{DE}} < 0$ 
in the FLRW acceleration equation into question. 

Emphasizing the importance of the fact that `negative pressure' 
is locally attractive in character -- rather than repulsive, as it is 
often regarded and popularly described -- and that normal gravitational 
attraction therefore represents (at least in a non-technical sense) a 
form of negative pressure, we have therefore made a virtue of necessity 
by recruiting cosmological structure formation itself, based upon 
ordinary gravitational forces, as the driver of the observed (possibly 
apparent) acceleration. This approach, known generally in the literature 
as ``backreaction'', removes all of the aforementioned problems by both 
eliminating the need for any Dark Energy species, and by solving the 
Coincidence Problem by naturally linking the onset of cosmic 
acceleration with the emergence of cosmic structure, which inevitably 
leads to the creation of observers such as ourselves just in time 
to see this `coincidence'. 

Noting that several versions of backreaction have been largely 
unsuccessful in their attempts to account for the observed acceleration, 
and that powerful arguments have been advanced which claim that 
backreaction as a paradigm {\it cannot} be made strong enough to 
succeed at this task, we have described how each of these arguments 
(and the backreaction mechanisms which they address) are functionally 
invalid. This lack of validity is not generally due to mathematical 
flaws within the arguments themselves, but due to their inapplicability 
to the real universe: all such ``no-go'' arguments or theorems are 
based upon one or more essential simplifications so restrictive 
that they eliminate from consideration all of the actual physics 
that is responsible for the crucial `accelerative' effects. 

Different arguments against backreaction fail for different reasons. 
No-go conclusions drawn from the Raychaudhuri equation are invalid 
because they neglect vorticity (or more precisely, the {\it square} 
of the vorticity, $\omega^{2}$), thus ironically dropping the dominant 
physical force in the universe which prevents the collapse of all 
(non-solid) structures into singularities, and which is a positive 
semi-definite quantity that does not go to zero regardless of 
how large a scale one uses for averaging it. Swiss-Cheese models 
(typically using Schwarzschild or LTB metrics for the holes) 
underestimate backreaction effects not only because they lack a 
mechanism for modeling vorticity and virialization, but more crucially 
because they partition the cluster-forming universe into a discrete 
set of non-intersecting, non-interacting volumes -- a construct 
completely at odds with how structure formation works in the real 
universe, where overdensities extend their reach far beyond their 
local domains to cause vast inflows, and where any given region of 
space feels the gravitational effects of many such influences 
{\it combined together}, rather than being impacted by one influence 
while being hermetically sealed off from all of the others. Lastly, 
there are no-go arguments based upon the generally Newtonian nature 
of metric perturbations and mass flows, apparently demonstrating 
that the sum total effects of backreaction must be small (or 
according to the formalism of Buchert and collaborators, identically 
zero); but such arguments are flawed due to the non-causal basis 
of the calculations (since they drop all signal travel-time delays, 
gravitomagnetic effects, and time derivatives of the perturbation 
potential), and because they depend upon `smallness' arguments that 
are irrelevant in the face of the combined perturbative effects of 
innumerable cluster-forming masses acting upon every single region 
of space within their causal grasp. 

The true utility of such no-go arguments, however, is that gaining an 
understanding of why and where they fail to apply can guide us in the 
construction of a more successful, working form of backreaction. The 
central challenge is figuring out how best to model the crucial cosmic 
phase transition from smooth matter distributed largely homogeneously 
throughout space, to most matter being concentrated into a fairly 
randomly-distributed collection of vorticity-stabilized, self-virialized, 
clumped structures. In this paper, we model the situation heuristically 
and simply by representing the dominant perturbation due to each `clump' 
as being (the Newtonian tail of) a Schwarzschild metric embedded in the 
expanding universe, superposed on top of the background FRW metric -- 
where we utilize the fact that the vorticity (or velocity dispersion) 
stabilizing each structure will add its own positive contribution to the 
volume expansion, as per the Raychaudhuri equation. We then combine the 
perturbative contributions of all such objects by summing them and averaging 
over location and direction, producing a ``smoothly-inhomogeneous'' universe 
model that neither possesses nor requires a local void or spatial variations. 
Instead, it possesses only a single clumping evolution function that 
increases over time, representing the ever-growing fraction of cosmic matter 
located in self-stabilized clusters, rather than remaining in the smooth 
background. Lastly, and perhaps most importantly, we introduce the mechanism 
of ``causal updating'' in order to represent the crucial process of how 
the information about this growth of clustering propagates through the 
universe to any given observer, coming in from the very limits of the 
causal horizon out to which that observer has had time to see clumped 
structure. 

Employing these physical considerations, we have constructed a formalism 
of ``causal backreaction'' for calculating how the smoothly-inhomogeneous 
metric evolves over time, given any pre-specified clumping evolution function 
$\Psi (t)$. From that metric solution, we can calculate a luminosity distance 
curve as a function of observed redshift $z^\mathrm{Obs}$, for comparison to 
residual Hubble curves measured via standard candles (particularly Type Ia 
supernovae), in order to check if an observed (possibly apparent) acceleration 
of the right magnitude and $z^\mathrm{Obs}$-dependence has been observed. 
Additionally, we can use our formalism to calculate several key cosmological 
observables produced by that function $\Psi (t)$ -- such as $t^\mathrm{Obs}_{0}$ 
(age of the universe), $\Omega^\mathrm{FRW}_\mathrm{M}$ (matter density as 
a fraction of the unperturbed FRW critical density), $w^\mathrm{Obs}_{0}$ 
(apparent Dark Energy equation of state), $j^\mathrm{Obs}_{0}$ (cosmic jerk 
parameter), and $l^\mathrm{Obs}_{\mathrm{A}}$ (characteristic angular scale 
of the CMB acoustic peaks) -- testing whether the essential results of an 
alternative cosmic concordance have been achieved for causal backreaction 
operating upon such a clumping evolution function. Altogether, this adds up 
to a sterner test than the mere production of an apparent acceleration. 

Armed with this formalism, and using a set of 60 clumping evolution functions 
designed for simplicity and using model input parameters determined from 
straightforward astrophysical considerations, we have found a $\sim$dozen-plus 
solutions that fit the SCP Union Compilation SNe data essentially as well as 
the best-fit $\Lambda$CDM model. Furthermore, about half of these models give 
good values (within reasonable theoretical and observational uncertainties) for 
all other calculated cosmological parameters. That is, these few `best' models 
(obtained without even conducting a full optimization search over the model 
parameter space) have successfully produced an alternative concordance, at 
least to the extent of solving the cosmic Age Problem, achieving spatial 
flatness (for the unperturbed, pre-clumping FRW universe), and being 
reasonably consistent with the CMB peak positions; and they achieve all of 
these goals -- along with explaining the observed acceleration -- in a 
matter-only universe, without any form of Dark Energy component. 

Now given that these important objectives -- particularly the difficult 
one of achieving flatness by bringing the total cosmic density from 
$\Omega _{0} \sim 0.3$ up to $\Omega _{0} \simeq 1$ -- have been major pillars 
in the argument for Dark Energy, our ability to achieve them with matter alone, 
simply by including the ordinary gravitational effects of causal backreaction, 
would seem to turn all forms of Dark Energy ($\Lambda$, Quintessence, etc.) 
into excess theoretical baggage. Nevertheless, since one cannot {\it prove} 
a theory with Occam's Razor alone, we have sought in this paper to find ways 
to distinguish our causal backreaction paradigm from any kind of Dark Energy 
(where possible), or at the very least to distinguish it from the simplest 
form of Dark Energy -- i.e., a Cosmological Constant. One generic implication 
of using backreaction to generate the cosmic acceleration is that this places 
an increased emphasis upon the possible existence of larger-scale anisotropies 
and inhomogeneities than those expected from the standard perturbed-FLRW 
cosmologies. To that end, we have commented above about potential 
observational signs of such deviations from the Cosmological Principle that 
have been seen over the past few years by other researchers, as well as 
referring to work by this author and others using Type Ia SNe data sets to 
search for direct signals of possible large-scale anisotropies. 

Additionally, there are potential discrepancies for $\Lambda$CDM in particular, 
including the cuspy CDM halo problem and the possible dearth of dwarf satellite 
galaxies, which might be naturally explainable due to clustering-induced 
feedback occurring in the causal backreaction paradigm. 
Finally, there would of course be alterations to virtually all other 
cosmological observables due to causal backreaction, providing any number 
of possible signals in the complementary cosmic data sets that can be used 
for testing our paradigm; though detailed calculations of most of those 
alterations are beyond the scope of this current paper. One altered parameter 
that we have in considered in detail here, however, is the jerk parameter 
$j^\mathrm{Obs}_{0}$, which for flat $\Lambda$CDM is required to be exactly 
unity. For our cosmologically successful simulation runs using the causal 
backreaction model, on the other hand, $j^\mathrm{Obs}_{0} > 1$ is clearly 
preferred, a general result which we have seen is supported reasonably well 
(though statistically very weakly) by trends in current cosmological data. 
Quantitatively speaking, the best runs in this paper predict an approximate 
range of $j^\mathrm{Obs}_{0} \sim 2.5 - 5.5$ (with a more narrow range of 
$j^\mathrm{Obs}_{0} \sim 2.6 - 3.8$ for the very best few runs), which is 
perhaps high compared even to the upper range of observational estimates; 
but as explained previously, such values computed here from our simulation 
runs are almost certainly higher than they should be due to current 
oversimplifications of our model (e.g., the absence of what we have 
termed ``recursive nonlinearities'', etc.), which must be remedied 
with further theoretical development of these calculations. 

Having obtained such favorable results for producing an alternative 
concordance without Dark Energy, we then considered the `ultimate' 
cosmic fates possible for a universe dominated by causal backreaction. 
Though the mathematical toy model introduced in this paper is still too 
simple at this point to yield detailed quantitative predictions about the 
universe in the far future, we expect the possibilities to be quite similar 
to the usual variety of potential fates for a universe dominated by 
various forms of Dark Energy with its typical designer equations of state. 
But there are two major differences, however: (1) With the acceleration 
(apparent or real) being provided by the detailed processes of structure 
formation in the matter itself, rather than from some `external' material 
component with its own independent equation of state evolution, it seems 
far more likely that a self-limiting, stop-and-go form of acceleration 
will be the ultimate cosmic fate, rather than any complete runaway 
scenario (such as the Big Rip of Phantom Energy); and: (2) Given the 
unwonted importance here of inhomogeneities at great cosmological 
distances for determining `local' gravitational behavior, causal 
backreaction is likely to lead to far more extreme and asymmetrical 
tidal forces on locally bound objects, eventually leading to a far 
messier and more chaotic universe than anything expected from `normal' 
models of inhomogeneity evolution, given any form of cosmic 
acceleration this side of a Big Rip. 

Expressed in one sentence, if one asks, ``What is the force behind 
the cosmic acceleration?'', the answer we would give is that it is 
not a `force' at all: rather, the motivating effect may be called 
``the Schwarz'' -- consisting of the total, summed effects of the 
Newtonian tails of Schwarzschild-like metric perturbations, produced 
by the virialization of innumerable self-stabilized structures filling 
the universe, with these influences propagating causally towards all 
observers from the extreme edges of their observable cosmic horizons.

\acknowledgments

I am grateful to Jacob Bekenstein, Krzysztof Bolejko, 
Varoujan Gorjian, Wayne Hu, Marek Kowalski, 
Edvard M\"{o}rtsell, David Rubin, Ran Sivron, 
David Wiltshire, and Ned Wright for brief but helpful 
communications; and I am especially grateful to 
Arthur Lue for several helpful and clarifying 
discussions.


\newpage

\begin{figure}
\begin{center}
\includegraphics[scale=0.75]{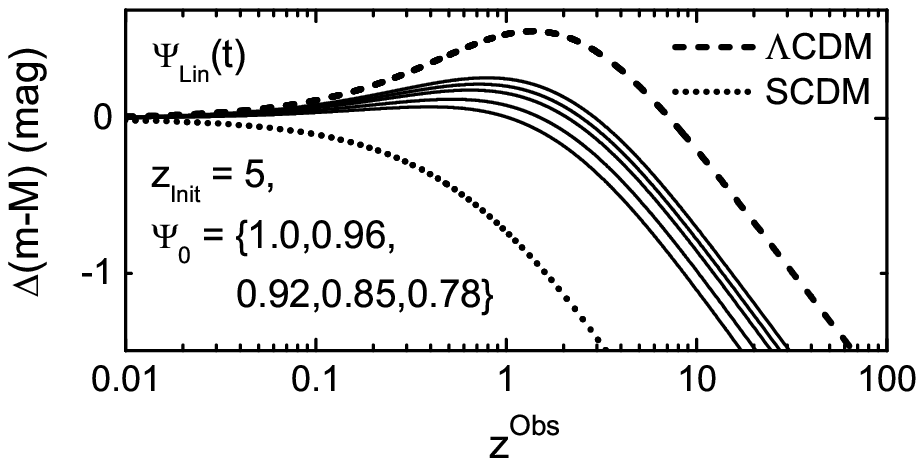}
\\ 
\includegraphics[scale=0.75]{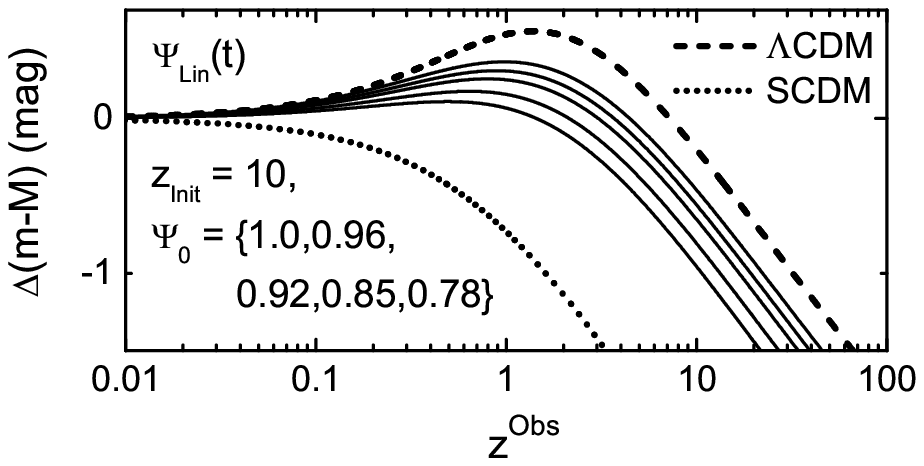}
\\ 
\includegraphics[scale=0.75]{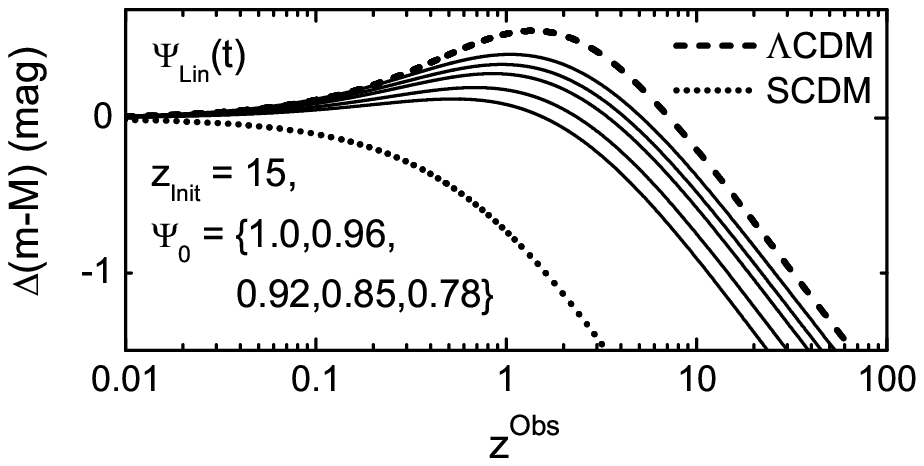}
\\
\includegraphics[scale=0.75]{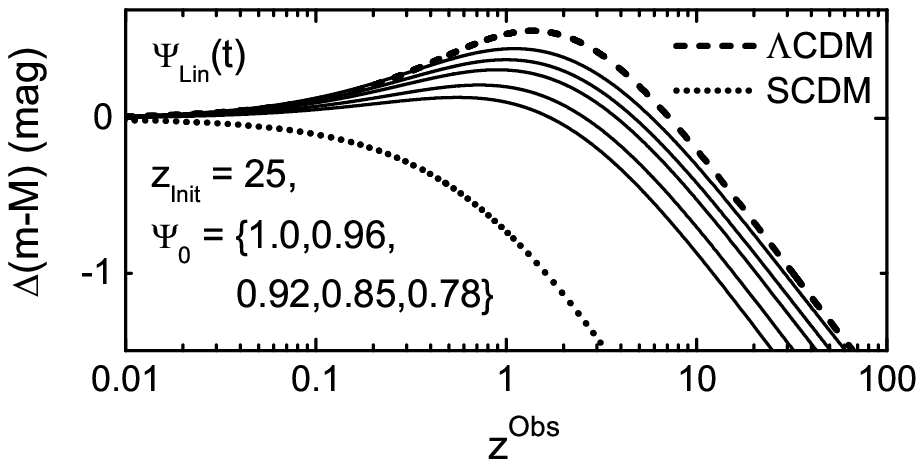}
\end{center}
\caption{Residual Hubble diagrams computed from 
our apparent acceleration model using 
$\Psi _{\mathrm{Lin}} (t)$ clumping functions. 
Different panels represent initial-clumping epochs 
of (respectively): $z_\mathrm{init} = (5,10,15,25)$. 
Each plot depicts the usual flat $\Lambda{\mathrm{CDM}}$ 
($\Omega _{\Lambda} = 0.73$) and SCDM cosmologies, shown 
versus five of our simulation runs using present-day 
clumping parameters of (plotted highest to lowest): 
$\Psi _{0} = (1.0,0.96,0.92,0.85,0.78)$.}
\label{FigLinPlots}
\end{figure}

\newpage

\begin{figure}
\begin{center}
\includegraphics[scale=0.75]{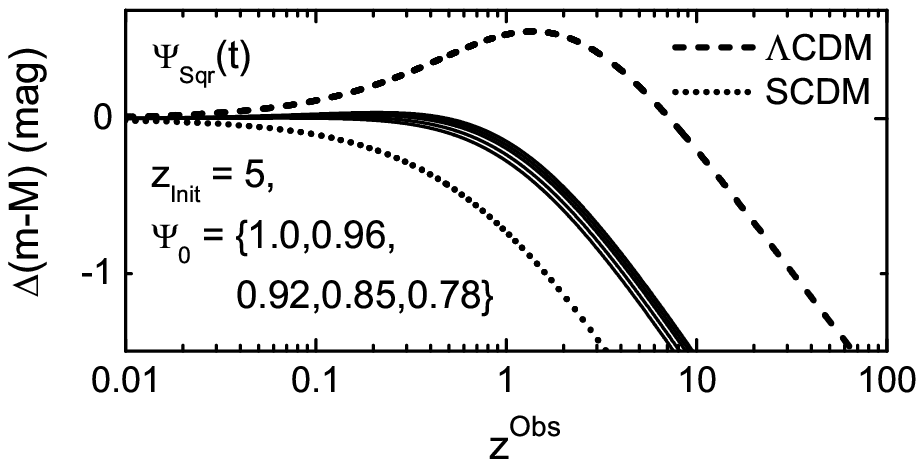}
\\ 
\includegraphics[scale=0.75]{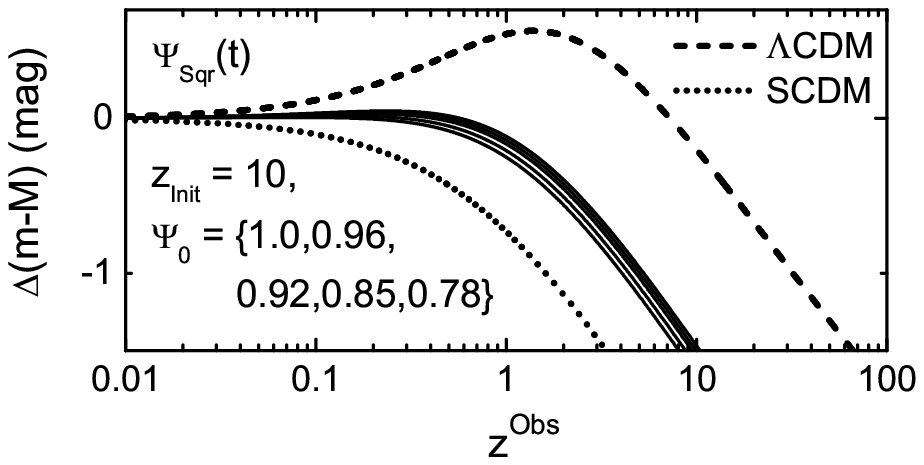}
\\ 
\includegraphics[scale=0.75]{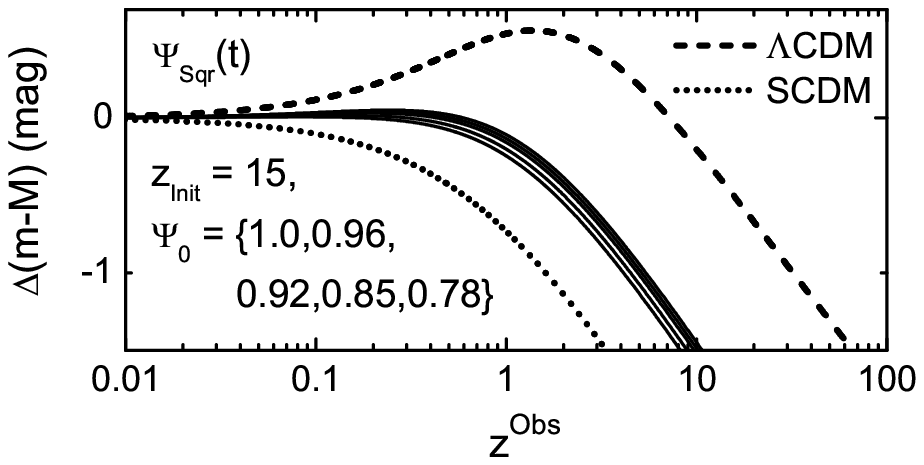}
\\
\includegraphics[scale=0.75]{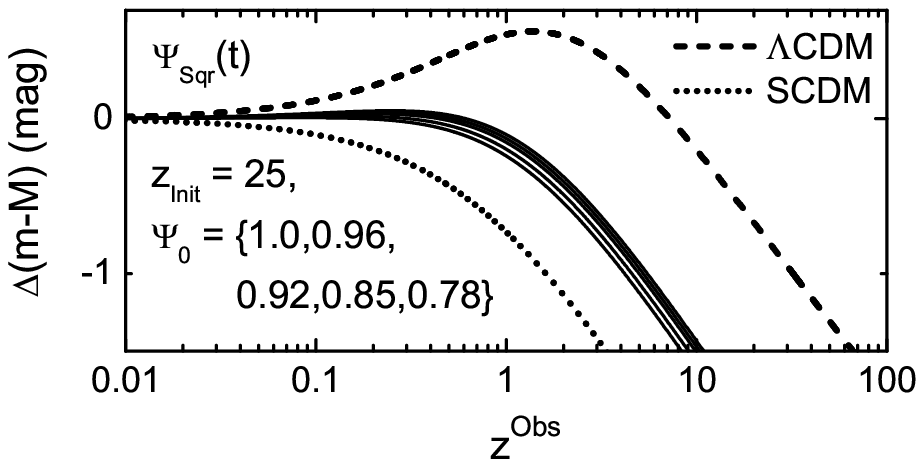}
\end{center}
\caption{Residual Hubble diagrams computed from 
our apparent acceleration model using 
$\Psi _{\mathrm{Sqr}} (t)$ clumping functions. 
Different panels represent initial-clumping epochs 
of (respectively): $z_\mathrm{init} = (5,10,15,25)$. 
Each plot depicts the usual flat $\Lambda{\mathrm{CDM}}$ 
($\Omega _{\Lambda} = 0.73$) and SCDM cosmologies, shown 
versus five of our simulation runs using present-day 
clumping parameters of (plotted highest to lowest): 
$\Psi _{0} = (1.0,0.96,0.92,0.85,0.78)$.}
\label{FigSqrPlots}
\end{figure}

\newpage

\begin{figure}
\begin{center}
\includegraphics[scale=0.75]{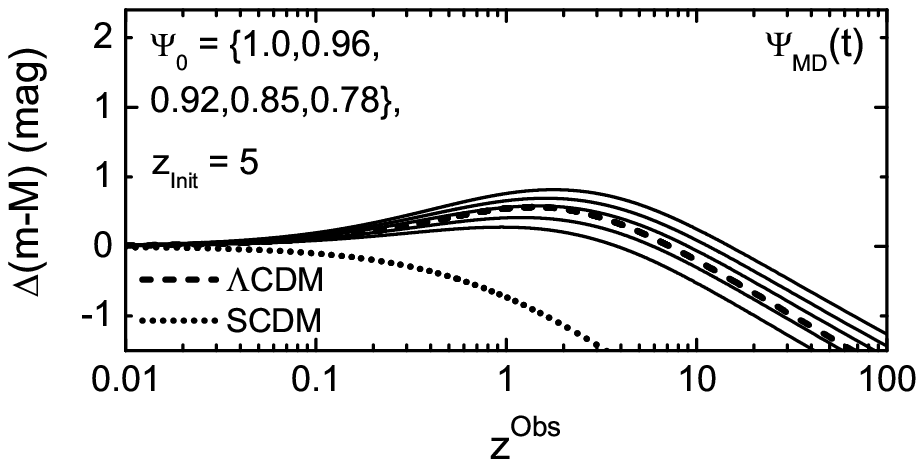}
\\ 
\includegraphics[scale=0.75]{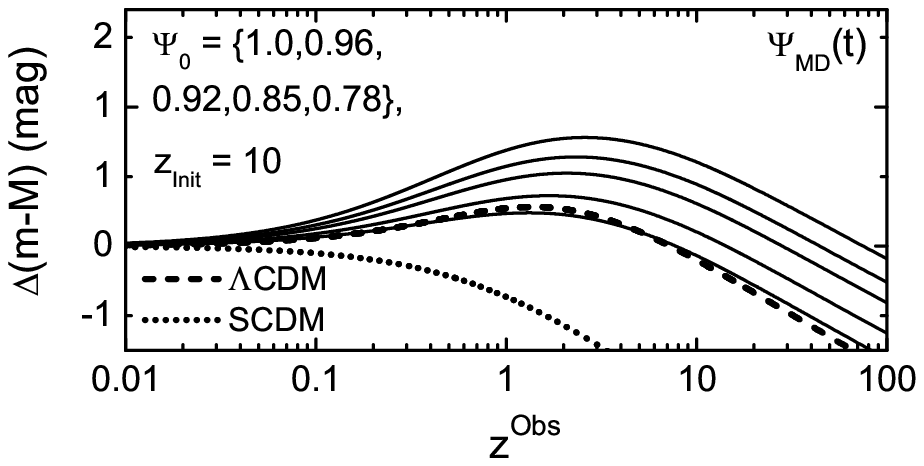}
\\ 
\includegraphics[scale=0.75]{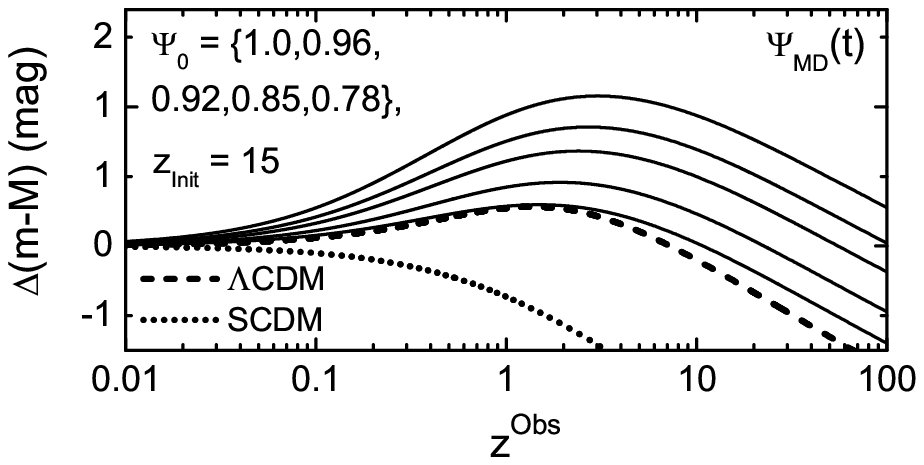}
\\
\includegraphics[scale=0.75]{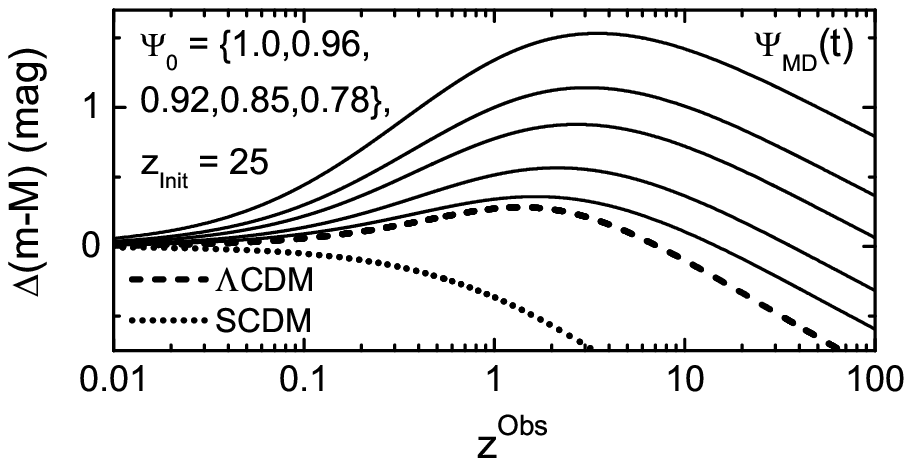}
\end{center}
\caption{Residual Hubble diagrams computed from 
our apparent acceleration model using 
$\Psi _{\mathrm{MD}} (t)$ clumping functions. 
Different panels represent initial-clumping epochs 
of (respectively): $z_\mathrm{init} = (5,10,15,25)$. 
Each plot depicts the usual flat $\Lambda{\mathrm{CDM}}$ 
($\Omega _{\Lambda} = 0.73$) and SCDM cosmologies, shown 
versus five of our simulation runs using present-day 
clumping parameters of (plotted highest to lowest): 
$\Psi _{0} = (1.0,0.96,0.92,0.85,0.78)$.}
\label{FigMDPlots}
\end{figure}

\newpage

\begin{figure}
\begin{center}
\includegraphics[scale=1.0]{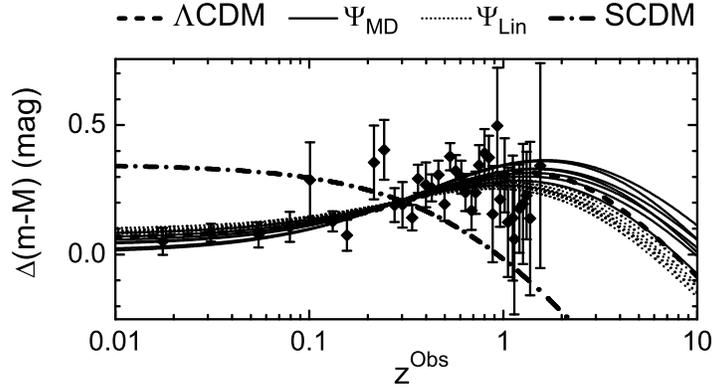}
\end{center}
\caption{Residual Hubble diagrams for the thirteen 
`best' simulation runs, as described in the text, of 
our inhomogeneity-perturbed apparent acceleration model, 
shown along with the best-fit flat SCDM and Concordance 
$\Lambda{\mathrm{CDM}}$ ($\Omega _{\Lambda} = 0.713$) 
cosmologies. From highest to lowest 
(at $z^\mathrm{Obs} \simeq 10$), the plotted 
$\Psi _{\mathrm{MD}} (t)$ curves have the 
parameters: 
$(\Psi _{0},z_\mathrm{init}) = (0.96,5)$, $(0.92,5)$, 
$(0.85,5)$, $(0.78,10)$, $(0.78,15)$, $(0.768,14)$, 
$(0.78,25)$. From highest to lowest 
(at $z^\mathrm{Obs} \simeq 10$), the plotted 
$\Psi _{\mathrm{Lin}} (t)$ curves have the 
parameters: $(\Psi _{0},z_\mathrm{init}) = (1.0,10)$, 
$(1.0,15)$, $(0.96,15)$, $(1.0,25)$, $(0.96,25)$, 
$(0.92,25)$. Shown along with these curves are the SCP Union 
SNe data, here binned and averaged for visual clarity (bin 
size $\Delta \mathrm{Log}_{10} [ 1 + z^\mathrm{Obs} ] = 0.01$). 
Each theoretical model is individually optimized 
in $H^\mathrm{Obs}_{0}$ to minimize its 
$\chi^{2}_{\mathrm{Fit}}$ for the full SCP Union SNe 
data set (see Table~\ref{TableSimRunsCosParams}). 
For simplicity, instead of moving the SNe data up or down 
for each different optimized $H^\mathrm{Obs}_{0}$ value, 
the optimization is depicted here by plotting the residual 
Hubble diagram of the SNe data versus a coasting universe 
of a single, fixed Hubble constant 
($H^\mathrm{Obs}_{0} = 72 ~ \mathrm{km} ~ \mathrm{s}^{-1} 
\mathrm{Mpc}^{-1}$), 
and then displacing each theoretical curve vertically, 
relative to the SNe data, as appropriate for each fit.
}
\label{FigBestCosmSims}
\end{figure}


\begin{table}
\begin{center}
\caption{Cosmological Parameters Derived from the `Best' Runs 
of our Set of Simulations
\label{TableSimRunsCosParams}}
\resizebox{16.5cm}{!} 
{
\begin{tabular}{cccccccccccc}
\tableline
\tableline
($\Psi _{0},z_\mathrm{init}$) &
$\chi^{2}_{\mathrm{Fit}}$ \tablenotemark{a} & 
$P_{\mathrm{Fit}}$ \tablenotemark{b} & 
$I_{0}$ \tablenotemark{c} &
$z^\mathrm{Obs}$ \tablenotemark{d} & 
$H^\mathrm{Obs}_{0}$ \tablenotemark{e} & 
$H^\mathrm{FRW}_{0}$ \tablenotemark{f} & 
$t^\mathrm{Obs}_{0}$ \tablenotemark{g} & 
$\Omega^\mathrm{FRW}_\mathrm{M}$ \tablenotemark{h} &
$w^\mathrm{Obs}_{0}$ & 
$j^\mathrm{Obs}_{0}$ & 
$l^\mathrm{Obs}_{\mathrm{A}}$ 
\\
\tableline
\multicolumn{12}{c}{{\it $\Psi _{\mathrm{Lin}}$ Clumping Model Runs}}\\
\tableline
(1.0,25) & 312.1 & 0.362 & 0.57 & 1.12 & 69.95 & 40.68 & 13.56 & 
1.037 & -0.817 & 3.45 & 284.2 \\
(1.0,15) & 313.3 & 0.344 & 0.55 & 1.12 & 69.64 & 41.63 & 13.45 & 
0.971 & -0.784 & 3.11 & 287.6 \\
(1.0,10) & 315.4 & 0.314 & 0.52 & 1.12 & 69.27 & 42.80 & 13.30 & 
0.897 & -0.746 & 2.76 & 291.3 \\
(0.96,25) & 314.8 & 0.323 & 0.55 & 1.11 & 69.38 & 41.57 & 13.38 & 
0.968 & -0.759 & 2.92 & 285.9 \\
(0.96,15) & 316.6 & 0.297 & 0.52 & 1.11 & 69.11 & 42.46 & 13.28 & 
0.911 & -0.732 & 2.67 & 289.1 \\
(0.92,25) & 319.0 & 0.265 & 0.53 & 1.11 & 68.85 & 42.44 & 13.21 & 
0.905 & -0.706 & 2.49 & 287.6 \\
\tableline
\multicolumn{12}{c}{{\it $\Psi _{\mathrm{MD}}$ Clumping Model Runs}}\\
\tableline
(0.78,10) & 312.1 & 0.362 & 0.63 & 1.12 & 69.96 & 38.25 & 13.88 & 
1.204 & -0.801 & 3.15 & 277.5 \\
(0.78,15) & 312.2 & 0.360 & 0.68 & 1.12 & 70.84 & 36.14 & 14.17 & 
1.409 & -0.895 & 4.17 & 270.5 \\
(0.78,25) & 316.8 & 0.295 & 0.72 & 1.12 & 71.80 & 34.22 & 14.46 & 
1.642 & -1.001 & 5.51 & 263.7 \\
(0.85,5) & 313.9 & 0.336 & 0.56 & 1.13 & 69.48 & 41.21 & 13.60 & 
0.991 & -0.747 & 2.59 & 288.7 \\
(0.92,5) & 312.1 & 0.363 & 0.60 & 1.14 & 70.71 & 39.41 & 14.00 & 
1.144 & -0.871 & 3.75 & 285.6 \\
(0.96,5) & 315.5 & 0.313 & 0.63 & 1.14 & 71.52 & 38.36 & 14.25 & 
1.248 & -0.954 & 4.67 & 283.8 \\
\tableline
\multicolumn{12}{c}{{\it Semi-Optimized \tablenotemark{i} 
$\Psi _{\mathrm{MD}}$ Clumping Model Run}}\\
\tableline
(0.768,14) & 311.7 & 0.369 & 0.66 & 1.12 & 70.37 & 36.91 & 14.03 & 
1.324 & -0.845 & 3.63 & 272.4 \\
\tableline
\multicolumn{12}{c}{{\it Comparison Values from 
Best-Fit \tablenotemark{j} 
flat $\Lambda{\mathrm{CDM}}$ Model 
$(\Omega _{\Lambda} = 0.713 = 1 - \Omega _\mathrm{M})$}} 
\\
\tableline
\nodata & 311.9 & 0.380 & \nodata & 1.0 & 69.96 & 69.96 & 13.64 & 
0.287 & -0.713 & 1.0 & 285.4 \\
\tableline
\multicolumn{12}{c}{{\it Comparison Values from 
Best-Fit \tablenotemark{k} flat SCDM Model 
$(\Omega _{\Lambda} = 0$, $\Omega _\mathrm{M} = 1)$}}\\
\tableline
\nodata & 608.2 & 3.4E-22 & \nodata & 1.0 & 61.35 & 61.35 & 10.62 & 
1.0 & 0.0 & 1.0 & 287.3 \\
\tableline
\end{tabular}
}
\tablenotetext{a}{$\chi^{2}_{\mathrm{Fit}}$ computed versus the 
SCP Union SNe data \citep{KowalRubinSCPunion}. As discussed 
in the text, the $\Psi _{\mathrm{MD}}$ and $\Psi _{\mathrm{Lin}}$ 
runs used here for study are not absolutely optimized 
(i.e., $\chi^{2}_{\mathrm{Fit}}$ minimized) with respect to 
their model parameters, ($\Psi _{0},z_\mathrm{init}$), as the 
$\Lambda{\mathrm{CDM}}$ model quoted here has been fully 
optimized with respect to $\Omega _{\Lambda}$.} 
\tablenotetext{b}{Each likelihood probability $P_{\mathrm{Fit}}$ 
is derived from the corresponding 
$\chi^{2}_{\mathrm{Fit}}$ using the $\chi^{2}_{N_{\mathrm{DoF}}}$ 
distribution with $N_{\mathrm{DoF}}$ degrees of freedom, where 
$N_{\mathrm{DoF}} = 304$ for our $\Psi _{\mathrm{Lin}}$ and 
$\Psi _{\mathrm{MD}}$ clumping models, $N_{\mathrm{DoF}} = 305$ 
for the flat $\Lambda{\mathrm{CDM}}$ model, 
and $N_{\mathrm{DoF}} = 306$ for flat SCDM.}
\tablenotetext{c}{The integrated (Newtonian) gravitational 
perturbation potential at $t_{0}$, as computed via 
Equation~\ref{EqnItotIntegration}.}
\tablenotetext{d}{Each $z^\mathrm{Obs}$ quoted here corresponds 
to $z^\mathrm{FRW} \equiv 1$.}
\tablenotetext{e}{The $H^\mathrm{Obs}_{0}$ value 
(given here in $\mathrm{km} ~ \mathrm{s}^{-1} \mathrm{Mpc}^{-1}$) 
for each run is found by minimizing its $\chi^{2}_{\mathrm{Fit}}$ 
with respect to the SCP Union SNe data.}
\tablenotetext{f}{Each $H^\mathrm{FRW}_{0}$ is computed relative 
to the corresponding optimized $H^\mathrm{Obs}_{0}$ value 
for that run.}
\tablenotetext{g}{All $t^\mathrm{Obs}_{0}$ values are listed here 
in GYr, and computed assuming {\it no radiation} 
(i.e., $\Omega _{R} \equiv 0$).}
\tablenotetext{h}{All $\Omega^\mathrm{FRW}_\mathrm{M}$ values given 
here for the $\Psi _{\mathrm{Lin}}$ and $\Psi _{\mathrm{MD}}$ models 
are normalized to $\Omega^\mathrm{Obs}_\mathrm{M} \equiv 0.27$.}
\tablenotetext{i}{This ``Semi-Optimized" $\Psi _{\mathrm{MD}}$ 
run is one chosen from a class of low-$\chi^{2}_{\mathrm{Fit}}$ 
models, as described in the text.}
\tablenotetext{j}{``Best-Fit" for the flat 
$\Lambda{\mathrm{CDM}}$ Model refers here to an optimization 
over $\Omega _{\Lambda}$ and $H^\mathrm{Obs}_{0}$.}
\tablenotetext{k}{``Best-Fit" for the flat 
SCDM Model refers here to an optimization 
over $H^\mathrm{Obs}_{0}$.}
\end{center}
\end{table}

\end{document}